\documentclass[longauth]{aa}  
\usepackage{graphicx}
\usepackage[varg]{txfonts}
\usepackage{hyperref}
\usepackage[pdftex,dfipsnames]{xcolor}
\usepackage{color}
\usepackage{xargs}
\usepackage{blindtext}
\usepackage{stackengine}
\usepackage[colorinlistoftodos,prependcaption,textsize=tiny]{todonotes}
\newcommandx{\change}[2][1=]{\todo[linecolor=blue,backgroundcolor=blue!25,bordercolor=blue,#1]{#2}}
\newcommand{\Qphi}{$Q_{\rm{\phi}}$}
\newcommand{\Uphi}{$U_{\rm{\phi}}$}
\newcommand{\Mjup}{$\mathrm{M_{Jup}}$}
\newcommand{\Rjup}{$\mathrm{R_{Jup}}$}
\newcommand{\Rsol}{$\mathrm{R_{\odot}}$}
\newcommand{\Msol}{$\mathrm{M_{\odot}}$}
\newcommand{\Lsol}{$\mathrm{L_{\odot}}$}
\newcommand{\mum}{$\mathrm{\mu}$m}

\newcommand{\comment}[1]{\ignorespaces}
\hypersetup{backref=true,pagebackref=true,hyperindex=true,breaklinks=true,colorlinks=true,urlcolor=blue,linkcolor=blue,citecolor=blue,pagecolor=red,bookmarks=true,bookmarksopen=true}

\hypersetup{final}
\begin{document} 

\title{Discovery of a planetary-mass companion within the gap of the transition disk around PDS~70 
   \thanks{
Based on observations performed with ESO Telescopes at the Paranal Observatory under programmes 095.C-0298, 095.C-0404, 096.C-0333, 097.C-0206, 097.C-1001, 099.C-0891.}
}
\titlerunning{Discovery of a planetary-mass companion around PDS~70}
\authorrunning{M. Keppler et al. }

\author{M. Keppler\inst{1}
         \and M. Benisty\inst{2,3}     
         \and A. M\"{u}ller\inst{1}
         \and Th. Henning\inst{1}
         \and R. van Boekel\inst{1}
         \and F. Cantalloube\inst{1}
         \and C. Ginski\inst{4,32} 
         \and R.G. van Holstein\inst{4}   
         \and A.-L. Maire\inst{1}
         \and A. Pohl\inst{1}
% ...
}

\institute{
$^{1}$  Max Planck Institute for Astronomy, K\"onigstuhl 17, D-69117 Heidelberg, Germany\\
$^{2}$  Univ. Grenoble Alpes, CNRS, IPAG, F-38000 Grenoble, France. \\
$^{3}$  Unidad Mixta Internacional Franco-Chilena de Astronom\'ia, CNRS/INSU UMI 3386 and Departamento de Astronom\'ia, Universidad de Chile, Casilla 36-D, Santiago, Chile\\
$^{4}$  Leiden Observatory, Leiden University, PO Box 9513, 2300 RA Leiden, The Netherlands\\
$^{5}$  Department of Physics, University of Oxford, Oxford, UK \\
$^{6} $ LESIA, Observatoire de Paris, Universit\'e PSL, CNRS, Sorbonne Universit\'e, Univ. Paris Diderot, Sorbonne Paris Cit\'e, 5 place Jules Janssen, 92195 Meudon, France
$^{7}$  INAF - Osservatorio Astronomico di Padova, Vicolo della Osservatorio 5, 35122, Padova, Italy\\
$^{8}$  Aix Marseille Univ, CNRS, CNES, LAM, Marseille, France\\
$^{9}$  CRAL, UMR 5574, CNRS, Universit\'e de Lyon, Ecole Normale Sup\'erieure de Lyon, 46 All\'ee d'Italie, F-69364 Lyon Cedex 07, France\\
$^{10}$ Institut f\"ur Astronomie und Astrophysik, Eberhard Karls Universit\"at T\"ubingen, Auf der Morgenstelle 10, 72076 T\"ubingen, Germany. \\
$^{11}$ Physikalisches Institut, Universit\"at Bern, Gesellschaftsstrasse 6, 3012 Bern, Switzerland\\
$^{12}$ Institute for Particle Physics and Astrophysics, ETH Zurich, Wolfgang-Pauli-Strasse 27, 8093 Zurich, Switzerland\\  
$^{13}$ N\'ucleo de Astronom\'ia, Facultad de Ingenier\'ia y Ciencias, Universidad Diego Portales, Av. Ejercito 441, Santiago, Chile\\
$^{14}$ Escuela de Ingenier\'ia Industrial, Facultad de Ingenier\'ia y Ciencias, Universidad Diego Portales, Av. Ejercito 441, Santiago, Chile \\
$^{15}$ University Observatory, Faculty of Physics, Ludwig-Maximilians-Universit\"at M\"unchen, Scheinerstr. 1, 81679 Munich, Germany\\
$^{16}$ Jet Propulsion Laboratory, California Institute of Technology, Pasadena, California 91109, USA\\
$^{17}$ Kavli Institute For Theoretical Physics, University of California, Santa Barbara, CA 93106, USA\\
$^{18}$ European Southern Observatory (ESO), Alonso de C\'{o}rdova 3107, Vitacura, Casilla 19001, Santiago, Chile\\
$^{19}$ Rice University, Department of Physics and Astronomy, Main Street, 77005 Houston, USA\\
$^{20}$ Department of Astronomy, Stockholm University, AlbaNova University Center, 106 91 Stockholm, Sweden \\
$^{21}$ Institute of Astronomy, Madingley Road, Cambridge CB3 0HA, UK\\
$^{22}$ Universit\'e C\^ote d'Azur, Observatoire de la C\^ote d'Azur, CNRS, Laboratoire Lagrange, France\\
$^{23}$ Instituto de F\'isica y Astronom\'ia, Facultad de Ciencias, Universidad de Valpara\'iso, Av. Gran Breta\~na 1111, Playa Ancha, Valpara\'iso, Chile \\ 
$^{24}$ N\'ucleo Milenio Formaci\'on Planetaria - NPF, Universidad de Valpara\'iso, Av. Gran Breta\~{n}a 1111, Valpara\'iso, Chile\\
$^{25}$ Max-Planck-Institute for Astronomy, Bonn, Germany\\
$^{26}$ Universidad de Chile, Departamento de Astronom\'ia, Camino El Observatorio 1515, Las Condes, Santiago, Chile\\
$^{27}$ Department of Astronomy/Steward Observatory, University of Arizona, 933 North Cherry Avenue, Tucson, AZ 85721, USA\\
$^{28}$ UMI-FCA, CNRS/INSU, France (UMI 3386), and Dept. de Astronom\'ia, Universidad de Chile, Santiago, Chile\\
$^{29}$ Monash Centre for Astrophysics (MoCA) and School of Physics and Astronomy, Monash University, Clayton Vic 3800, Australia\\
$^{30}$ Geneva Observatory, University of Geneva, Chemin des Mailettes 51, 1290 Versoix, Switzerland\\
$^{31}$ Institute for Astronomy, University of Hawaii at Manoa, Honolulu, HI 96822, USA\\
$^{32}$ Anton Pannekoek Institute for Astronomy, Science Park 904, NL-1098 XH Amsterdam, The Netherlands\\
$^{33}$ European Southern Observatory (ESO), Karl-Schwarzschild-Str. 2, 85748 Garching, Germany\\
$^{34}$ INCT, Universidad De Atacama, calle Copayapu 485, Copiap\'{o}, Atacama, Chile \\ 
$^{35}$ INAF - Osservatorio Astrofisico di Arcetri, Largo E. Fermi 5, I-50125 Firenze, Italy\\
$^{36}$ Zentrum f\"ur Astronomie der Universit\"at Heidelberg, Landessternwarte, K\"onigstuhl 12, 69117 Heidelberg, Germany \\
$^{37}$ INAF - Osservatorio Astronomico di Capodimonte, Salita Moiariello 16, 80131 Napoli, Italy\\ 
$^{38}$ Geneva Observatory, University of Geneva, Chemin des Mailettes 51, 1290 Versoix, Switzerland \\
$^{39}$ ONERA -- The French Aerospace Lab, F-92322 Ch\^atillon, France\\
$^{40}$ NOVA Optical Infrared Instrumentation Group, Oude Hoogeveensedijk 4, 7991 PD Dwingeloo, The Netherlands\\
$^{41}$ Konkoly Observatory, Research Centre for Astronomy and Earth Sciences, Hungarian Academy of Sciences, P.O. Box 67, H-1525 Budapest, Hungary\\
$^{42}$ The University of Michigan, Ann Arbor, MI 48109, USA\\
}

\date{Received --; accepted --}

\abstract
    {Young circumstellar disks are the birthplaces of planets. Their study is of prime interest to understand the physical and chemical conditions under which planet formation takes place. Only very few detections of planet candidates within these disks exist, and most of them are currently suspected to be disk features.}
    {In this context, the transition disk around the young star PDS~70 is of particular interest, due to its large gap identified in previous observations, indicative of ongoing planet formation. We aim to search for the presence of an embedded young planet and search for disk structures that may be the result of disk-planet interactions and other evolutionary processes.}
    {We analyse new and archival near-infrared (NIR) images of the transition disk PDS~70 obtained with the VLT/SPHERE, VLT/NaCo and Gemini/NICI instruments in polarimetric differential imaging (PDI) and angular differential imaging (ADI) modes.  }
    {We detect a point source within the gap of the disk at about 195 mas ($\sim$22 au) projected separation. The detection is confirmed at five different epochs, in three filter bands and using different instruments. The astrometry results in an object of bound nature, with high significance. The comparison of the measured magnitudes and colours to evolutionary tracks suggests that the detection is a companion of planetary mass. The luminosity of the detected object is consistent with that of an L-type dwarf, but its IR colours are redder, possibly indicating the presence of warm surrounding material. Further, we confirm the detection of a large gap of $\sim$54 au in size within the disk in our scattered light images, and detect a signal from an inner disk component. We find that its spatial extent is very likely smaller than $\sim$17 au in radius, and its position angle is consistent with that of the outer disk. The images of the outer disk show evidence of a complex azimuthal brightness distribution which is different at different wavelengths and may in part be explained by Rayleigh scattering from very small grains. }
    {The detection of a young protoplanet within the gap of the transition disk around PDS~70 opens the door to a so far observationally unexplored parameter space of planetary formation and evolution. Future observations of this system at different wavelengths and continuing astrometry will allow us to test theoretical predictions regarding planet-disk interactions, planetary atmospheres and evolutionary models. }


\maketitle


\section{Introduction}
More than two decades after the first detection of an extrasolar planet \citep{1995Natur.378..355M}, we are facing an extraordinary diversity of planetary system architectures \citep{2015ARA&A..53..409W}. Exploring which of the properties of these systems are imprinted by the initial conditions of the disks and which develop through a variety of dynamical interactions is crucial for understanding the planet population. It is therefore of high importance to study planets and their environments at the stage during which these objects are formed. \\
Transition disks (TDs) are of key interest in this context, as many of them are believed to bear direct witness to the process of planet formation. These objects were initially identified by a significantly reduced near-infrared (NIR) excess compared to the median spectral energy distribution (SED) of young stars with disks \citep{1989AJ.....97.1451S}. In the meantime, recent high-resolution imaging observations of TDs at different wavelengths have revealed large gaps \citep[e.g.][]{2010ApJ...718L..87T,2011ApJ...742L...5A,2014ApJ...781...87A,2015ApJ...798...85P,2017A&A...605A..34P,2018MNRAS.475L..62H}, azimuthal asymmetries \citep[e.g.][]{2013Natur.493..191C,2013Sci...340.1199V,2014ApJ...783L..13P}, spirals \citep[e.g.][]{2012ApJ...748L..22M,2013ApJ...762...48G,2017ApJ...840...32T}, and multiple rings \citep[e.g.][]{2016A&A...595A.114D,2016A&A...595A.112G,2016ApJ...820L..40A,2017ApJ...837..132V,2017ApJ...850...52P,2018MNRAS.474.5105B}, as well as shadowed regions and brightness dips, some of them varying with time \citep[e.g.][]{2015A&A...584L...4P,2017ApJ...849..143S,2017ApJ...835..205D}. In many cases, cavities and substructures are present within these young, gas-rich disks, which have often been interpreted as tracers of ongoing planet formation and are suspected to originate from planet-disk interactions (see \citealp{2014prpl.conf..497E} for a review). \\
\begin{table}[b]
\caption{Photometry and stellar parameters of the PDS~70 system used in this study.}
\label{table:stellar_properties}
\begin{tabular}{lc}
\hline \hline
Parameter                       &Value          \\
\hline 
$V$                              &12.233$\pm$0.123 mag          $^{(a)}$ \\
$J$                              &9.553$\pm$0.024 mag           $^{(b)}$ \\
$H$                              &8.823$\pm$0.04  mag           $^{(b)}$ \\
$K$                              &8.542$\pm$0.023 mag           $^{(b)}$ \\
$L'$                             &7.913$\pm$0.03  mag           $^{(c)}$ \\ 
Distance                         &$113.43\pm0.52$\ pc           $^{(d)}$ \\ 
$\boldsymbol{\mu_{\alpha}\times}\mathrm{cos}\boldsymbol{(\delta)}$ &-29.66$\pm$0.07 mas/yr $^{(d)}$ \\
$\boldsymbol{\mu_{\delta}}$      &-23.82$\pm$0.06 mas/yr        $^{(d)}$ \\ 
Spectral type                    &K7                            $^{(e)}$ \\
$T_{\mathrm{eff}}$               &3972$\pm$36 K                 $^{(e)}$ \\ 
Radius                           &1.26$\pm$0.15 \Rsol           $^{(f)}$ \\              
Luminosity                       &0.35$\pm$0.09 \Lsol           $^{(f)}$ \\              
Mass                             &0.76$\pm$0.02 \Msol           $^{(g)}$ \\
Age                              &$5.4\pm1.0$ Myr               $^{(g)}$ \\ 
Visual extinction $A_V$ &0.05 $^{+0.05}_{-0.03}$ mag            $^{(g)}$ \\ 
\hline
\end{tabular}
\tablefoot{\footnotesize{(a) \cite{2015AAS...22533616H}; (b) \cite{2003yCat.2246....0C}; (c) $L'$-band magnitude obtained by logarithmic interpolation between the WISE W1 and W2-band magnitudes from \cite{2014yCat.2328....0C}; (d) \cite{2016AA...595A...1G,2018arXiv180409365G}; (e) \cite{2016MNRAS.461..794P}}; (f) derived from \cite{2016MNRAS.461..794P}, scaled to a distance of 113.43 pc. (g) \cite{mueller2018} }
\end{table}
The characterisation of TDs is therefore of prime interest to understand the physical and chemical conditions under which planet formation takes place. Observations of forming planets within those disks are extremely challenging as the disk often outshines the planet, requiring observations at high contrast and angular resolution. Such detections have been reported for a few targets only; namely HD~100546 \citep{2013ApJ...766L...1Q,2014ApJ...791..136B,2015ApJ...807...64Q,2015ApJ...814L..27C}, LkCa15 \citep{2012ApJ...745....5K,2015Natur.527..342S}, HD~169142 \citep{2013ApJ...766L...2Q,2014ApJ...792L..22B,2014ApJ...792L..23R}, and MWC 758 \citep{2018A&A...611A..74R}, but most of the detections are presently being challenged \citep[e.g.][Sissa et al. subm.]{2017AJ....153..244R,2018MNRAS.473.1774L}. The point sources identified as planet candidates in high-contrast imaging observations with the angular differential technique could be confused with brightness spots in asymmetric disks.  \\ 

The aim of this paper is to study the pre-main sequence star PDS~70 (V* V1032 Cen) with high-contrast imaging. PDS~70 is a K7-type member of the Upper Centaurus-Lupus subgroup (UCL), part of the Scorpius-Centaurus association \citep{2006A&A...458..317R,2016MNRAS.461..794P}, at a distance of $113.43\pm0.52$ pc \citep{2016AA...595A...1G,2018arXiv180409365G}. Strong lithium absorption and the presence of a protoplanetary disk provide evidence of a young age ($\lesssim$ 10 Myr) for PDS~70 \citep{GregorioHetem:2002co,2004ApJ...600..435M}, which is confirmed through the comparison to theoretical model isochrones \citep[][]{2016MNRAS.461..794P}. Taking into account the recent Gaia DR2 data release, which provides for the first time a stellar parallax and therefore a first precise distance estimation for PDS~70, we derive an age of 5.4$\pm$1.0 Myr \citep[see][and Appendix \ref{stellar_parameters}]{mueller2018}. The stellar properties are summarised in Table~\ref{table:stellar_properties}. \\
The first evidence of the presence of a disk was provided by the measurement of infrared (IR) excess in the SED \citep{GregorioHetem:2002co,2004ApJ...600..435M}. Modelling of the SED predicted that PDS~70 hosts a disk whose inner region is substantially cleared of dust, but with a small optically thick inner disk emitting in the NIR \citep{2012ApJ...758L..19H,2012ApJ...760..111D}. The first spatially resolved image of the disk was obtained by \cite{2006A&A...458..317R} together with the detection of a companion candidate at 2.2" to the North of the host star using the NaCo instrument in the Ks filter. The companion candidate was later identified as a background source \citep{2012ApJ...758L..19H}. The gap was resolved in NIR scattered light observations using Subaru/HiCIAO \citep{2012ApJ...758L..19H} as well as in the dust continuum observed with the Submillimeter Array (SMA) at 1.3 mm \citep{2015ApJ...799...43H} and most recently with the Atacama Large Millimeter/submillimeter Array (ALMA) at 870 $\mu$m \citep{2018ApJ...858..112L}. The latter dataset showed evidence of the presence of an inner disk component extending out to a radius of several au that appears to be depleted of large grains. These recent ALMA observations also show that the surface brightness at sub-millimeter wavelengths, tracing large dust grains, peaks at a radial distance further out ($\sim$ 0.7\arcsec, see Fig. 5 of \citealp{2018ApJ...858..112L}) than the location of the cavity wall\footnote{Following \cite{2013A&A...560A.111D}, the NIR cavity wall is defined as the radial position where the flux equals half the value between the minimum flux at the bottom of the gap and the flux maximum at the wall. } ($\sim$ 0.39\arcsec; see \citealp{2012ApJ...758L..19H}, and Fig.~\ref{cuts} of this work) measured in scattered light imaging, tracing small micron-sized dust grains in the disk surface layer. This segregation in the spatial distribution of dust grains with various sizes is thought to be generated by a radial pressure gradient in the disk, and has already been observed in several systems \citep[e.g.][]{2015A&A...584L...4P,2018MNRAS.475L..62H}. Several mechanisms have been proposed to be able to create such pressure bumps, such as magnetohydrodynamic effects or planets carving the gap \citep[e.g.][]{2015A&A...573A...9P,2016A&A...596A..81P}. The combination of the presence of an inner disk with the spatial segregation of dust grains makes PDS~70 a prime candidate for hosting planets that are carving the gap. \\

In this paper, we present an extensive dataset on PDS~70 using the high-contrast imager SPHERE \citep[Spectro-Polarimetric High-contrast Exoplanet REsearch;][]{2008SPIE.7014E..18B}, complemented with datasets obtained with VLT/NaCo and Gemini/NICI. Our observations include both angular differential imaging and polarimetric observations at multiple wavelengths and epochs, from the optical to the NIR, covering a time period of more than four years. We report on the robust detection of a point source within the gap of the disk, which is interpreted as a planetary-mass companion. Furthermore, we detect for the first time scattered light emerging from the inner disk. We analyse the well-known outer disk with respect to its morphological appearance and investigate its structure with a radiative transfer model. \\
This paper is structured as follows: the observational setup and reduction strategy of the different datasets are described in Sect. \ref{sect:obs}. The analysis of the disk is presented in Sect. \ref{sect:disk}, and Sect. \ref{modeling} describes our modelling efforts. Section \ref{sect:CC} is dedicated to the analysis of the point source, and our findings are summarised in Sect. \ref{sect:conclusions}. 

\begin{table*}[t]
\centering
\caption{Observing log of data used within this study. }
\label{table:obslog}
\begin{tabular}{llllllcccccc}
\hline \hline
Date      &ProgID       & Instrument  & Mode$^{(a)}$   &Filter& R$^{(b)}$ &Coronagraph &  $\Delta$$\theta^{ (c)}$ & DIT$^{(d)}$ [s]& $\Delta t^{(e)}$[min]& $\epsilon$["]$^{ (f)}$ \\
\hline
2012-03-31&GS-2012A-C-3 &NICI$^{(g)}$ &ADI   &L'   &-- & --           &99.4$^{\circ}$  &0.76     &118&--       & \\  
2015-05-03&095.C-0298(A)&IRDIS        &ADI   &H2H3 &-- & ALC\_YJH\_S  &52.0$^{\circ}$  &64.0     &70 &0.7      & \\
2015-05-03&095.C-0298(A)&IFS          &ADI   &YJ   &54& ALC\_YJH\_S   &52.0$^{\circ}$  &64.0     &70 &0.7      & \\
2015-05-31&095.C-0298(B)&IRDIS        &ADI   &H2H3 &-- & ALC\_YJH\_S  &40.8$^{\circ}$  &64.0     &70 &1.1      & \\
2015-05-31&095.C-0298(B)&IFS          &ADI   &YJ   &54& ALC\_YJH\_S   &40.8$^{\circ}$  &64.0     &70 &1.1      & \\
2015-07-09&095.C-0404(A)&ZIMPOL       &PDI   &VBB  &-- & --           &--              &40.0     &114&1.1      & \\ 
2016-03-25&096.C-0333(A)&IRDIS        &PDI   &J    &-- & ALC\_YJ\_S   &--              &64.0     &94 &1.9      & \\ 
2016-05-14&097.C-1001(A)&IRDIS        &ADI   &K1K2 &30&--             &16.9$^{\circ}$  &0.837    &22 &1.0      & \\ 
2016-05-14&097.C-1001(A)&IFS          &ADI   &YJH  &-- &--            &16.9$^{\circ}$  &4.0      &23 &1.0      & \\ 
2016-06-01&097.C-0206(A)&NaCo         &ADI   &L'   &-- & --           &83.7$^{\circ}$  &0.2      &155 &0.5     & \\
2017-07-31&099.C-0891(A)&IRDIS        &PDI   &J    &-- & --           &--              & 2.0     &36 &0.7      & \\
\hline
\hline
\end{tabular}
\tablefoot{$^{(a)}$Observing mode: Angular Differential Imaging (ADI) or Polarimetric Differential Imaging (PDI); $^{(b)}$Spectral resolution; $^{(c)}$total field rotation, after frame selection; $^{(d)}$detector integration time; $^{(e)}$total time on target (including overheads); $^{(f)}$mean MASS/DIMM seeing; $^{(g)}$archival data, published in \cite{2012ApJ...758L..19H}.  
}
\end{table*}

\section{Observations and data reduction}\label{sect:obs}
This section gives an overview of the observations and data-reduction strategy of the datasets used in this study. Table \ref{table:obslog} summarises the observation setups and conditions. 

\begin{figure*}[tbh]
    \begin{centering}
    \begin{minipage}[t]{1.0\textwidth}
    \includegraphics[width=0.32\textwidth]{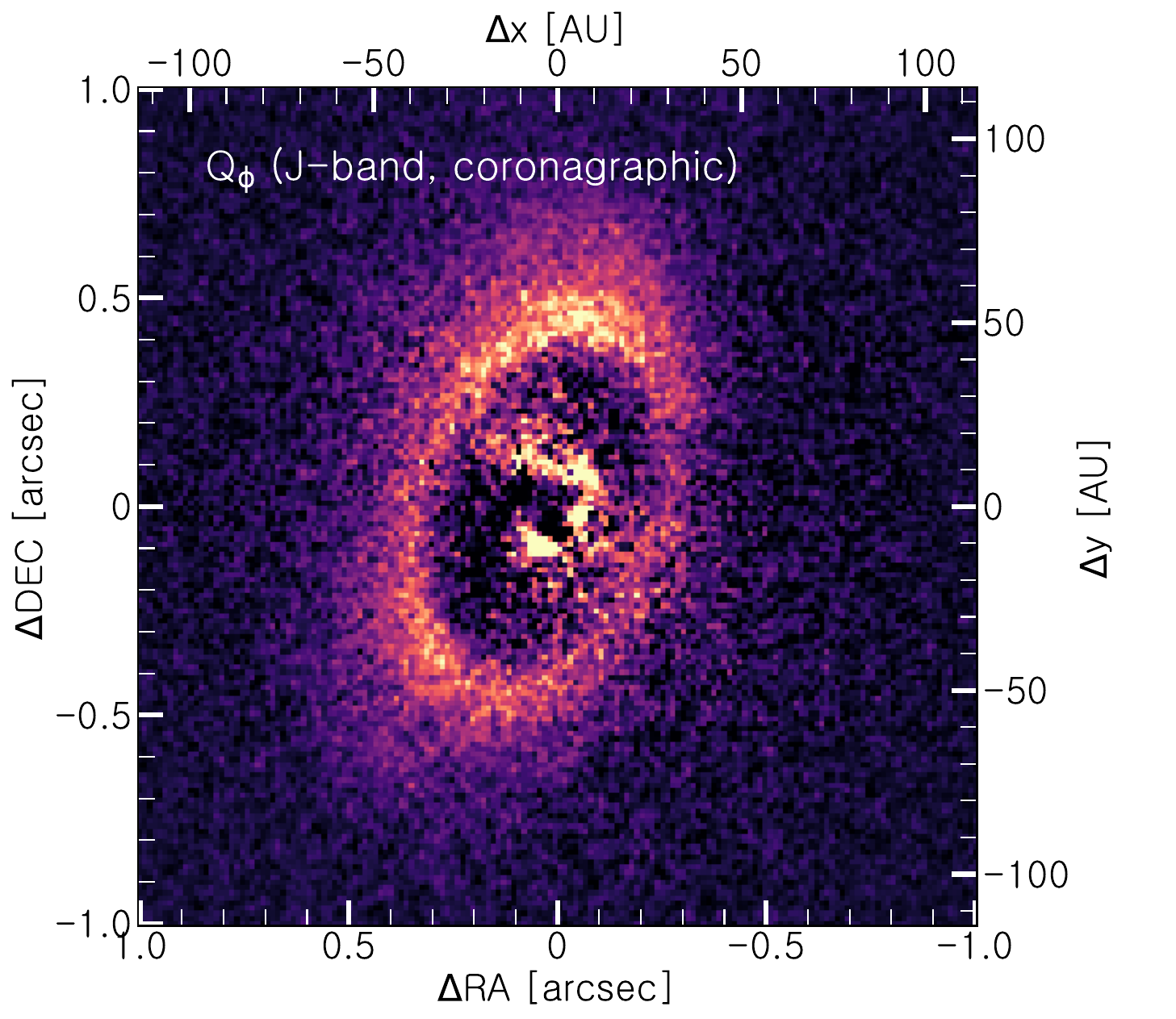}
    \includegraphics[width=0.32\textwidth]{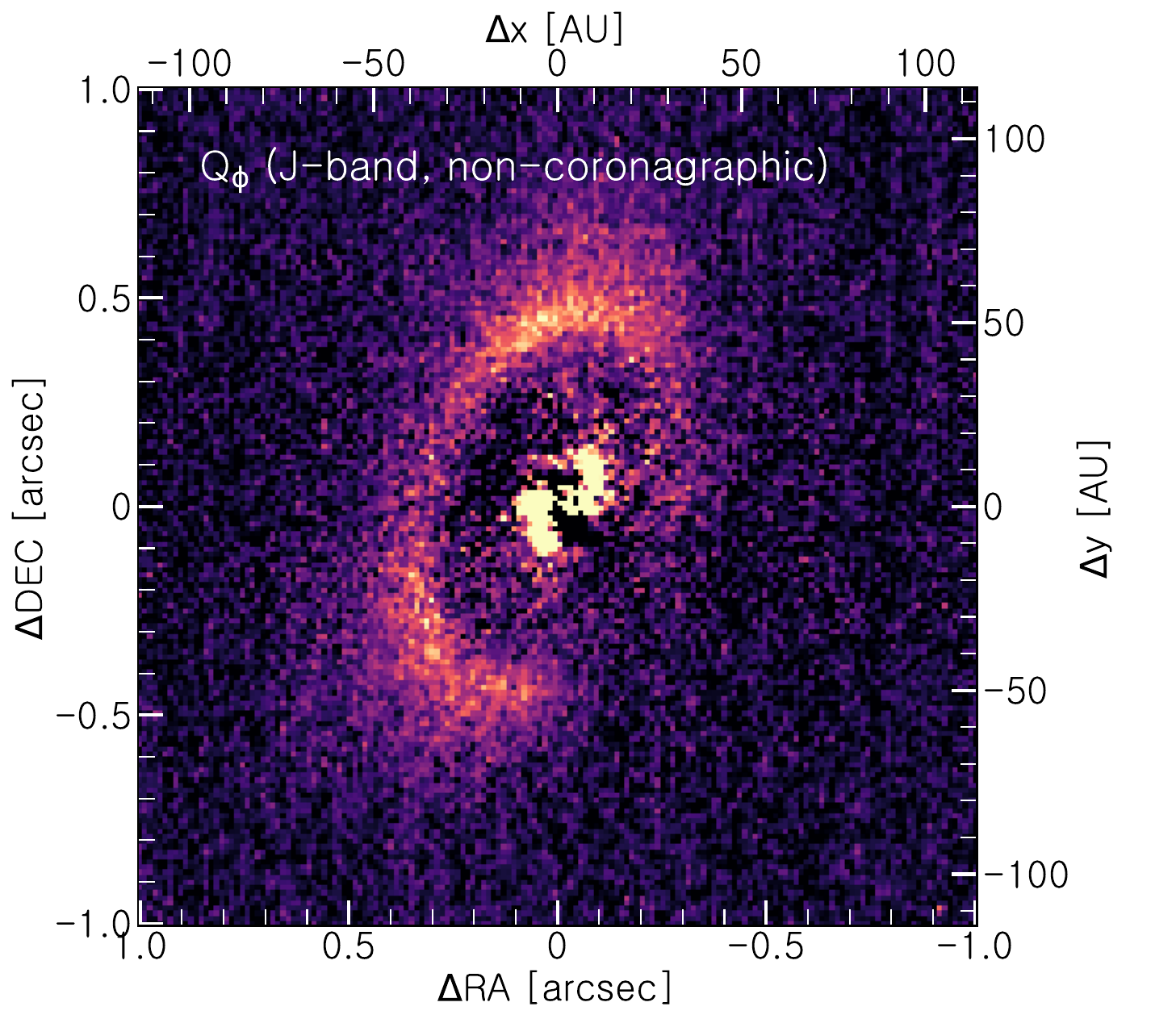}
    \includegraphics[width=0.36\textwidth]{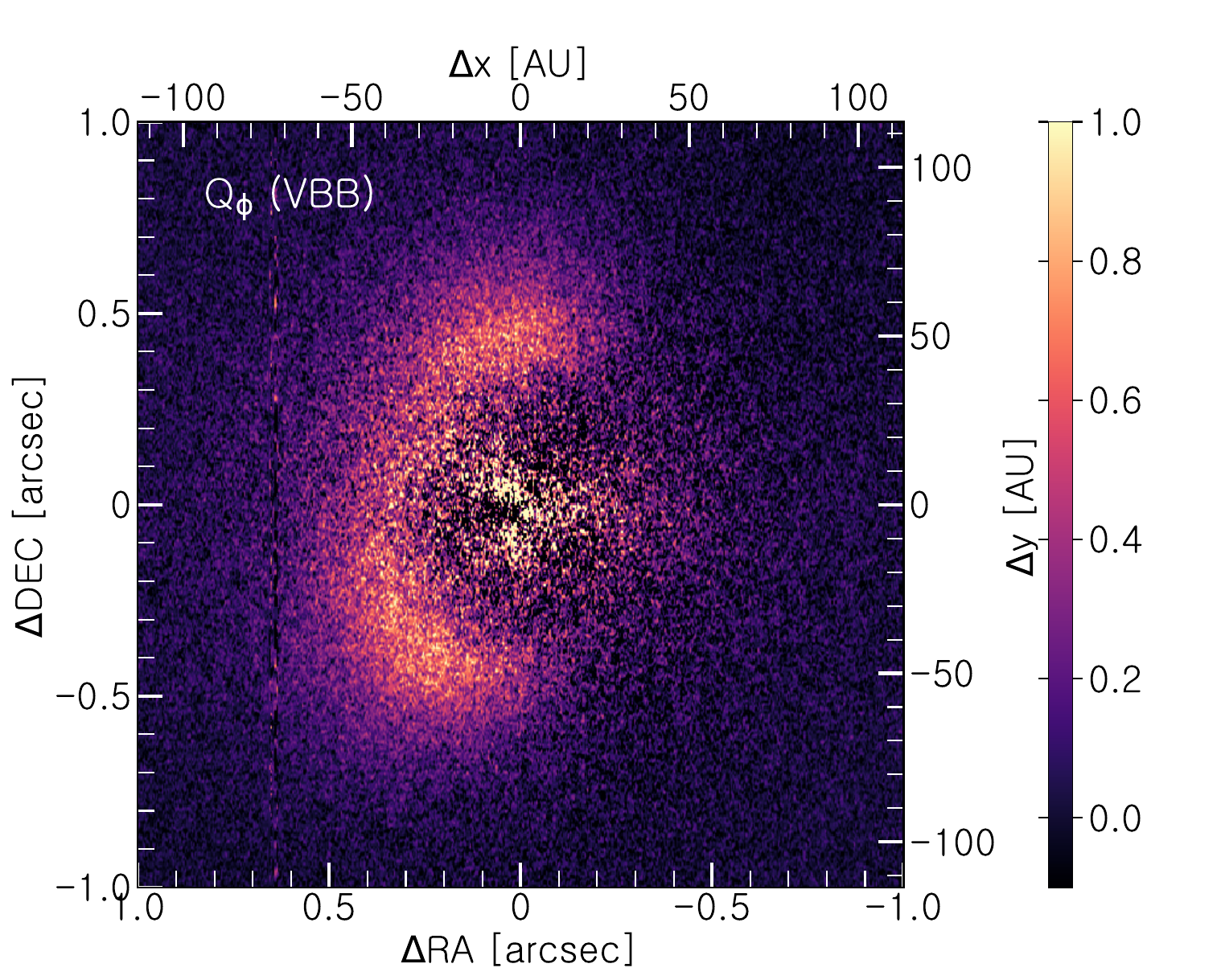}
    \end{minipage}
    \begin{minipage}[t]{1.0\textwidth}
    \includegraphics[width=0.32\textwidth]{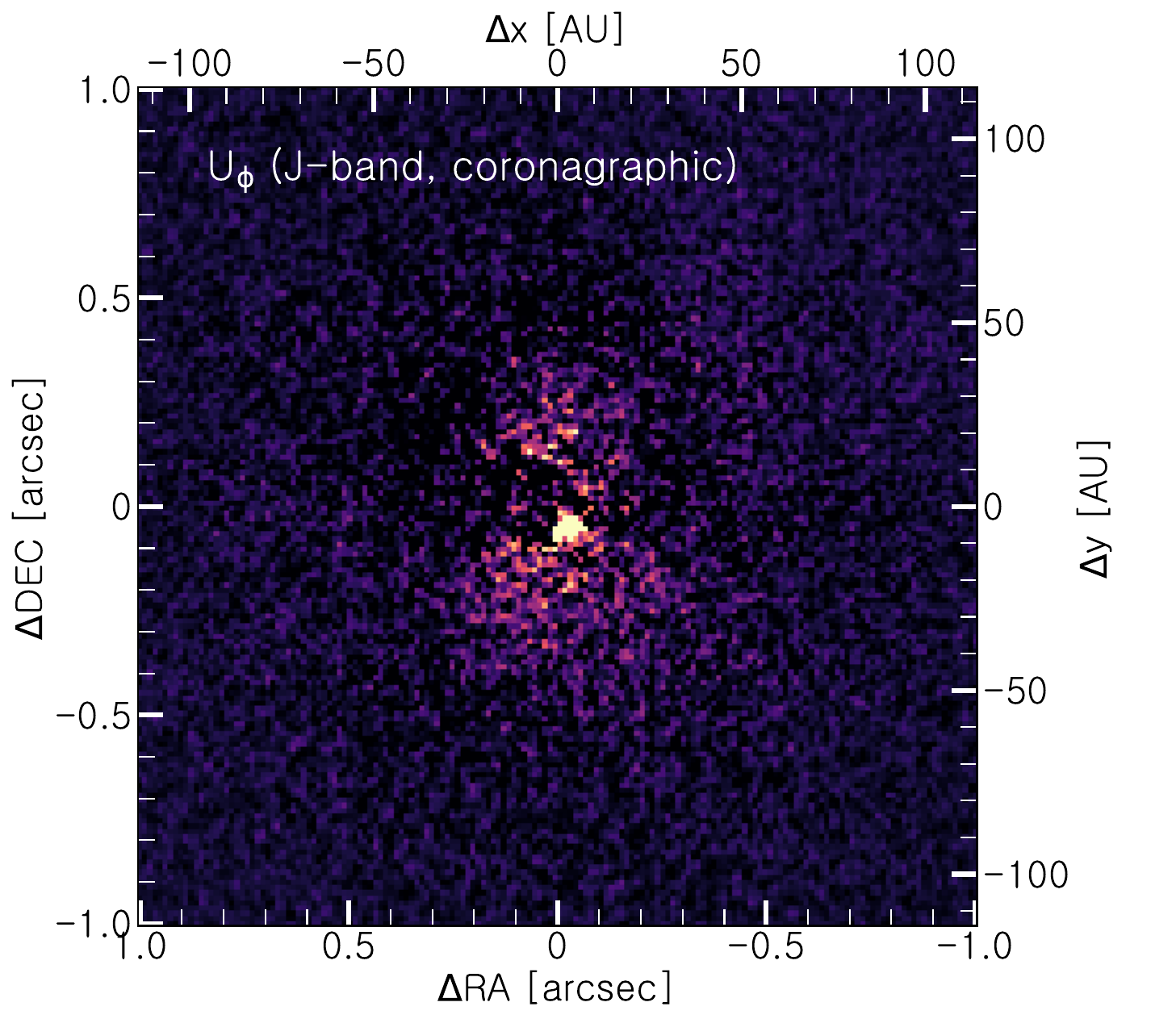}
    \includegraphics[width=0.32\textwidth]{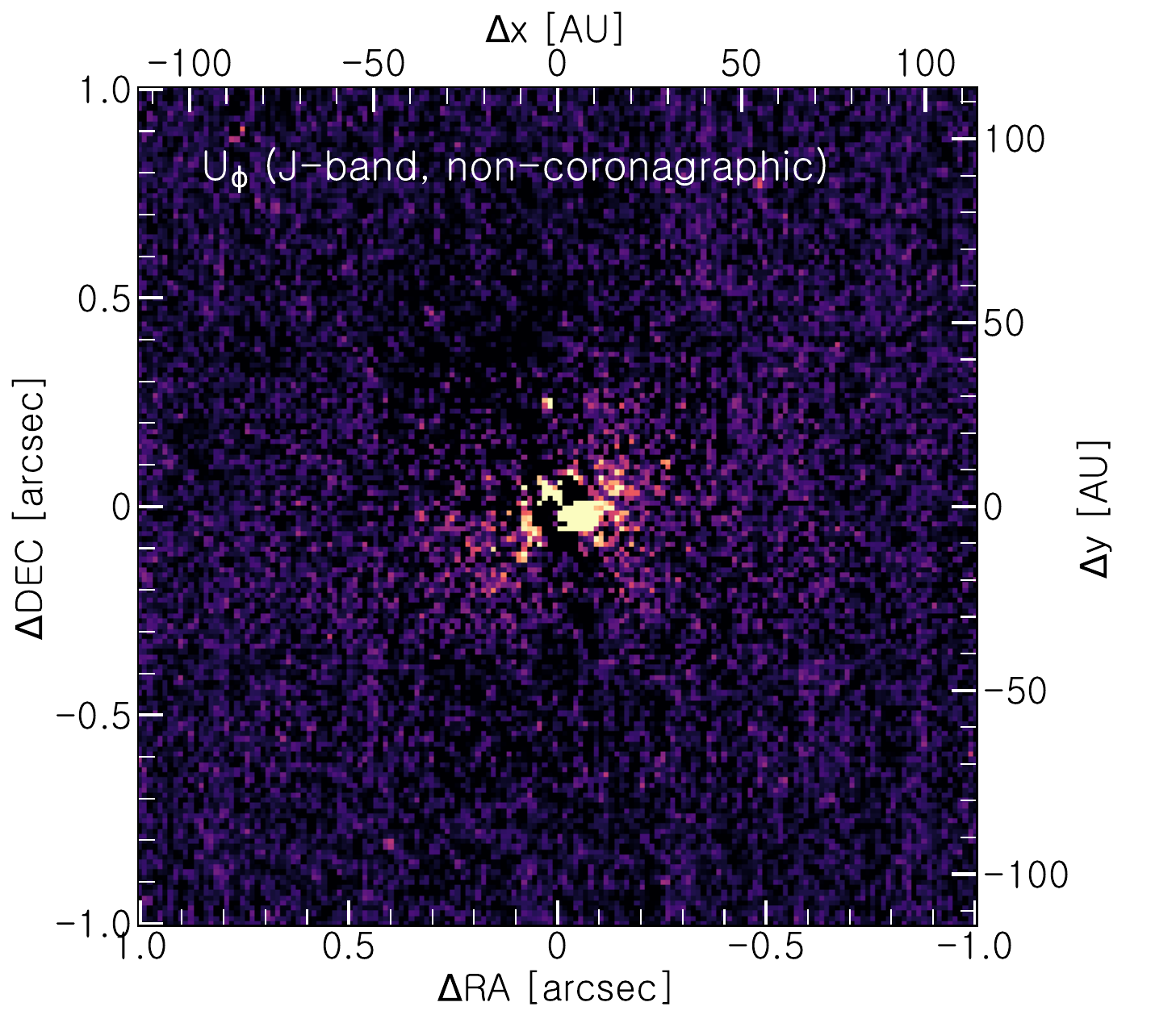}
    \includegraphics[width=0.36\textwidth]{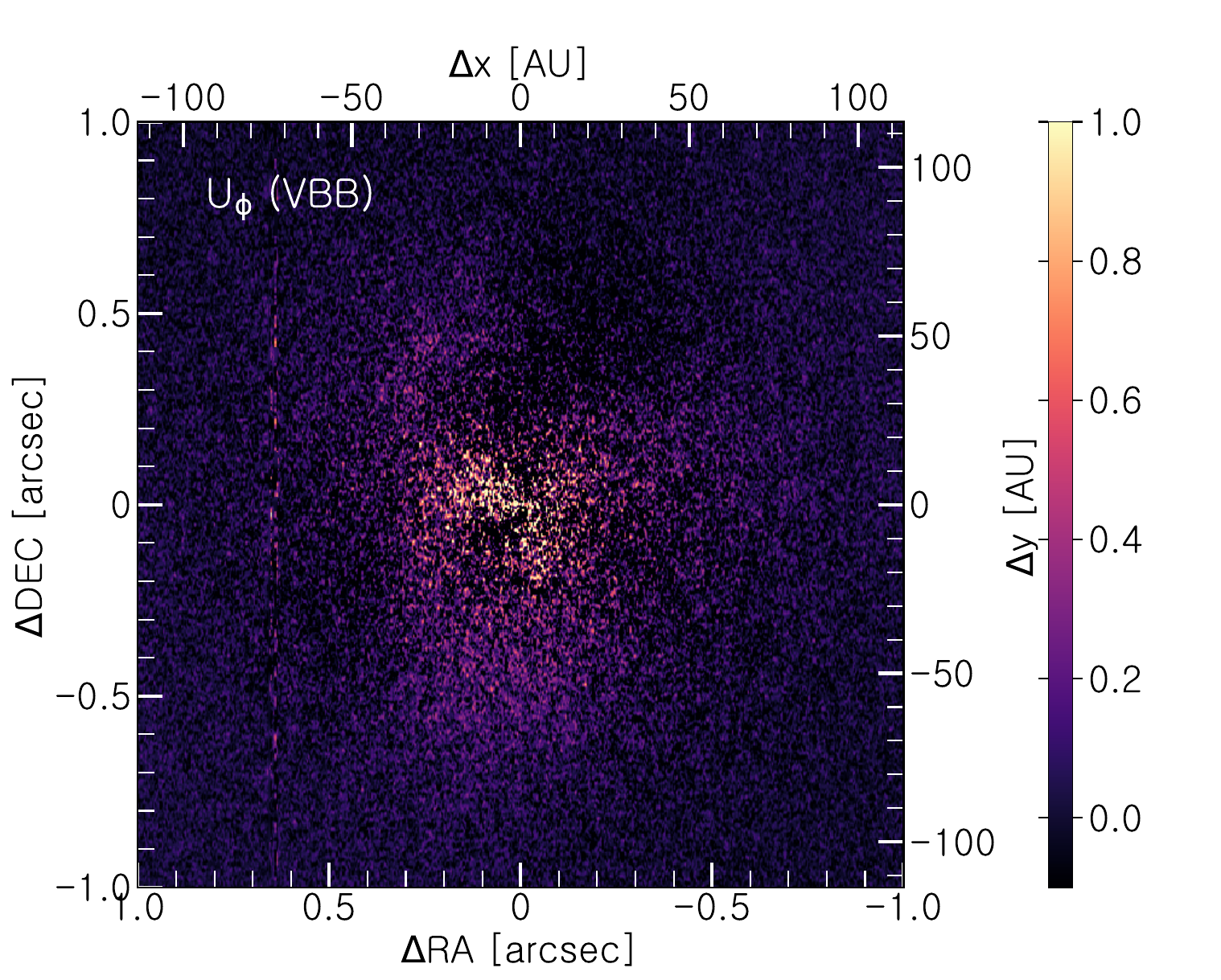}
    \end{minipage}
    \end{centering}
   \caption{SPHERE PDI observations. The first row shows the \Qphi \ images, the second row the \Uphi \ images. The left and middle columns correspond to the IRDIS J-band observations, taken with coronagraph (March 25, 2016), and without coronagraph (July 31, 2017), after correcting for the instrumental polarisation and subtraction of the central source polarisation. The right column presents the ZIMPOL observations (July 9, 2015). The colour scale was chosen arbitrarily but is the same for each pair of \Qphi\ and \Uphi. We note that negative values are saturated to enable a better visual contrast. North is up and east is to the left.  
}\label{fig:obs_PDI}
\end{figure*}

\subsection{IRDIS polarimetric observations (J-band)}\label{sect:obs_IRDIS_setup}
Polarimetric Differential Imaging (PDI) is a highly effective method to image faint disks close to their host stars by suppressing the stellar signature \citep{2001ApJ...553L.189K}. This method takes advantage of the fact that the light from the star itself is mostly unpolarised, whereas light scattered by dust grains becomes polarised.\\
During the nights of March 25, 2016, and July 31, 2017, we obtained two datasets of PDS~70 with the IRDIS instrument \citep{2008SPIE.7014E..3LD} in PDI mode \citep[][]{2014SPIE.9147E..1RL} using the J-band ($\lambda_{\mathrm{J}}$=1.25 $\mu$m).  
In this mode, the beam is split into two beams with orthogonal polarisation states. The direction of polarisation to be measured can be tuned with a half-wave plate (HWP). In our observation setup, one polarimetric cycle consisted of rotating the HWP to four different angles in steps of $\mathrm{22.5^{o}}$. \\
During the first epoch (March 25, 2016) we employed an apodized Lyot coronagraphic mask \citep[N\_ALC\_YJ\_S, diameter $\sim$145 mas; ][]{2009A&A...495..363M,2011ExA....30...39C}.  
We obtained eight polarimetric cycles with two long exposures (64 seconds) per HWP position. Near the end of the sequence, the seeing degraded, and the telescope guiding was lost such that we had to discard the last polarimetric cycle. Before and after each sequence, we obtained a sequence of short, non-saturated images of the star outside the coronagraphic mask. In addition, to enable an accurate determination of the stellar position behind the coronagraph, we took calibration images in which four satellite spots in a square pattern centered on the star were generated by introducing a sinusoidal modulation onto the deformable mirror. \\ 
The second epoch (July 31, 2017) was obtained without a coronagraph in order to provide access to the innermost disk regions. We therefore chose very short exposures (2 secs) to prevent the star from saturating the detector. The observations consisted of seven polarimetric cycles, with 20 exposures taken per HWP position. We aligned the images by fitting two-dimensional (2D) Gaussians to the target star in each frame. \\
The first step of data reduction consisted of dark subtraction, flatfielding, interpolation of bad pixels, and recentring of the frames. The rotation of the HWP to four different angles allowed us to determine a set of four linear polarisation components $\mathrm{Q^+, Q^-, U^+, U^-}$, from which we then obtained the clean Stokes Q and U frames using the double-difference method \citep{2011A&A...531A.102C}, and the normalised double-difference method (van Holstein et al., in prep.) for the non-coronagraphic and coronagraphic dataset, respectively. \\
One of the most important steps in the reduction of polarimetric data is the subtraction of instrumental polarisation (IP). Standard techniques to correct for the IP, as described by \cite{2011A&A...531A.102C}, estimate the IP directly from the data itself. Here, the main assumption is that the central unresolved source, consisting of the stellar photosphere plus thermal emission and/or scattered light from the inner disk, if the latter is present, is unpolarised. This implies that they cannot differentiate between intrinsic polarisation of the central source and the IP. In this case, correcting for the IP would mean subtracting any (physical) polarisation of the central unresolved source. In contrast to that, to correct our measurements for the IP and cross-talk effects, we applied the detailed Mueller matrix model \citep[][van Holstein et al. in prep., de Boer et al. in prep.]{2017SPIE10400E..15V}. This method models the complete optical path that the beam traverses on its way from entering the telescope to the detector, and has already been applied in the analysis of several circumstellar disks observed with SPHERE \citep{2017A&A...605A..34P,2018A&A...610A..13C}. The incident Stokes Q and U images are recovered by solving a set of equations describing every measurement of Q and U for each pixel individually. These Stokes Q and U images correspond to the images as they enter the telescope (star and disk convolved with telescope PSF and noise). With this method, we can therefore make a model prediction of the IP, and correct for the IP alone without subtracting the polarised signal from the central source. After the correction for the IP, any remaining polarised signal at the location of the central source, induced by, for example, unresolved material close to the star, such as an inner disk, would then become visible in the form of a central butterfly pattern in the $Q_{\phi}$ image (see definition of \Qphi \ below) and can affect the signal of the outer disk. This leftover signal can then be chosen to be subtracted following the method by \cite{2011A&A...531A.102C}. On the other hand, it can also be kept, if one desires to study the unresolved inner disk (see van Holstein et al., in prep.). We prepared two reductions, one for which we subtracted the central source polarisation (allowing us to study the morphology of the outer disk; see Sect. \ref{subsect:PDI_outer_disk}), and one where we did not subtract it (for analysing the inner disk; see Sect. \ref{sect:PDI_disk_inner}). We corrected our data for true north and accounted for the instrument anamorphism \citep{2016SPIE.9908E..34M}. We measured a PSF full width at half maximum (FWHM) of $\sim$52 and $\sim$49 milli-arcseconds (mas) on the unsaturated flux frames of the coronagraphic observations, and of the total intensity frame for the non-coronagraphic data, respectively. \\
We transform our Stokes images into polar coordinates ($Q_{\phi}$, $U_{\phi}$), according to the following definition from \cite{2006A&A...452..657S}: 
\begin{equation}\label{Stokes_equation}
\begin{split}
Q_{\phi} &= + Q \ \mathrm{cos } (2\phi) + U \ \mathrm{sin } (2\phi) \\
U_{\phi} &= -Q \ \mathrm{sin} (2\phi) + U \ \mathrm{cos} (2\phi), 
\end{split}
\end{equation}
where $\phi$ denotes the position angle measured east of north with respect to the position of the star. 
In this formulation, a positive signal in the $Q_{\phi}$-image corresponds to a signal that is linearly polarised in azimuthal direction, whereas radially polarised light causes a negative signal in $Q_{\phi}$. Any signal polarised in the direction $\pm$ 45\degr with respect to the radial direction is contained in \Uphi. Therefore, in the case of low or mildly inclined disks, almost all scattered light is expected to be contained as positive signal in $Q_{\phi}$. However, due to the non-negligible inclination of the disk around PDS~70 \citep[49.7$^{\circ}$,][]{2012ApJ...758L..19H}, and because the disk is optically thick such that multiple scattering processes cannot be neglected, we expect some physical signal in $U_{\phi}$ \citep{Canovas:2015bz}. $U_{\phi}$ can therefore only be considered as an upper limit to the noise level. 

\subsection{ZIMPOL polarimetric observations (VBB band)}
PDS~70 was observed during the night of July 9, 2015, with the SPHERE/ZIMPOL instrument \citep{2008SPIE.7014E..3FT}. These non-coronagraphic observations were performed in the SlowPolarimetry readout mode (P2), using the Very Broad Band (\textit{VBB}, 590-881 nm) filter, which covers the wavelength range from the R- to the I-band. Especially in the second half of the sequence, the conditions were poor (seeing above 1\arcsec), resulting in a PSF FWHM of $\sim$159 mas. 
Since the detailed Mueller matrix model for the correction of the instrumental polarisation by van Holstein et al. (in prep.) only applies to the IRDIS data,  
our correction for instrumental polarisation effects was performed by equalising the ordinary and extraordinary beams for each frame, as described by \cite{2014ApJ...781...87A}. We interpolated two pixel columns in the image that were affected by readout problems. 

\subsection{IRDIFS angular and spectral differential imaging observations (Y-H-band)}\label{GTO}
During the guaranteed time observations (GTO) of the SPHERE consortium, PDS~70 was observed twice within the SHINE \citep[SpHere INfrared survey for Exoplanets;][]{2017sf2a.conf..331C} program on the nights of May 3, 2015, and May 31, 2015. The data were taken in the IRDIFS observing mode, with IRDIS working in the dual-band imaging mode \citep[][]{2010MNRAS.407...71V} making use of the H2H3 narrow-band filter pair ($\lambda_{H2}$ = 1.593 $\mu$m, $\lambda_{H3}$ = 1.667 $\mu$m), and with IFS operating simultaneously in the wavelength range of the Y and J broadband filter (0.95 - 1.33 $\mu$m) with a spectral resolution of R$\sim$50 \citep{2008SPIE.7014E..3EC}. We made use of the N\_ALC\_YJH\_S coronagraphic mask (apodized Lyot, diameter 185 mas). The observations were performed in pupil-tracking mode to allow for angular differential imaging \citep[ADI;][]{2006ApJ...641..556M}. Before and after the sequences, we obtained calibration frames for measuring the location of the star behind the coronagraph and unsaturated images of the star without coronagraph for photometric calibration. Each sequence consisted of 64 exposures, from which we removed 10 and 14 bad quality frames for the May 3 and May 31 epochs, respectively. \\
After basic reduction steps applied to the IRDIS data (flat fielding, bad-pixel correction, sky subtraction, frame registration, frame selection to remove poor-quality frames based on the frame-to-frame photometric variability of the background object north from the star, correction of the instrument distortion, and correction of the flux calibration for the neutral density filter transmission) we used several different strategies to model and subtract the stellar speckle pattern. First of all, we applied the cADI method \citep[classical Angular Differential Imaging;][]{2006ApJ...641..556M}. We then ran a sPCA (smart Principal Component Analysis) algorithm, adapted from \cite{2013A&A...559L..12A}, which itself is based on the KLIP algorithm of \cite{2012ApJ...755L..28S}. Further, we used the ANDROMEDA package \citep{2015A&A...582A..89C} which applies a statistical approach to search for point sources. Further, we applied the PCA and TLOCI \citep[Template Locally Optimized Combination of Images;][]{2014SPIE.9148E..0UM} approach using the SpeCal implementation \citep{2018arXiv180504854G}. The main difference between our two PCA reductions is that the SpeCal PCA implementation does not select the frames for building the PCA library, and is therefore considered to be more aggressive than the former one. Finally, to obtain a non-ADI view of the disk morphology, we simply derotated and stacked the frames and applied a Laplacian filter, here referred as `gradient reduction', in order to enhance low spatial frequencies (i.e. fine disk structures) in the image. However, since this reduction is not flux conservative, we used it only for a qualitative analysis of the outer disk structures.  \\ 
Concerning the IFS data, the basic data reduction was performed using the Data Reduction and Handling software \citep{2008SPIE.7019E..39P} and custom IDL routines adapted from \cite{2015MNRAS.454..129V} and \cite{2015A&A...576A.121M}. We post-processed the data using the cADI, ANDROMEDA and PCA-SpeCal algorithms. The IRDIS and IFS data were astrometrically calibrated following the methods in \cite{2016SPIE.9908E..34M}.

\subsection{IRDIFS\_EXT angular and spectral differential imaging observations (Y-K-band)}
The star was also observed with SPHERE in IRDIFS\_EXT mode during the night of May 14, 2016. In this mode, the IRDIS K1K2 narrow-band filter pair is used ($\lambda_{K1}$ = 2.11 $\mu$m, $\lambda_{K2}$ = 2.25 $\mu$m), whereas IFS is operating in the wavelength range of the YJH broad-band filter (0.97-1.64 $\mu$m) at a spectral resolution of R$\sim$30. No coronagraph was used during the observations, and short detector integration times (DIT; 0.837 s for IRDIS, 4 s for IFS) were chosen to prevent any saturation of the detector. As the observing conditions were relatively stable during the sequence, no frame selection was performed. The data-reduction strategy was identical to the one in Sect. \ref{GTO}, and we post-processed the data with sPCA, PCA-SpeCal, TLOCI and ANDROMEDA. 

\subsection{NaCo angular differential imaging observations (L'-band)}
We also made use of observations of PDS~70 carried out with VLT/NaCo within the ISPY (Imaging Survey for Planets around Young stars; Launhardt et al. in prep.) GTO program on the night of June 1, 2016. The sequence was obtained in pupil-stabilised mode making use of the L'-band filter (3.8 $\mu$m) and the 27.1 mas/pixel pixel scale. No coronagraph was employed. A DIT of 0.2 s was used to prevent any saturation during the sequence. 
The seeing was rather stable during the observations (average seeing 0.5\arcsec), and seven frames were rejected.
The data reduction and post-processing strategy was identical to the one in Sect. \ref{GTO}. 

\subsection{Archival NICI angular differential imaging observations (L'-band)}
Finally, we used archival Gemini/NICI non-coronagraphic data taken on March 31, 2012 using the L' filter. The data together with the observing conditions and strategy were published in \cite{2012ApJ...758L..19H}. As for the NaCo observations, no coronagraph was applied. Thirty-seven out of 144 frames were sorted out. We re-reduced and post-processed the data with the same approach as presented in Sect. \ref{GTO}.

\section{Disk analysis}\label{sect:disk}

\begin{figure}[tb]
    \noindent
    \includegraphics[width = .5\textwidth]{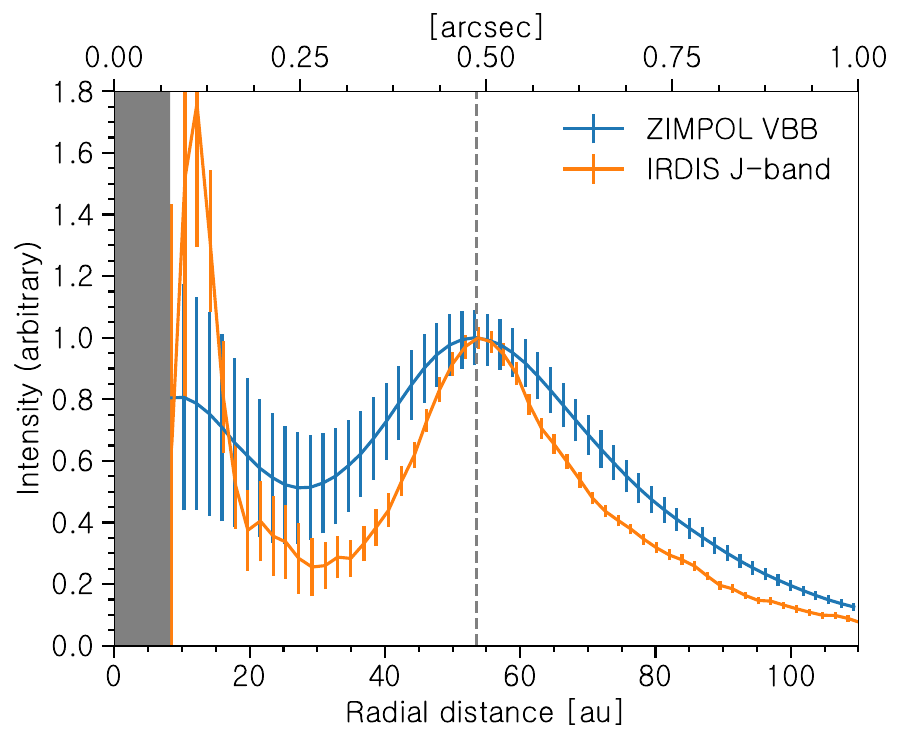}
    \caption{Radial profile of the VBB (blue) and J-band (orange) deprojected, azimuthally averaged $\mathrm{Q_{\phi}}$ images. The profiles were normalized according to the brightness peak of the outer disk, whose location ($\sim$ 54 au) is indicated by the grey dashed line. The grey shadow indicates the radius of the coronagraph in the J-band observations ($\sim$8 au).}\label{cuts}
\end{figure}
\begin{figure}[tb]
    \includegraphics[width = .5\textwidth]{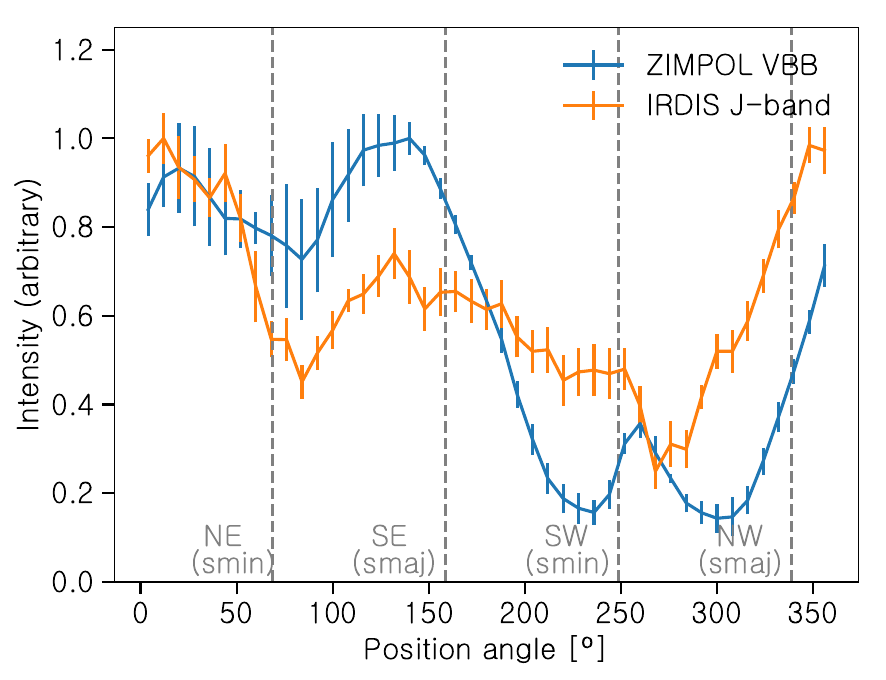}
    \caption{Azimuthal profile of the $Q_{\phi}$ images averaged over an annulus of 35-70 au. The grey lines mark position angles of the SE and NW semi-major axes (`smaj'), as well as NE and SW disk minor axis (`smin'), respectively.}\label{azimuthal_profile}
\end{figure}

\subsection{The outer disk in polarised scattered light}\label{subsect:PDI_outer_disk}
In total, we have three observations of the disk taken in PDI mode: two observations in IRDIS J-band (coronagraphic and non-coronagraphic), as well as one observation with the ZIMPOL VBB filter. Figure \ref{fig:obs_PDI} shows the respective $\mathrm{Q_{\phi}}$ and $\mathrm{U_{\phi}}$ images. As expected from previous observations, we detect the disk in all three datasets as an elliptical ring. Because the setup of the non-coronagraphic IRDIS data was not optimal to detect the outer disk (very short DITs, hence lower S/N), we focus here on the IRDIS coronagraphic dataset, and use the reduction where we subtracted the central source polarisation (see Sect. \ref{sect:obs_IRDIS_setup}). The \Qphi \ images show evidence of residual signal that is contained within a region smaller than $\sim$12 pixels (17 au). In both the IRDIS and ZIMPOL data there is some signal in \Uphi, but this signal is mostly detected at radii inward of the outer disk and does therefore not impact our analysis of the outer disk. \\
\begin{figure*}[tbh]
    \begin{center}
    \begin{minipage}[t]{1.0\textwidth}
    \centering
    \includegraphics[width=0.31\textwidth]{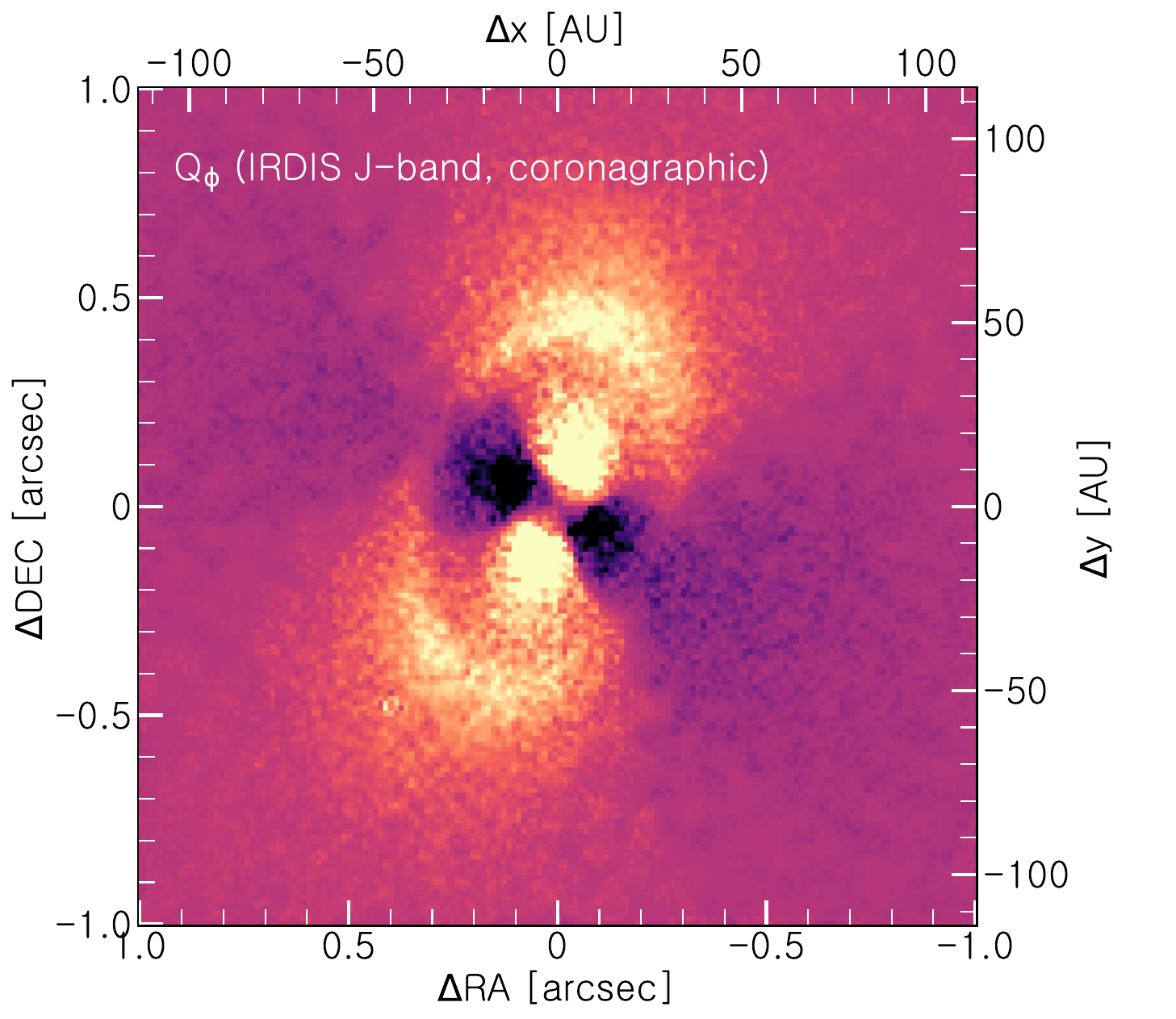}
    \includegraphics[width=0.31\textwidth]{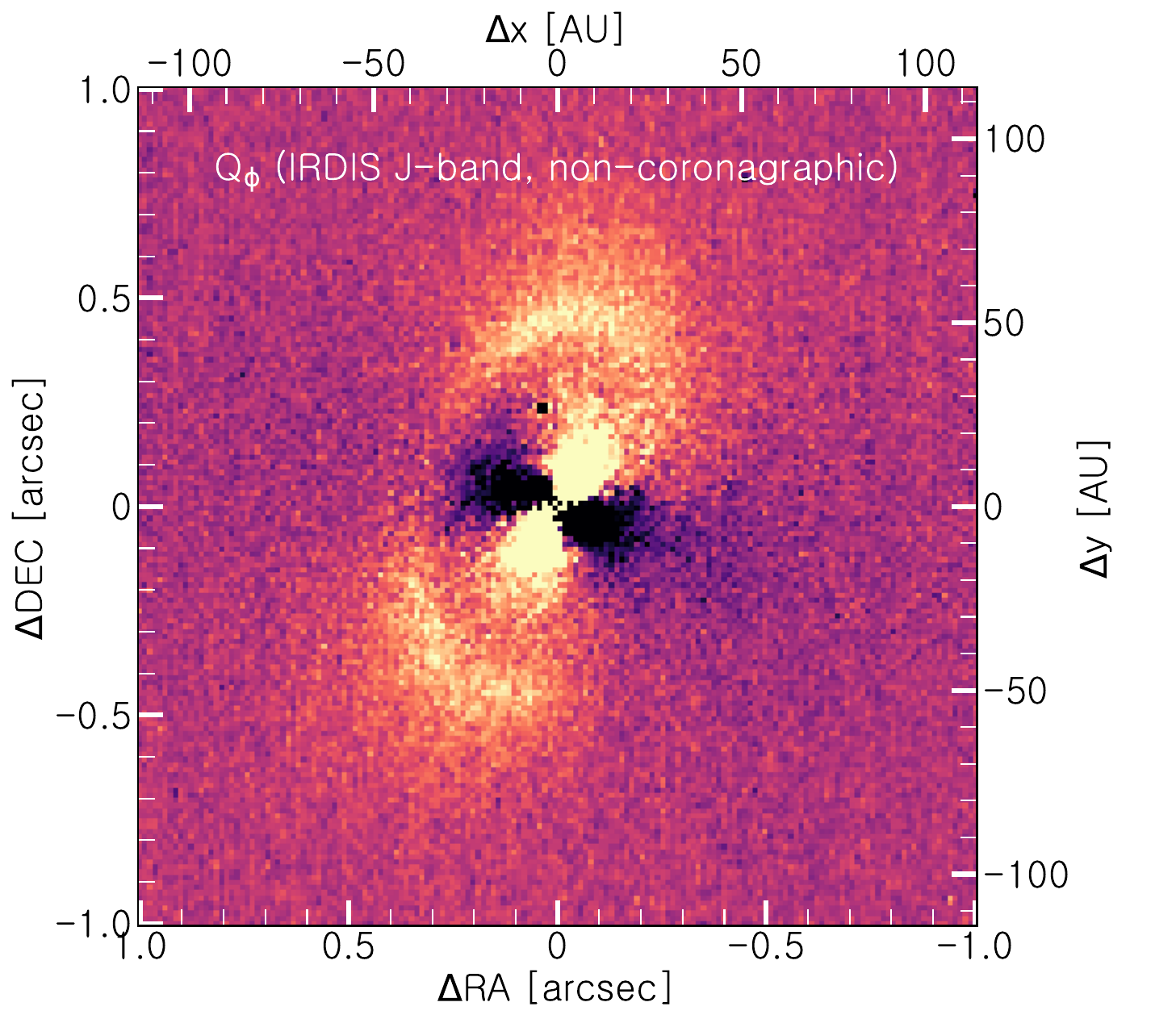}
    \includegraphics[width=0.36\textwidth]{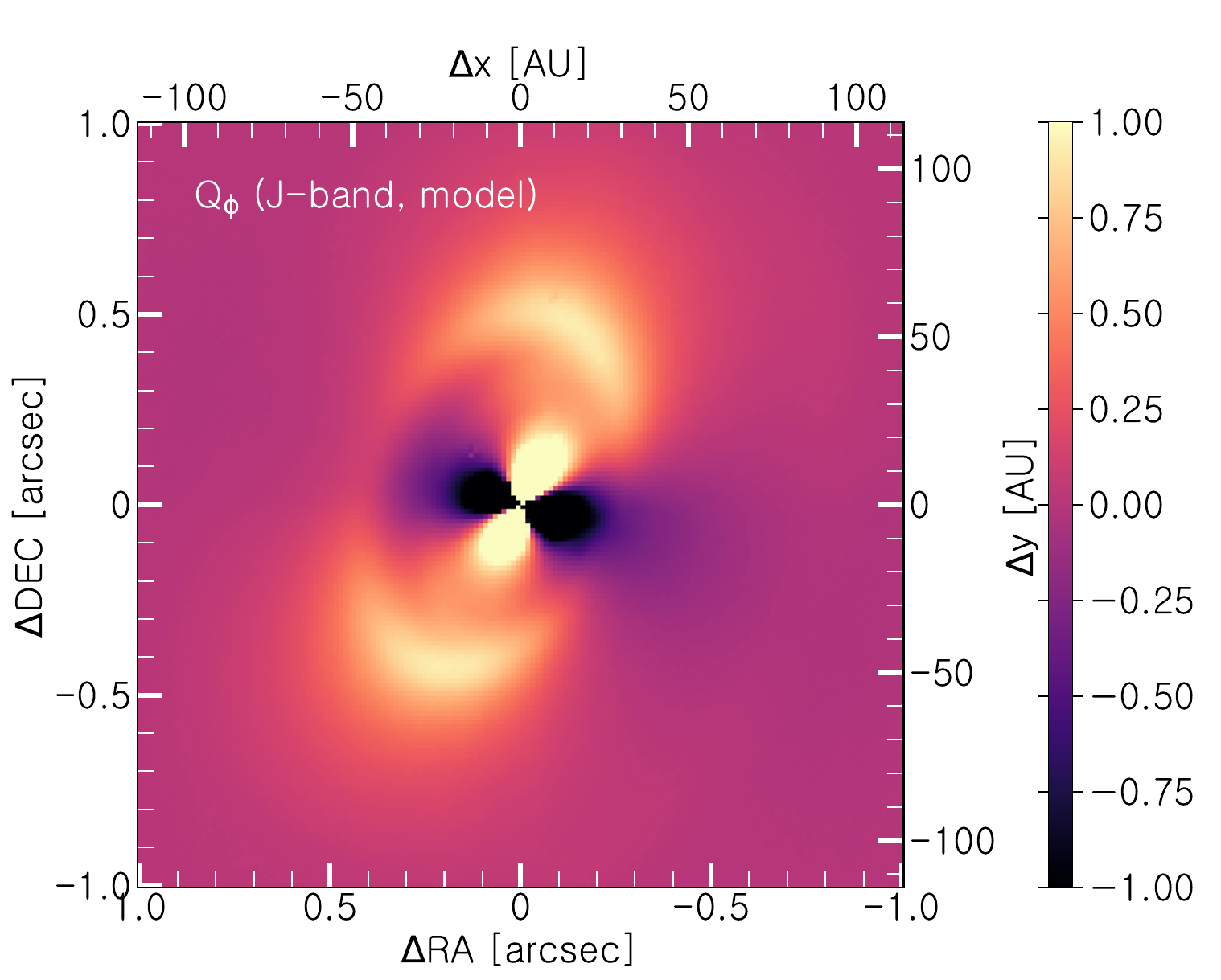}
    \end{minipage}
    \begin{minipage}[t]{1.0\textwidth}
    \centering
    \includegraphics[width=0.31\textwidth]{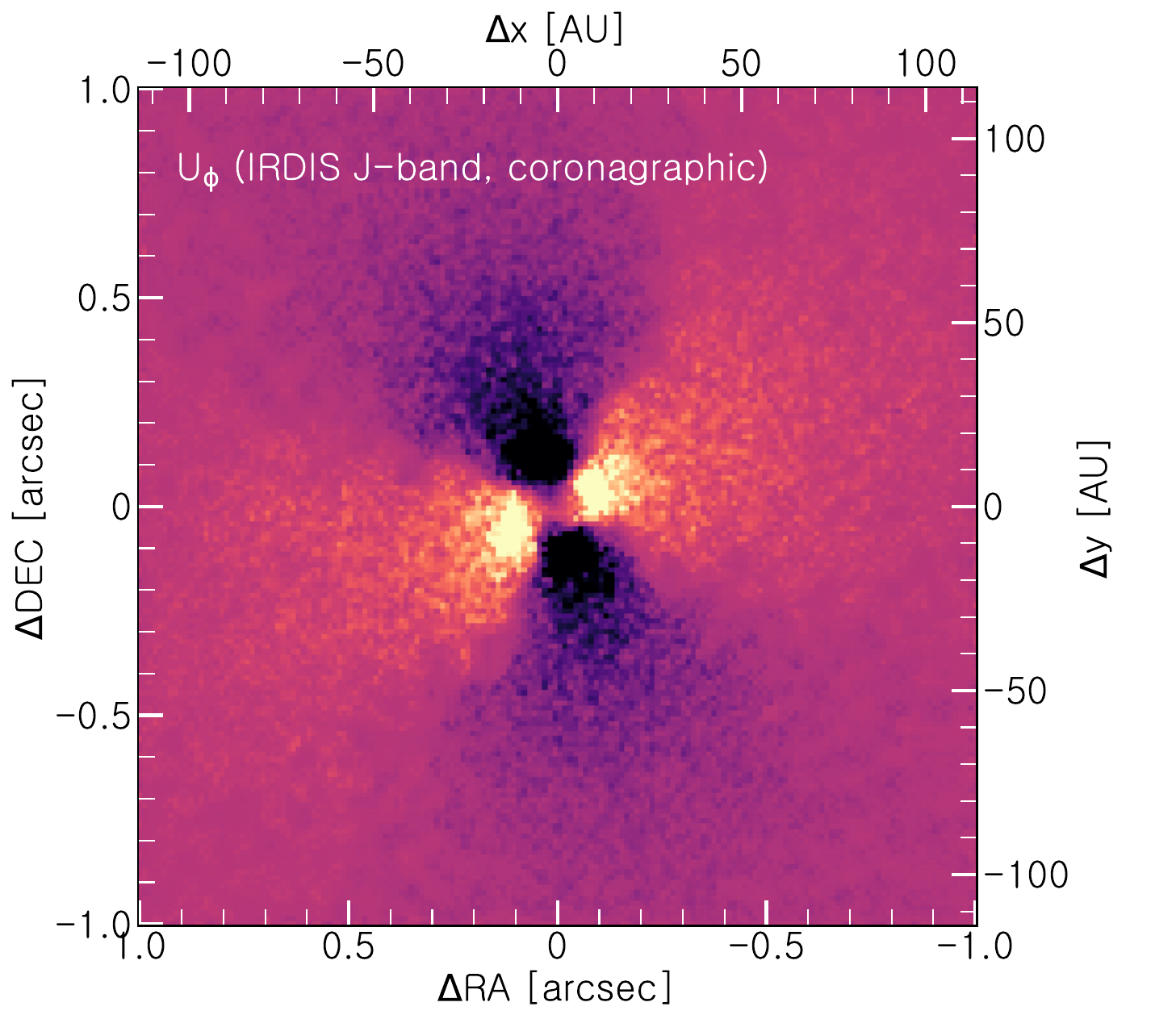}
    \includegraphics[width=0.31\textwidth]{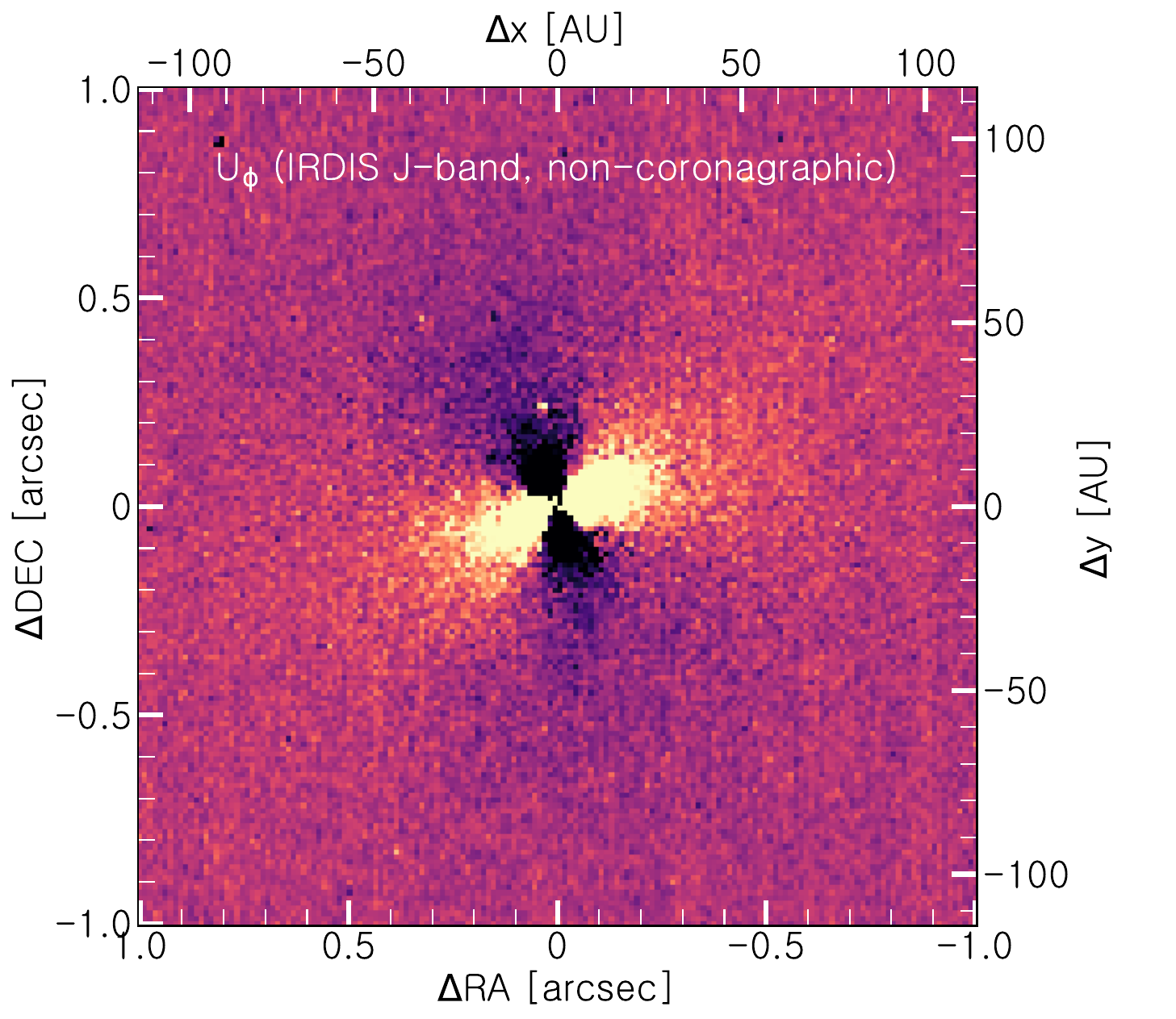}
    \includegraphics[width=0.36\textwidth]{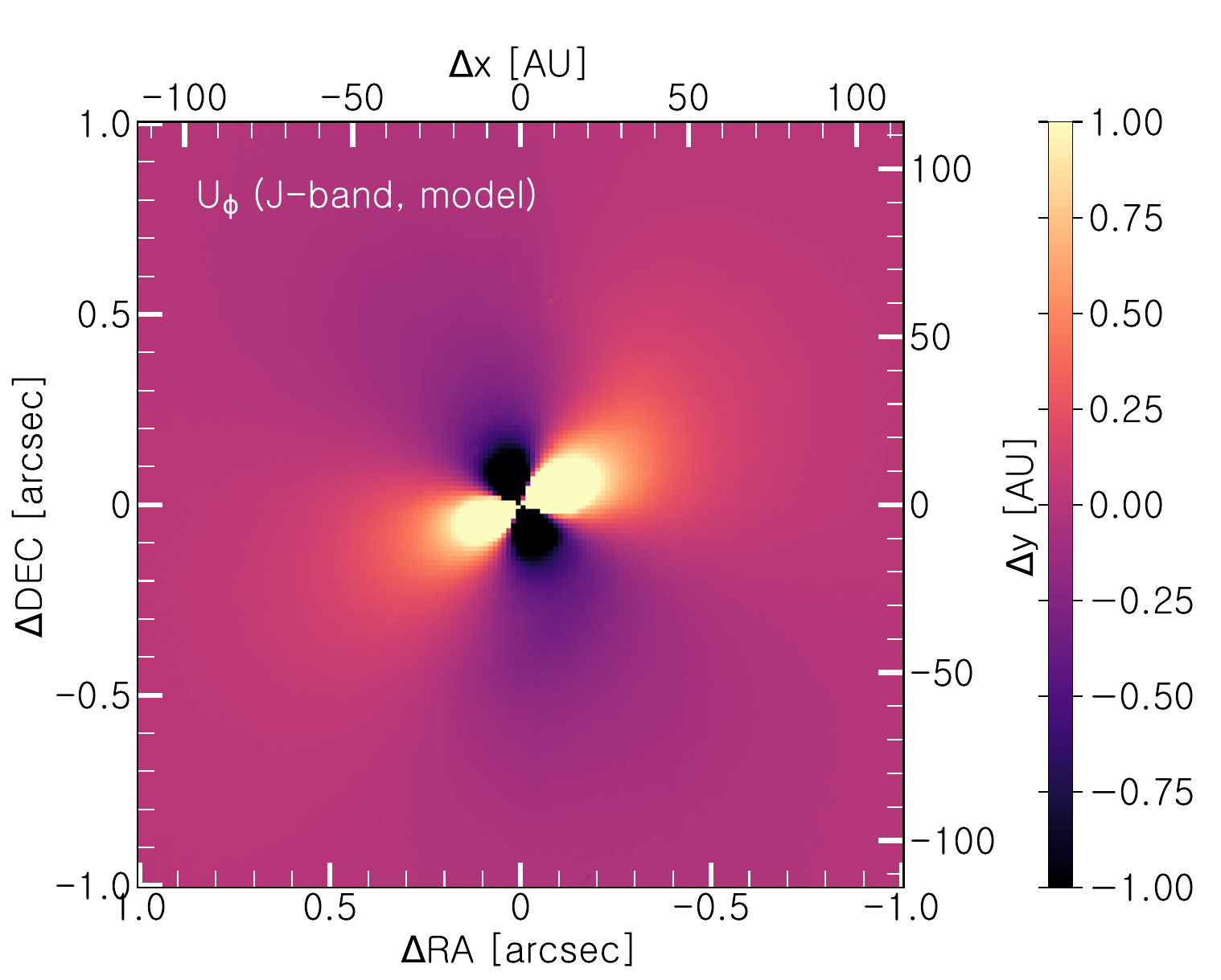}
    \end{minipage}
    \end{center}
   \caption{IRDIS coronagraphic (left column) and non-coronagraphic (middle column) PDI images, corrected for the instrumental polarisation but without subtracting the central source polarisation. The right column shows the model image as a comparison, including an inner disk with an outer radius of 2 au. The first row corresponds to the $\mathrm{Q_{\phi}}$ images, the second row to the $\mathrm{{U_{\phi}}}$ images. North is up and east is to the left. 
}\label{fig:butterfly}
\end{figure*}
In Fig. \ref{cuts} we present the azimuthally averaged radial brightness profile after deprojecting the disk \citep[using a position angle of 158.6\degr and an inclination of 49.7\degr, as determined by][]{2012ApJ...758L..19H}. To avoid effects from small-scale noise we smoothed our images with a small Gaussian kernel with a FWHM of 50$\%$ of the measured image resolution. The uncertainties were computed from the standard deviation of $U_{\phi}$ in the corresponding radial bins, divided by the square-root of the number of resolution elements fitting in that bin. We remind the reader that especially towards the region close to the star, \Uphi\ might contain physical signals, and therefore the error bars only indicate an upper limit for the noise. The mean radius of the disk brightness peak is determined to be $\sim$54 au. The outer disk ring appears wider in the VBB profile than in the J-band profile. We note however that the PSF FWHM was about three times larger during the VBB observations than during the J-band observations. Inside $\sim$25 au, the profile rises towards the centre which is associated to emission from the inner disk (see Sect. \ref{sect:PDI_disk_inner}). The slope is much stronger in the J-band than in the VBB-band. This can be explained by the fact that the regions close to the star in PDI observations are affected by PSF damping effects which are more strongly pronounced at shorter wavelengths where the Strehl ratio is significantly lower \citep[][]{2014ApJ...790...56A,2018arXiv180310882A}. In Fig. \ref{azimuthal_profile}, we plot the azimuthal profile of the deprojected \Qphi\ images. The profile was derived by averaging over azimuthal bins with a size of 8\degr \ between 35 and 70 au in radial direction. We note an azimuthal brightness modulation for both datasets. In each of them, the east side (corresponding to PA $\lesssim$ 160$^{\circ}$) appears on average brighter than the west side. By averaging the brightness in azimuthal bins of size $\pm$20$^{\circ}$ around the PA of the semi-major axes, we derive a brightness contrast of $\sim$1.8 on the brightness maxima along the semimajor axis in the SE and NW in the VBB and $\sim$ 0.8 in the J-band images. We also found a disk brightness ratio of $\sim$3.1 and $\sim$1.4 along the minor axes, in the NE and SW for the VBB and J-band, respectively. Therefore, the brightness ratio shows the same trend along the minor axis in the VBB and J-band, with the NE side being brighter than the SW side in both the VBB and the J-band. However, along the major axis, the trend is opposite between the two bands, with the SE side being brighter than the NW side in the VBB band, but fainter than the NW side in the J-band. By fitting an ellipse to the disk, \cite{2012ApJ...758L..19H} showed that the center of such an ellipse is offset towards the east side with respect to the star. This is due to the flaring geometry of the disk and implies that the east side of the disk corresponds to the far side. 

\begin{figure*}[t]
    \begin{minipage}[tbh]{1.0\textwidth}
    \includegraphics[width=0.305\textwidth]{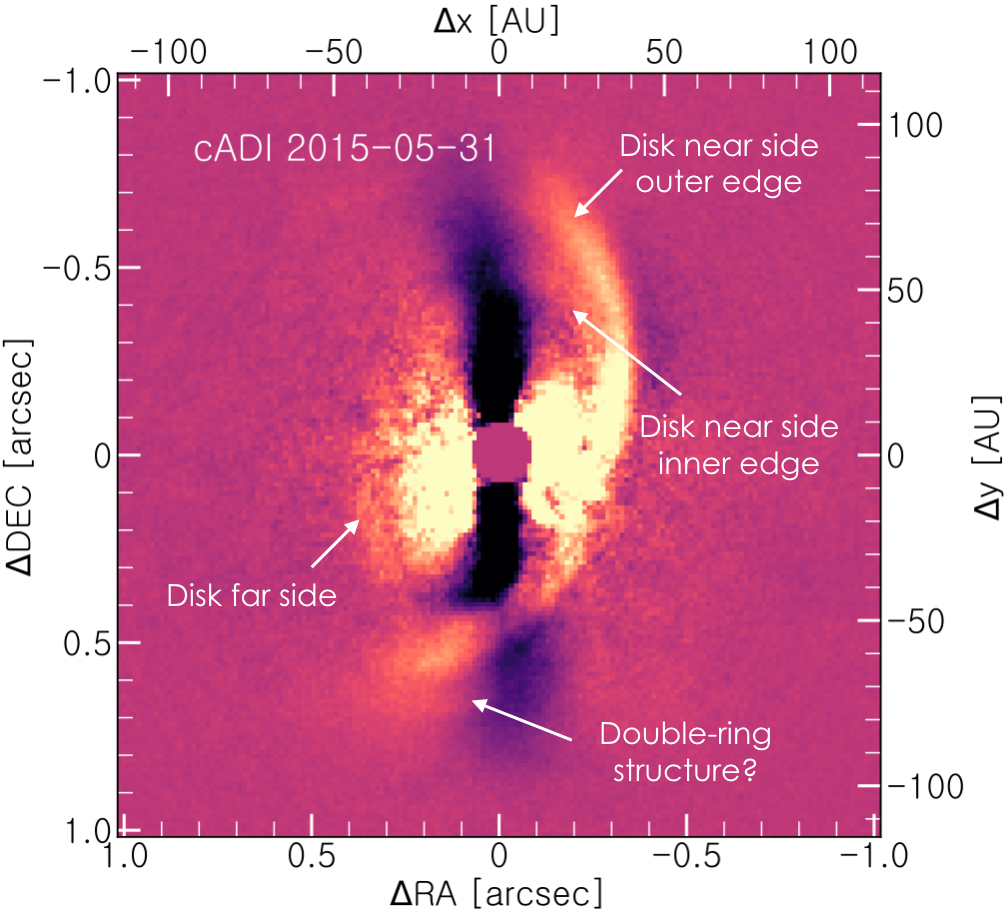}
    \includegraphics[width=0.32\textwidth]{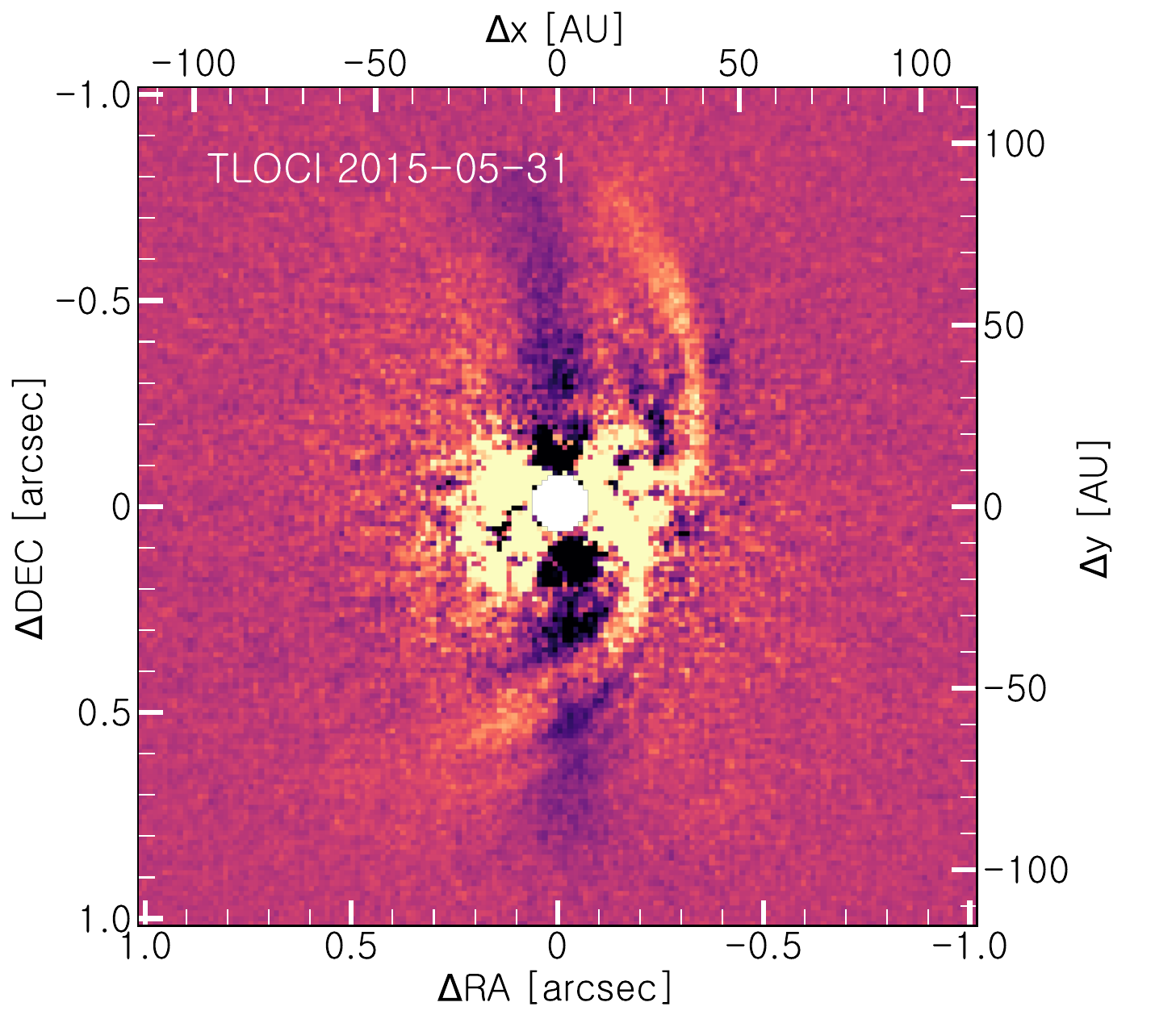}
    \includegraphics[width=0.36\textwidth]{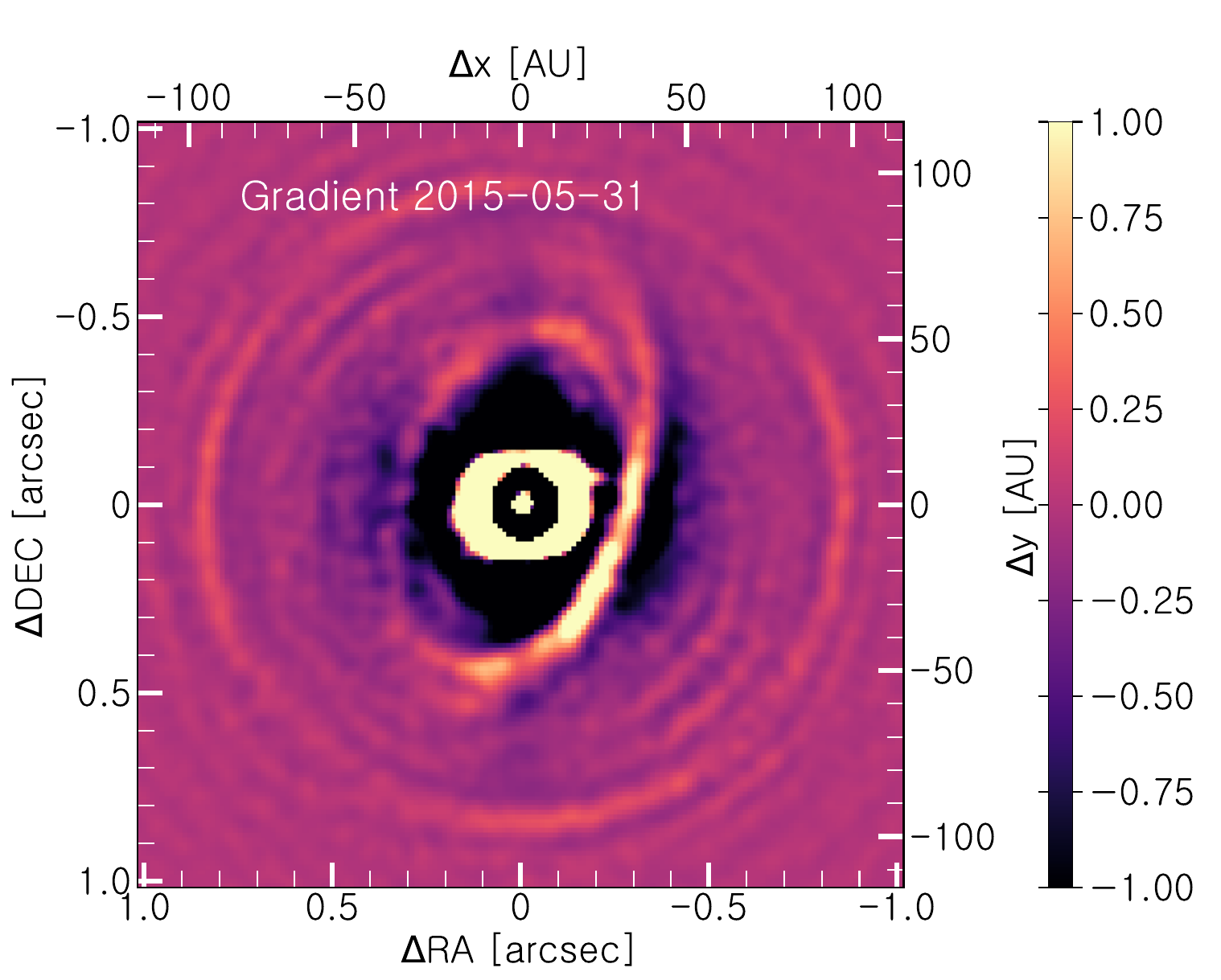}
    \end{minipage}
   \caption{SHINE IRDIS observations of May 31, 2015, showing the cADI (left), TLOCI (middle), and gradient reduction (right). North is up and east is to the left. }\label{fig:obs_ADI}
\end{figure*}

\subsection{Detection of the inner disk in polarised light}\label{sect:PDI_disk_inner}
The presence of an optically thick inner disk in the innermost few astronomical units was predicted from SED fitting of the NIR excess \citep{2012ApJ...758L..19H,2012ApJ...760..111D}. The ALMA observations by \cite{2018ApJ...858..112L} (beam size of 0.19~\arcsec$\times$0.15~\arcsec) detected thermal emission from an inner disk component, which appears to be depleted of large grains.  However, this inner disk component has not been directly detected in scattered light until now. \\ 
For the analysis of the inner region, we mainly rely on the IRDIS polarimetric non-coronagraphic dataset, as it allows us to probe regions closer to the star than does the coronagraphic dataset. As seen in Sect. \ref{subsect:PDI_outer_disk}, the outer disk is well recovered in the IRDIS PDI images by subtracting the remaining central source polarisation. To study the inner disk region, we therefore choose to focus on the dataset corrected for the IP, but without subtracting the central source polarisation. The corresponding $Q_{\phi}$ and $U_{\phi}$ images (coronagraphic and non-coronagraphic) are presented in Fig. \ref{fig:butterfly} (left and middle column). \\
We detect a strong butterfly pattern in both \Qphi\ and \Uphi. This pattern affects the outer disk. We determine a central source polarisation degree of 0.75$\pm$0.2$\%$, and an angle of linear polarisation of 66$\pm$11$^{\circ}$. Hence, the polarisation direction in the inner region is approximately perpendicular to the disk semi-major axis. This butterfly pattern can be explained by the fact that we detect signal from the inner disk which is unresolved. If the inner disk is oriented in the same direction as the outer disk \citep[PA of $\sim$158.6$^{\circ}$,][]{2012ApJ...758L..19H}, the majority of this signal will be polarised in perpendicular direction to the semi-major axis, because the polarisation degree is highest along the semi-major axis (scattering angle of $\sim$90$^{\circ}$). This signal gets smeared out when convolved with the instrument PSF, causing a large butterfly pattern in the resulting \Qphi \ and \Uphi \ images. The fact that the unresolved signal is polarised perpendicular to the disk semi-major axis implies that the inner and outer disks are approximately aligned. We note that although we cannot determine the inclination of the inner disk, we can infer that it must be larger than zero, because had the inner disk been seen face-on, its signal would have cancelled out due to axial symmetry. Further, the absence of shadows on the outer disk indicates that the inclination of the inner disk should be similar to that of the outer disk. \\
We note that even after subtracting the central source polarisation from the non-coronagraphic (as well as the coronagraphic) data, a signal in $Q_{\phi}$ is detected inside about 17 au. The subtraction of the central source polarisation removes almost all signal from an \textit{unresolved} source, and the leftover signal could originate from a partially resolved inner disk (larger than the resolution element). We therefore suspect that the disk is slightly larger than the resolution element, but not extending farther than 17 au, because otherwise we would have detected larger residuals after subtracting the central source polarisation.  \\
We note that the polarisation of the central source is unlikely to be caused by interstellar dust due to the low extinction measured towards PDS~70 \citep[$A_V$=0.05 $^{+0.05}_{-0.03}$ mag,][]{mueller2018}. We further study the inner disk characteristics by comparison with our radiative transfer models (Sect. \ref{modeling}). 

\subsection{The disk in IRDIS angular differential imaging}
We considered the SHINE IRDIS ADI observations for the characterisation of the disk. In comparison to the observations presented in the previous section, they trace the total intensity and were taken at a longer wavelength (H-band). However, whereas the ADI technique is optimised for detecting point sources (see Sect. \ref{sect:CC}), when applied to extended sources as disks, the images suffer from self-subtraction effects. Figure \ref{fig:obs_ADI} shows the resulting cADI image, TLOCI, and gradient images, and the sPCA, PCA-SpeCal images are shown in Fig. \ref{fig:obs_TLOCI}. In all reductions and epochs, the disk's west side is clearly detected. As previously mentioned, this side corresponds to the near side of the disk. The disk's extension along the semi-major axis is larger in the ADI images than in polarised light. When overlaying the two images, it appears that the signal we see in ADI in fact corresponds to the \textit{outer} region of the disk as seen in polarised light (see Fig. \ref{fig:obs_TLOCI}). Furthermore, in some reductions, the inner edge of the disk's far side appears to be detected. This is especially true in the gradient image, but also in the cADI reduction. \\
The gradient image exhibits many circular artifacts while the disk signal deviates from this circular symmetry, and the inner edge of the outer disk is well detected at all position angles. The cADI, TLOCI and both PCA reductions show a feature near the outer edge of the south-west side of the disk that looks like a double-ring structure beyond the main dust ring. This feature is detected at position angles in the range $\sim$170-300\degr. It follows the same shape as the main disk, but with an offset of roughly 125 mas. Although it is detected in four different reductions (sPCA, PCA-SpeCal, TLOCI, cADI), it is not clear if this feature is real because these observations might suffer from the generation of artificial (sub-)structures due to the self-subtraction. It is also a concern that the structure, if it were real, is not detected in the gradient reduction, or in the PDI images. However, this double ring could be too faint to be detected in the PDI (since disks are in general much fainter in polarised light than in total intensity) and in the gradient image (which is affected by the circular artifacts). 

\section{Radiative transfer modeling}\label{modeling}

\subsection{Model setup}

\begin{table}[b]
\caption{Parameters for our RT model.}
\begin{tabular}{l|l|l|}
\hline \hline
Parameter                         & inner disk            &outer disk \\
\hline
R$_{in}$ [au]                     &0.04                   &60     \\
R$_{out}$ [au]                    &[2,4,8,12,16,20]       &120    \\
R$_{c}$ [au]                      &40                     &40     \\
\hline
h$_{100,small}$[au]               &13                     &13     \\
h$_{100,big}$[au]                 &2                      &2      \\
$\beta $                          &1.25                   &1.25    \\
$\delta_{\mathrm{disk}}$          &[0.05,0.1,0.25,1.0]    &1        \\
\hline
disk inclination $i$ [$^{\circ}$] &49.7                   &49.7   \\
disk pos. angle $PA$ [$^{\circ}$] &158.6                  &158.6  \\
\hline
\end{tabular}
\tablefoot{Parameters for our model. R$_{in}$ and R$_{out}$ denote the inner and outer radius of the inner disk, whereas $R_c$ is the characteristic radius (truncation radius of exponential cutoff). $\delta_{\mathrm{disk}}$ corresponds to the depletion factor applied to the surface density of the inner and outer disk.  $h_{100}$ quantifies the scale height of small and big grains at 100 au, and $\beta$ the flaring index, respectively. }\label{model_setup}
\end{table}

\begin{figure*}[hbt]
    \begin{minipage}[t]{1.0\textwidth}
    \includegraphics[width = 0.33\textwidth]{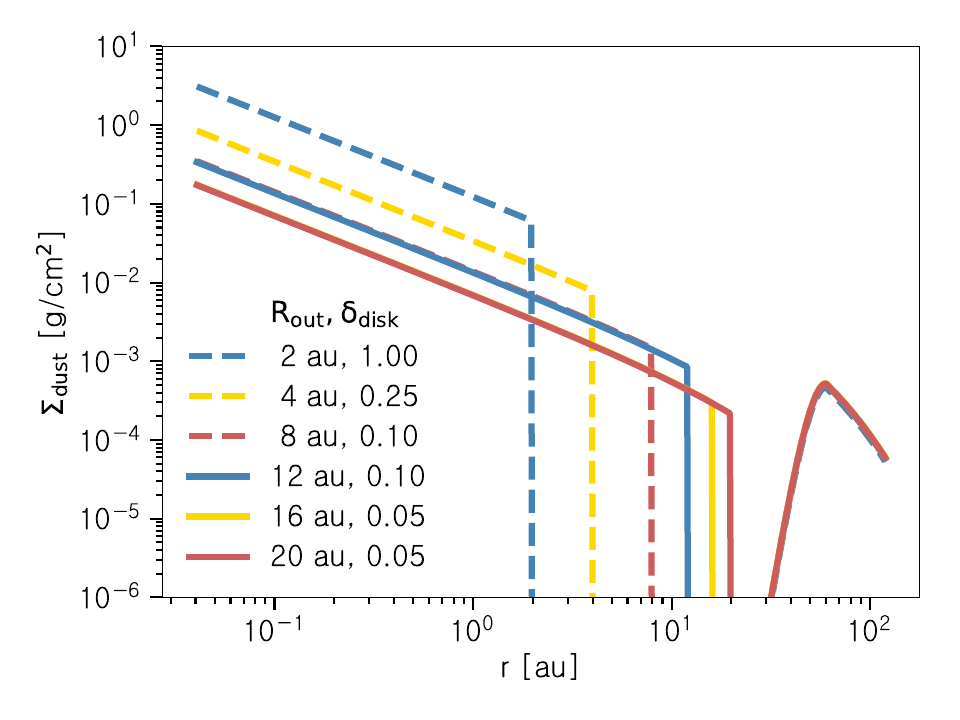}
    \includegraphics[width = 0.33\textwidth]{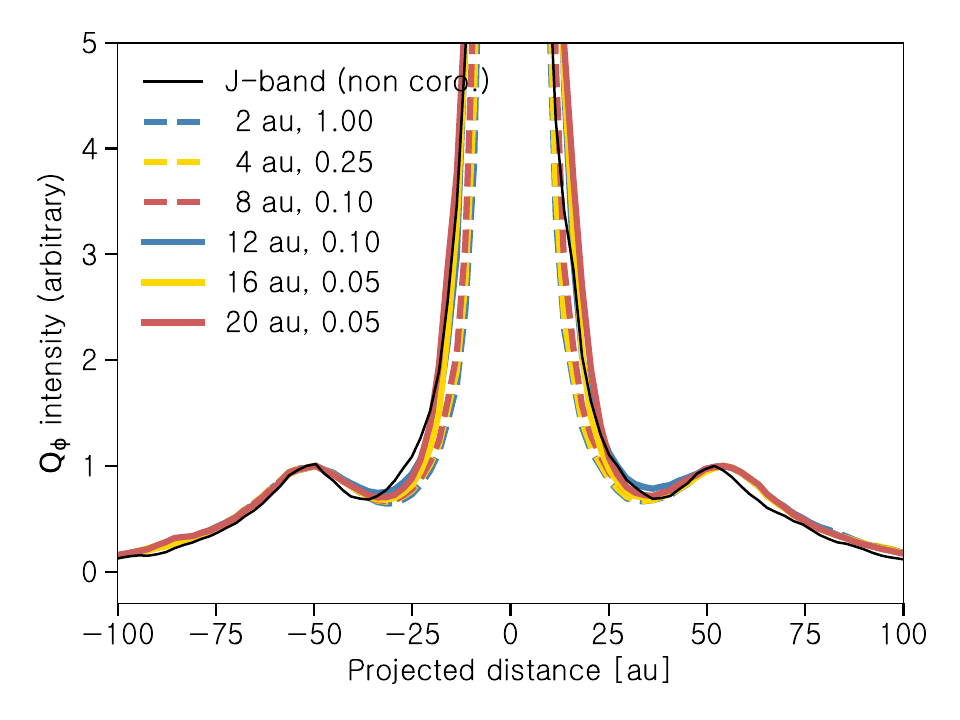}
    \includegraphics[width = 0.34\textwidth]{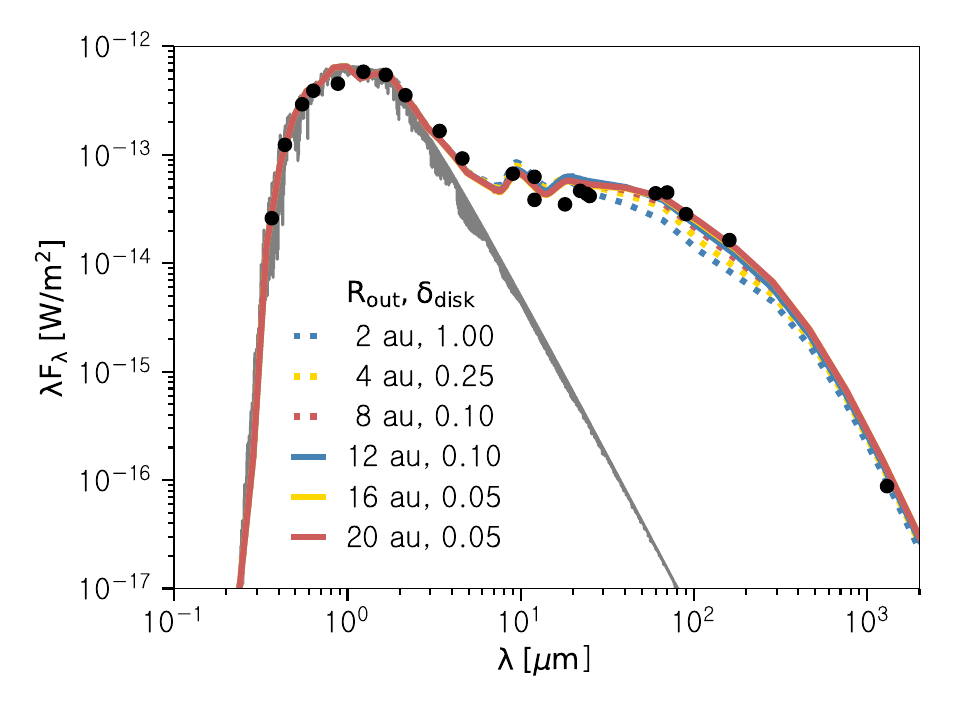}
\caption{\textit{Left:} Surface density of our model for the small grains using different outer radii and depletion factors for the inner disk. \textit{Middle: } Cuts along the semi-major axis of model $Q_{\phi}$ images for different extents and depletion factors of the inner disk, in comparison with the observations (black line). \textit{Right:} 
Comparison of photometric measurements of PDS~70 (black points), and the synthetic spectra (coloured lines). The photometry was taken from \cite{1992AJ....103..549G}, \cite{2003yCat.2246....0C}, and \cite{2012ApJ...758L..19H}. Due to the low optical extinction towards PDS 70 \citep{2016MNRAS.461..794P}, no dereddening was applied to the optical and 2MASS photometric data.
The grey line corresponds to the stellar model spectrum (K7 type) that we used for our RT calculations.}
\label{fig:cuts_butterfly}
    \end{minipage}
\end{figure*}

To compare our multiwavelength observations with a physical model, we built a radiative transfer (RT) model, where the basic parametric approach by \cite{2012ApJ...760..111D} is taken as a starting point. We used the RT Monte-Carlo code RADMC-3D \citep{2012ascl.soft02015D}. Our aim is to find a plausible model which reproduces our observations and the SED. We note that we are not looking for a globally best fitting model. RADMC-3D computes the thermal structure of the disk and produces images in scattered polarised light and total intensity by ray-tracing at any desired wavelength. Our grid has an inner radius of 0.04 au and an outer radius of 120 au. The surface density is proportional to $r^{-1}$ and is truncated by a tapered edge with a characteristic radius $R_c$. We radially parametrize the dust surface density by:\\
\begin{equation} 
\Sigma_{disk} (r) = \Sigma_{0} \frac{R_c}{r} \exp\left(\frac{-r}{R_c}\right), 
\end{equation} 
where $\Sigma_{0}$ scales the amount of dust contained within the disk.   \\
We assume a Gaussian distribution profile in the vertical direction and parametrize the density distribution in terms of the scale height $h$ as:
\begin{equation}
\rho (r,z) = \frac{\Sigma (r)}{\sqrt{2 \pi} h} \exp(-z^{2}/2 h^{2})\,.
\end{equation}
The disk is assumed to be flared with a constant power law index $\beta$, such that the radial dependence of the scale height can be written as
\begin{equation}
h(r) = h_{100}\times \left( \frac{r}{100 \mathrm{\  au}}\right) ^{\beta}, 
\end{equation}
where $h_{100}$ is the scale height at 100 au. To mimic the gap in the disk, we heavily deplete (by a factor of $\mathrm{10^{-15}}$) the surface density between the outer radius of the inner disk and the inner radius of the outer disk. To ensure a smooth transition from the gap to the outer disk, we considered an outer disk radius of 60 au, inwards of which we multiplied the surface density with a Gaussian profile parametrized by a standard deviation of 8 au. The surface density of the inner disk is multiplied with a depletion factor $\delta_{\mathrm{disk}}$ with respect to that of the outer disk. The general shape of the surface density is plotted in Fig. \ref{fig:cuts_butterfly} (left panel). We consider two grain-size distributions (small and large grains), whose number density follows a power law as a function of the grain size $a$ with $n(a)\propto a^{-3.5}$. The population of small grains ranges from 0.001 to 0.15 $\mu$m, and the large grains from 0.15 to 1000 \mum \ in size. To mimic dust settling in a simplified way, we assign a lower scale height (2 au at a radial distance of 100 au) to the disk of big grains, for both the inner and outer disk parts. The relative mass fraction of small to large grains is 1/31. We determine the optical properties for spherical, compact dust grains according to the Mie theory using the BHMIE code \citep{1983asls.book.....B}. A total dust mass of $3.0\times10^{-5}\mathrm{\ M_{\odot}}$ was used. The dust mixture is composed of 70$\%$ astronomical silicates \citep{2003ApJ...598.1026D}, and 30$\%$ amorphous carbon grains \citep{1996MNRAS.282.1321Z}. The opacity mixture was generated according to the Bruggeman mixing rule. The scattering mode in RADMC-3D was set to anisotropic scattering with full treatment of polarisation. We computed the Stokes $Q$ and $U$ frames, as well as images in total intensity at the wavelengths of interest (0.7 $\mu$m and 1.25 $\mu$m), and convolved our images with the total intensity frames obtained during the corresponding observations. We then computed the \Qphi\ and \Uphi\ frames according to Eq.\ref{Stokes_equation} as well as a synthetic SED. Table \ref{model_setup} summarises the parameters used for our model.  

\subsection{Modelling the inner disk}
We test different models containing an inner disk and compare the butterfly pattern in the model with the observations. For this purpose, we consider models with different inner disk configurations in terms of two free parameters: outer radius $R_{out}$ of the inner disk, and depletion factor $\delta_{\mathrm{disk}}$ of the surface density of the inner disk. We compare cuts through the convolved $Q_{\phi}$ model images along the semi-major axis with the non-coronagraphic observations (without subtracting the central source polarisation; see Fig. \ref{fig:cuts_butterfly} middle panel). The depletion factor was varied between 0.05 and 1.0, and the inner disk outer radius between 2 au (corresponding to the completely unresolved case) and 20 au, by keeping the total dust mass constant. For each radius value, we identified the best matching depletion factor. Our modelling appears to be degenerate since we could obtain, for both small and large radii values, model images reproducing reasonably well the observations. The range of solutions includes configurations with a small inner disk with a high surface density, as well as those with a larger inner disk with lower surface density. In any case, the inferred surface density outside 20 au  in all configurations is very low, indicating that only a small amount of material is left at the location of the companion ($\sim$22 au, see Sect. \ref{sect:CC}). \\ 

\subsection{Discussion of brightness asymmetries in the disk}

\begin{figure*}[hbt]
    \noindent
    \centering
    \begin{minipage}{1.0\textwidth}
    \centering
    \includegraphics[width = 0.41\textwidth]{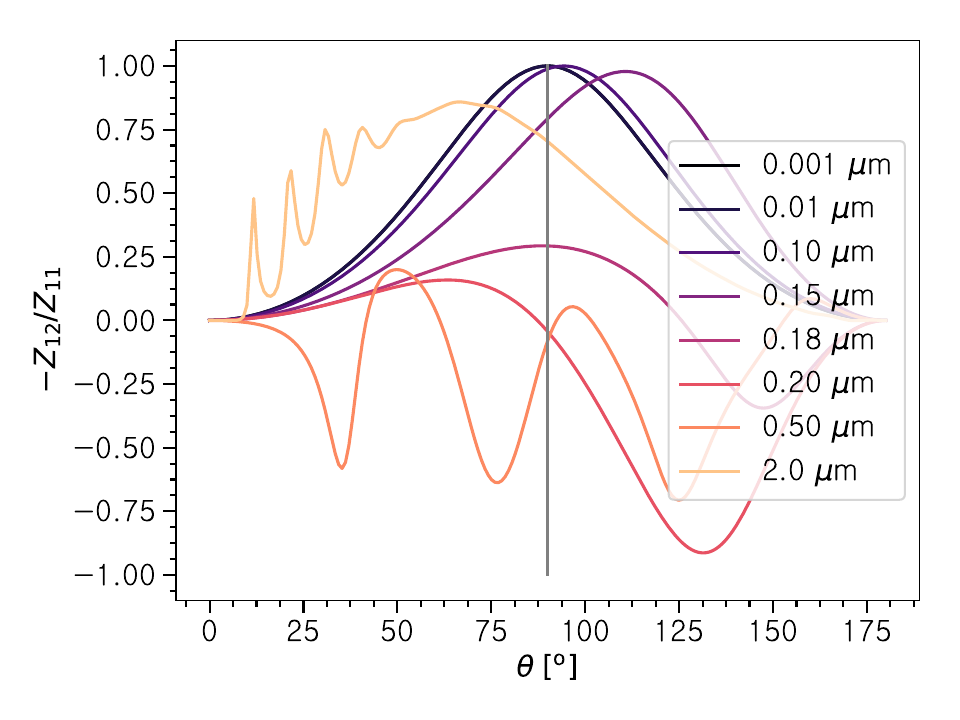}
    \includegraphics[width = 0.41\textwidth]{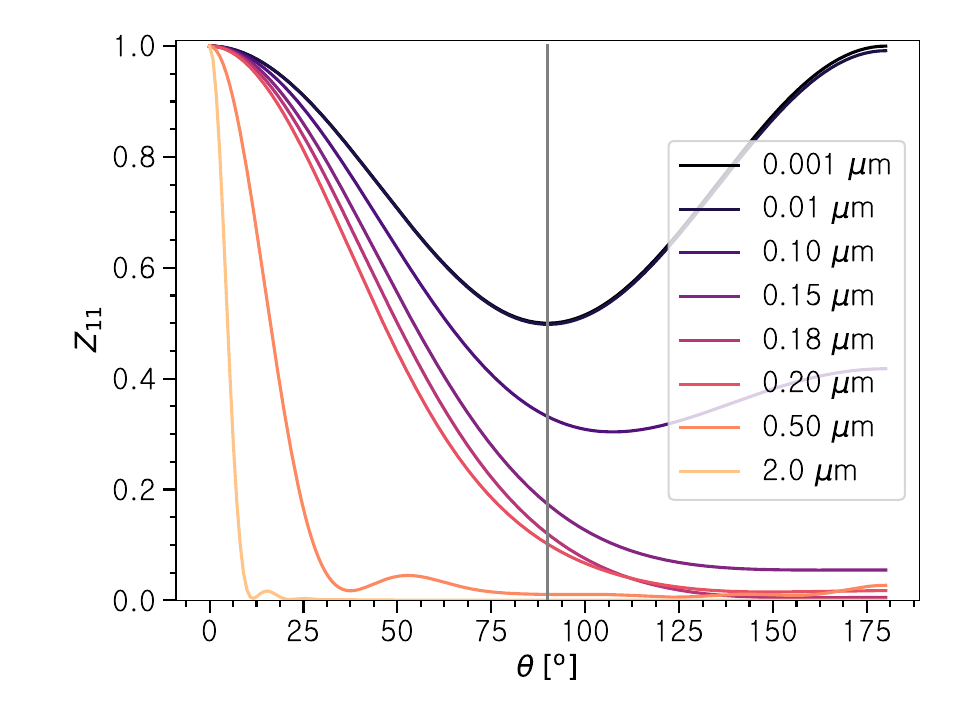}
    \end{minipage}      
    \caption{Linear polarisation degrees for silicates of different (semi-mono dispersive) grain sizes (left), and their phase function (right, normalised to $Z_{11}$(0)). The curves were computed for an observing wavelength of 0.7 $\mu$m.  }\label{fig:phf_ZIMPOL}
\end{figure*}

\begin{figure*}[hbt]
    \noindent
    \centering
    \begin{minipage}{1.0\textwidth}
    \centering
    \includegraphics[width = 0.32\textwidth]{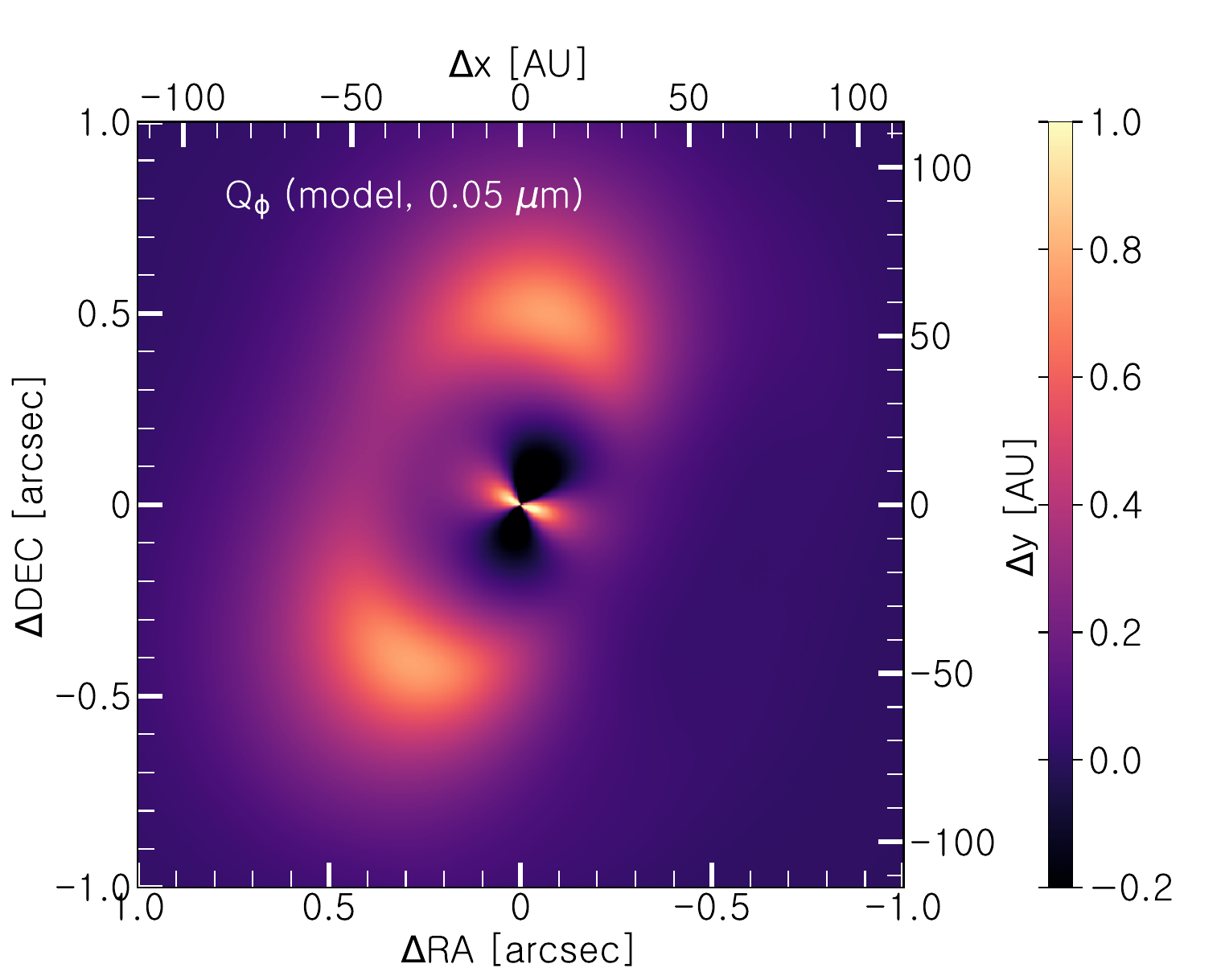}
    \includegraphics[width = 0.32\textwidth]{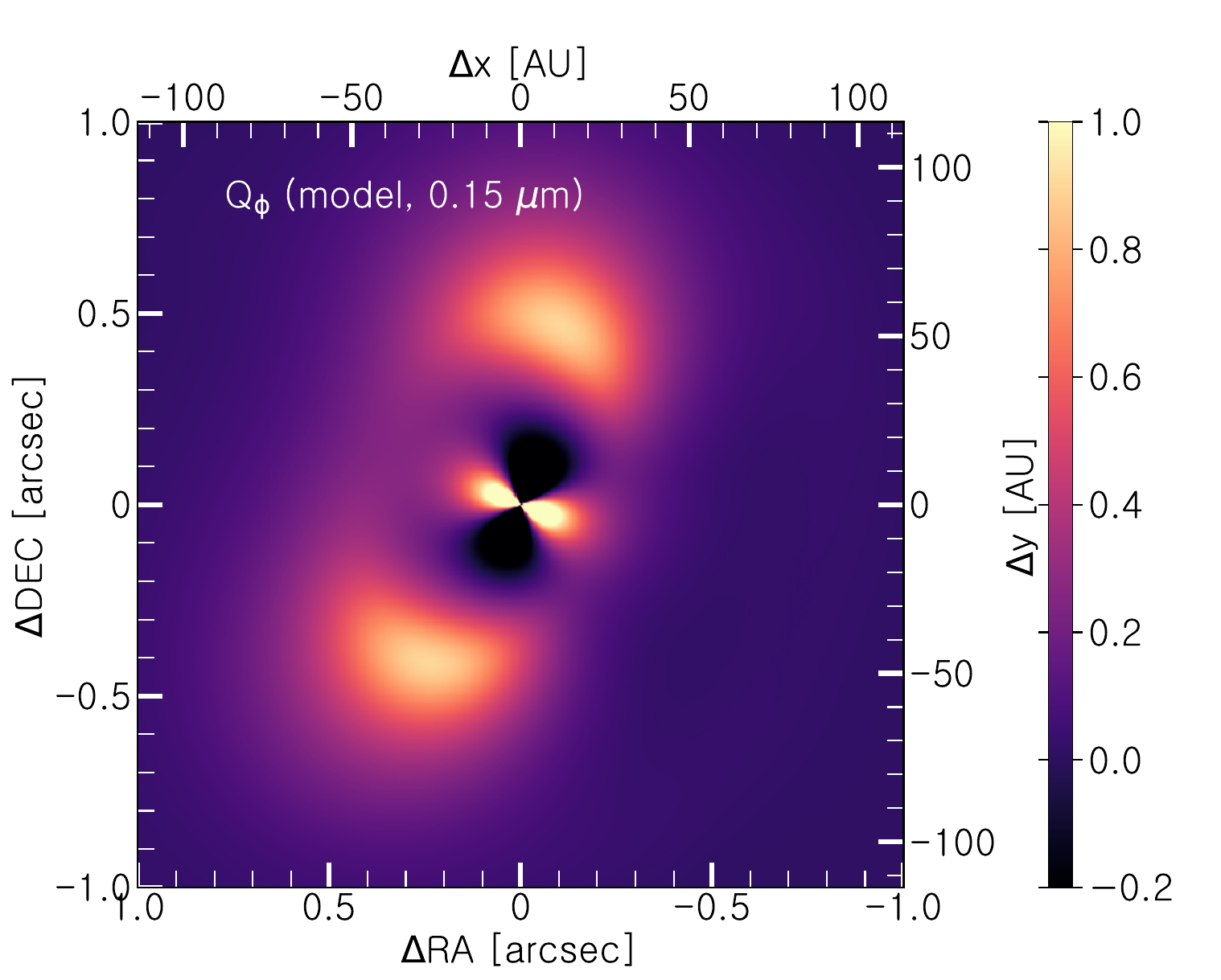}
    \includegraphics[width = 0.32\textwidth]{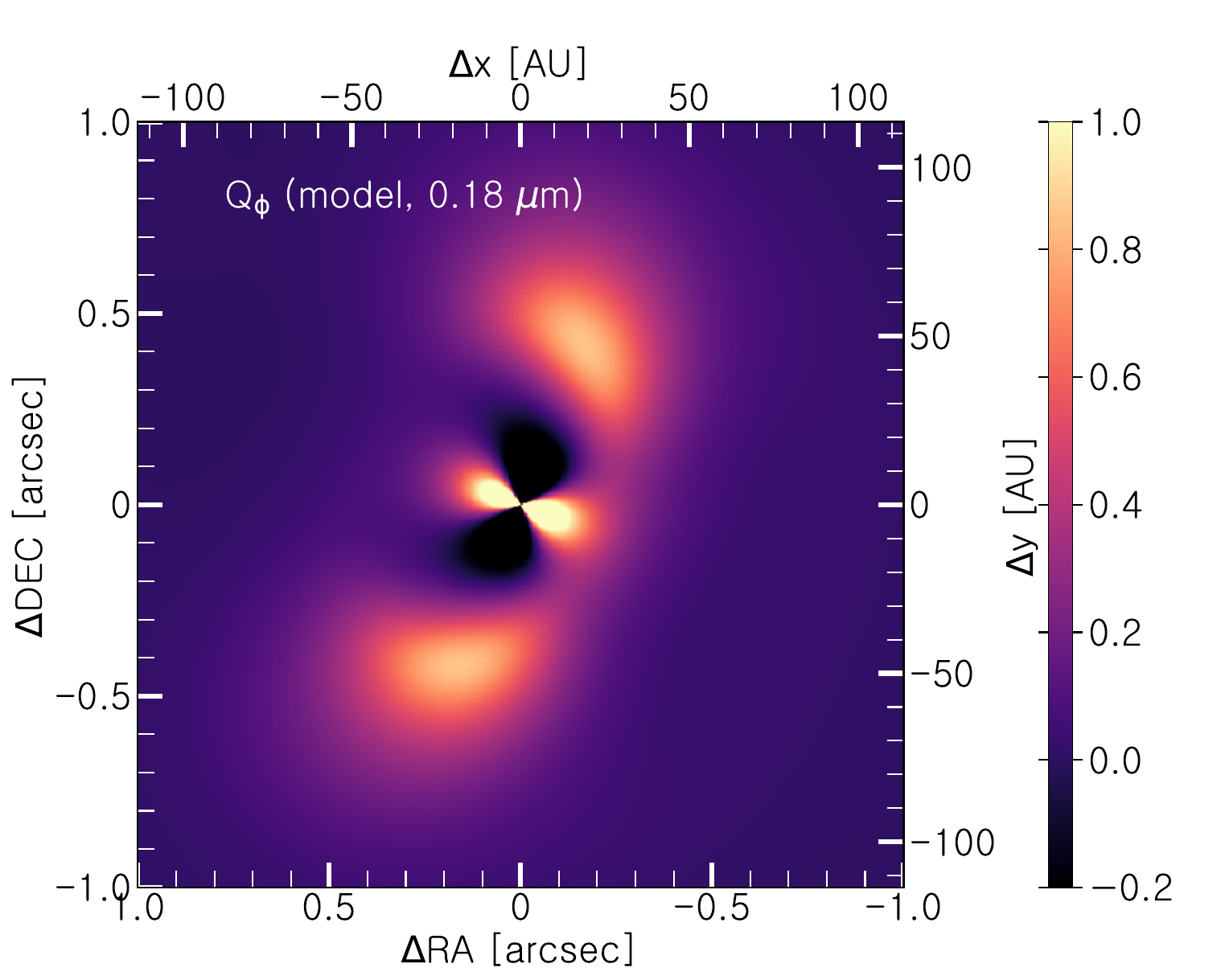}
    \includegraphics[width = 0.32\textwidth]{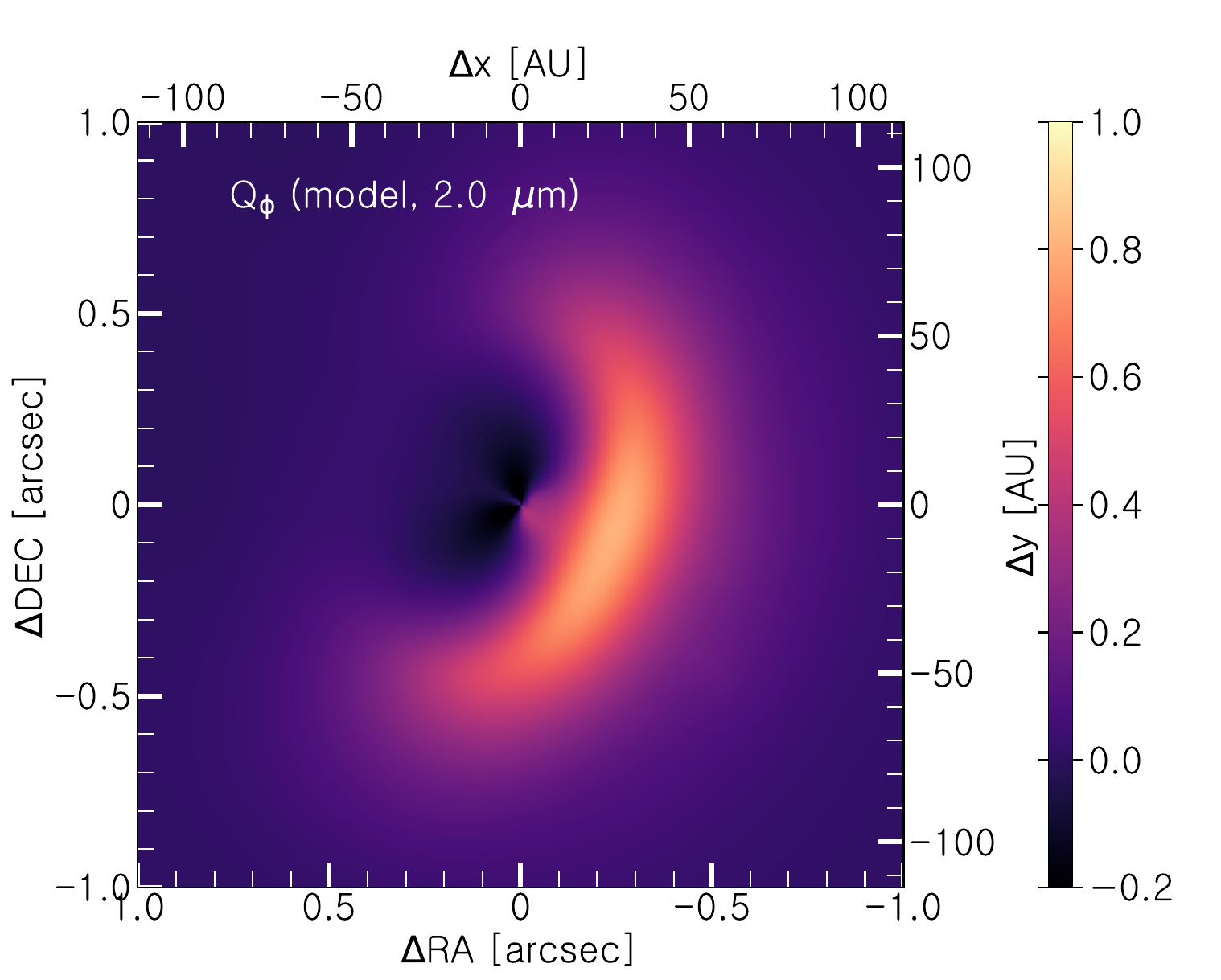}
    \includegraphics[width = 0.32\textwidth]{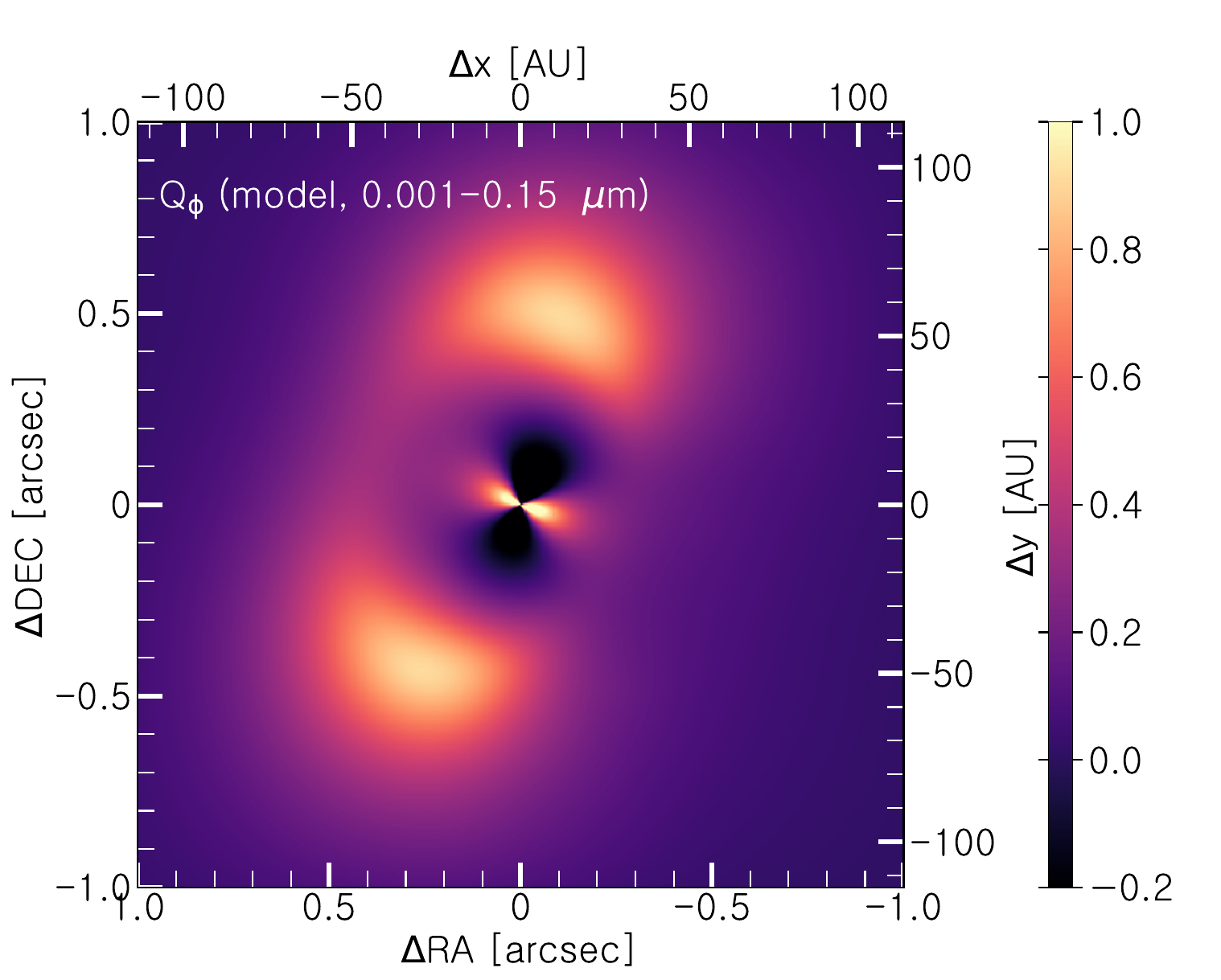}
    \includegraphics[width=0.32\textwidth]{./Qphi_ZIMPOL.pdf}
    \end{minipage}      
    \caption{Convolved ray-traced $Q_{\phi}$-images evaluated at 0.7 $\mu$m using different grain size distributions. Successively, the central source polarisation emerging from the unresolved inner disk (radius 2 au) was subtracted using a \Uphi-minimisation. The lower-right panel shows the VBB observation for comparison. North is up and east is to the left. }\label{fig:phf_ZIMPOL_models}
\end{figure*}

In Sect. \ref{subsect:PDI_outer_disk}, we discussed the outer disk asymmetries and found that the east side is brightest in polarised light but is marginally detected in total intensity. Furthermore, along the north-south axis, the brightness maxima differ between the optical and NIR images taken a year apart. \\
The observed asymmetries between the east and west side might be connected to the scattering properties of the dust. The polarised intensity is the product of the degree of polarisation and total intensity. The phase function of both depends on particle size and observing wavelength. In Fig. \ref{fig:phf_ZIMPOL}, (left) we plot the linear polarisation degree expressed by the Mueller matrix elements ($-Z_{12}/Z_{11})$ as a function of scattering angle. For this test, we computed the scattering matrix using Mie theory for different grain sizes. To smooth out the resonances in the phase function appearing when considering perfect spheres of a single size, instead of using a mono-dispersive grain size distribution, we generated for each grain size a narrow Gaussian distribution centred around the considered grain size and with a FWHM of 10$\%$ of the grain size. The scattering matrix was evaluated at 0.7 $\mu$m which corresponds approximately to the central wavelength of the VBB filter. The plot shows that for very small grains (0.001-0.01 $\mu$m), the polarisation degree is symmetric around 90$^{\circ}$. For larger grains, the polarisation degree has several minima, until at grain sizes much larger than the observing wavelength ($\gtrsim$ 2 $\mu$m), the maximum of the polarisation degree is shifted towards angles smaller than 90$^{\circ}$. Regarding the scattering phase function (Fig. \ref{fig:phf_ZIMPOL}, right panel), particles smaller than $\lesssim$~0.1 \mum \ scatter symmetrically around 90\degr. Above that value, the phase function becomes asymmetric and forward scattering is increasingly pronounced. \\
\begin{table*}[tb]
\centering
\caption{Astrometric calibrations used for the ADI datasets. }
\begin{tabular}{llccccc}
\hline \hline
Date       &Instrument&Filter   &True North corr-             &Rotator offset                 &Pixel scale    \\
           &          &         &rection angle [$^{\circ}$]   &position  [$^{\circ}$]         &[mas/px]        \\
\hline
2012-03-31 & NICI     & L'      &  0.0  $\pm$0.1  $^{(b)}$    & 180.0$\pm$0.1$^{(f)}$          &17.95 $\pm$0.01 $^{(d)}$   \\ 
2015-05-03 & IRDIS    & H2      & -1.700$\pm$0.076$^{(a)}$    &-135.99$\pm$0.11                &12.255$\pm$0.021$^{(a)}$   \\ 
2015-05-03 & IRDIS    & H3      & -1.700$\pm$0.076$^{(a)}$    &-135.99$\pm$0.11                &12.250$\pm$0.021$^{(a)}$  \\ 
2015-05-31 & IRDIS    & H2      & -1.700$\pm$0.076$^{(a)}$    &-135.99$\pm$0.11                &12.255$\pm$0.021$^{(a)}$    \\
2015-05-31 & IRDIS    & H3      & -1.700$\pm$0.076$^{(a)}$    &-135.99$\pm$0.11                &12.250$\pm$0.021$^{(a)}$   \\
2016-05-14 & IRDIS    & K1      & -1.675$\pm$0.080$^{(a)}$    &-135.99$\pm$0.11                &12.243$\pm$0.021$^{(a)}$   \\ 
2016-05-14 & IRDIS    & K2      & -1.675$\pm$0.080$^{(a)}$    &-135.99$\pm$0.11                &12.238$\pm$0.021$^{(a)}$   \\ 
2016-06-01 & NaCo     & L'      &  0.518$\pm$0.120$^{(e)}$    &89.5$\pm$0.1$^{(c)}$           &27.195$\pm$0.063$^{(e)}$    \\
\hline
\end{tabular}
\tablefoot{$^{(a)}$ \cite{2016SPIE.9908E..34M} $^{(b)}$ assumed value (no astrometric measurement around the NICI epoch available). $^{(c)}$ adopted from \cite{2012A&A...542A..41C} $^{(d)}$ adopted from \cite{2014PASP..126.1112H} and \cite{2014A&A...567A..34W} $^{(e)}$ Launhardt et al. in prep. $^{(f)}$ Cassegrain rotator position angle. There is no information on the uncertainty available; we therefore adopt an uncertainty of 0.1\degr. 
} 
\label{table:astrometric_uncertainties}
\end{table*}
\begin{table*}[tb]
\centering
\caption{Properties of the point-like source, as derived from the sPCA reduction. }
\begin{tabular}{llccccccccc}
\hline \hline
Date       & Instr.& Filter  & $\Delta$RA [mas] &$\Delta$DEC [mas] & Sep[mas] & PA[deg] & $\Delta$ mag & mag$_{\mathrm{app}}$ & S/N  \\
\hline
2012-03-31 & NICI  & L'& 58.7$\pm$10.7& -182.7$\pm$22.2& 191.9$\pm$21.4& 162.2$\pm$3.7 &6.59$\pm$0.42&  14.50$\pm$0.42 & 5.6 \\
2015-05-03 & IRDIS & H2& 83.1$\pm$3.9 & -173.5$\pm$4.3 & 192.3$\pm$4.2 & 154.5$\pm$1.2 &9.35$\pm$0.18&  18.17$\pm$0.18 & 6.3 \\
2015-05-03 & IRDIS & H3& 83.9$\pm$3.6 & -178.5$\pm$4.0 & 197.2$\pm$4.0 & 154.9$\pm$1.1 &9.24$\pm$0.17&  18.06$\pm$0.17 & 8.1 \\
2015-05-31 & IRDIS & H2& 89.4$\pm$6.0 & -178.3$\pm$7.1 & 199.5$\pm$6.9 & 153.4$\pm$1.8 &9.12$\pm$0.24&  17.94$\pm$0.24 & 11.4\\
2015-05-31 & IRDIS & H3& 86.9$\pm$6.2 & -174.0$\pm$6.4 & 194.5$\pm$6.3 & 153.5$\pm$1.8 &9.13$\pm$0.16&  17.95$\pm$0.17 & 6.8 \\
2016-05-14 & IRDIS & K1& 90.2$\pm$7.3 & -170.8$\pm$8.6 & 193.2$\pm$8.3 & 152.2$\pm$2.3 &7.81$\pm$0.31&  16.35$\pm$0.31 & 5.5 \\
2016-05-14 & IRDIS & K2& 95.2$\pm$4.8 & -175.0$\pm$7.7 & 199.2$\pm$7.1 & 151.5$\pm$1.6 &7.67$\pm$0.24&  16.21$\pm$0.24 & 3.6 \\
2016-06-01 & NaCo  & L'& 94.5$\pm$22.0& -164.4$\pm$27.6& 189.6$\pm$26.3& 150.6$\pm$7.1 &6.84$\pm$0.62&  14.75$\pm$0.62 & 2.7 \\
\hline
\hline
\end{tabular}
\label{table:cc_parameters}
\end{table*}
The observed polarised intensity is now an interplay between the polarisation degree and total intensity, and the disk's geometry. Due to the disk's flared geometry, the scattering angles at the far side are closer to values of 90$^{\circ}$ than at the near side \citep[the scattering angles are symmetric around $\theta = 90^{\circ}-\psi-i$ and $\theta = 90^{\circ}-\psi+i$ on the near and far sides, respectively, when $i$ denotes the inclination and $\psi$ the opening angle of the disk, see ][]{2012A&A...537A..75M}. For smaller grains that are in the Rayleigh scattering regime and therefore not strongly forward scattering, this could therefore make the far side appear brighter than the near side \citep[e.g.][]{2010A&A...518A..63M,2012A&A...537A..75M}. On the other hand, the larger the grains that are considered, the more forward scattering they are, and the more the polarised intensity is dominated by the phase function of the total intensity. The near side then becomes the brighter side, the same in total intensity as in polarised intensity. \\
To test this hypothesis, we computed ray-traced \Qphi\ images at 0.7 \mum\ using different grain sizes. We subtracted the central source polarisation after convolving the Stokes $Q$ and $U$ images from the $Q_{\phi}$ frame using a $U_{\phi}$ minimisation. In this procedure, scaled versions of the total intensity I frame are subtracted from the Stokes Q and U model images, such that the absolute value of U in a defined region is minimised. Figure \ref{fig:phf_ZIMPOL_models} compares the resulting images. As expected, the disk model with larger grains ($\gtrsim$ 0.18 $\mu$m) shows strong forward scattering even in polarised light, whereas for the disk containing small grains, the far side appears brighter. As a mono-dispersive grain size distribution would be unrealistic, we also tested a grain size distribution of small grains (0.001-0.15 \mum), and still find that the far side appears brighter than the near side (Fig. \ref{fig:phf_ZIMPOL_models}, lower middle panel). We conclude that if the brightness asymmetry between the east and west sides is real, and not dominated by effects from poor seeing conditions and reduction artifacts, we need predominantly small sub-micron-sized grains ($\lesssim$ 0.15 \mum) to reproduce the observations in a qualitative way. Although we are able to reproduce the qualitative behaviour of the brightness asymmetries, we are not able to reproduce contrast ratios as large as in the observations between the two sides. Furthermore, our brightness contrast is very similar in the VBB-band and J-band (contrary to the observations). \\
In total intensity, on the other hand, the near side is always expected to be brighter than the far side, as even in the Rayleigh scattering regime (where the scattering phase function is symmetric around a minimum at 90\degr \ and where grains are not forward scattering), the scattering angles at the near side are farther from 90\degr \ than at the far side, corresponding to a higher scattering efficiency.  \\
We note that our model is based on Mie scattering. It is certainly worth testing the impact of particles that are not spherical and homogeneous, or that are of a somewhat different chemical composition; for example including water ice mantles that are neglected in our modelling approach. However, this is beyond the scope of this study. \\ 
Summarising, by retaining only small ($\lesssim$ 0.15 \mum) grains in the disk surface layer, we are able to reproduce the brightness contrasts between the east and west sides in a qualitative (although not quantitative) way. The existence of the north/south brightness asymmetry and its different behaviour in the VBB- and J-band on the other hand cannot be explained solely with grain scattering properties, as the scattering angles are expected to be symmetric with respect to the semi-minor axis. One could speculate that the grain properties are different in the north and south region, but more complex modelling would be needed to explain this behaviour. \\
It should be mentioned that the strong butterfly patterns detected in $Q_{\phi}$ and $U_{\phi}$ after correcting for instrumental polarisation effects affect the outer disk. In the $Q_{\phi}$ image, this adds positive signal along the disk semi-major axis, and subtracts signal along the semi-minor axis. The subtraction of this central component is likely imperfect and some residuals may be present in the images of the outer disk in Fig.\ref{fig:obs_PDI}. In addition, all our polarimetric observations suffered from rather poor seeing conditions, which might further influence the apparent brightness distribution in the disk. Therefore, we cannot rule out that artifacts from the data reduction and/or weather conditions affect the azimuthal brightness distribution of the polarimetric datasets.\\

\begin{figure*}[hbt]
    \noindent
    \centering
    \begin{minipage}[t]{1.0\textwidth}
    \includegraphics[width=1.0\textwidth]{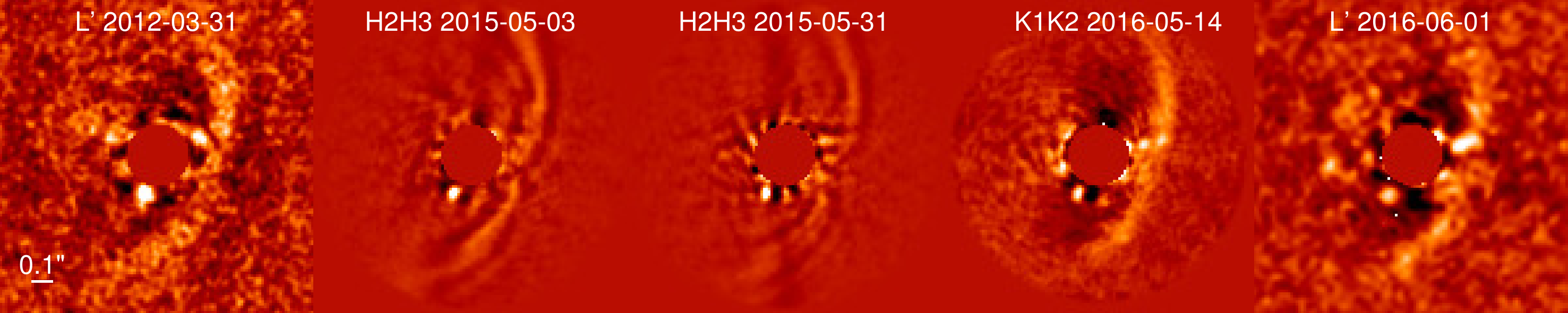}
    \end{minipage}
    \begin{minipage}[t]{1.0\textwidth}
    \end{minipage}
    \caption{Images of the point source detection as retrieved with the sPCA reduction (from left to right): NICI L'-band (2012-03-31), IRDIS H2H3-band (2015-05-03), IRDIS H2H3-band (2015-05-31), IRDIS K1K2-band (2016-05-14), NaCo L'-band (2016-06-01). North is up and east is to the left.  The images were smoothed with a Gaussian kernel of size 0.5$\times$FWHM. 
}\label{obs:CC}
\end{figure*}

\section{Detection of a planetary mass companion}\label{sect:CC}
In the two SHINE IRDIS epochs, we identified a point source at a separation of about 195 mas and a position angle of about 155$^{\circ}$ from PDS~70. Upon analysis of the IRDIS K12 Open Time data, as well as of the NaCo data, the point source was detected around the same location as in the IRDIS data. In addition, we reanalysed archival data taken with Gemini/NICI in L' band on March 31, 2012, published by \cite{2012ApJ...758L..19H}. The authors reported a non-detection of companion candidates (apart from the previously mentioned background source to the north), but their analysis considered only regions outside of $\sim$200 mas. Upon our re-reduction of this dataset, we detected the point source around the expected location. The final detection images obtained with sPCA are shown in Fig.\ref{obs:CC}. The corresponding images retrieved using ANDROMEDA, PCA-SpeCal, and TLOCI are presented in Fig. \ref{obs:CC_algos}. We detected the point source in all our available ADI epochs, spanning a total time range of four years. \\
We also noted another structure present in some of the PCA and TLOCI processed images, located at a similar separation and a  position angle of about 280 degrees. To check the point-like nature of this structure, we processed the data with the ANDROMEDA algorithm, which is optimised for the retrieval of point sources \citep{2015A&A...582A..89C}. Figure \ref{fig:ANROMEDA_SNR} shows the corresponding S/N maps, which are a result of forward modelling under the assumption of the presence of a point-like source. It can be seen that this structure is not consistently recovered by ANDROMEDA in the different epochs at significant S/N. Especially towards longer wavelengths, no other source apart from the above-described source at about 155 degrees is detected with any significance. This implies that the structure found at about 280 degrees is not point-like and, if physical, we can associate this structure with a disk feature. The latter interpretation is supported by the projected proximity to the outer disk ring. 
\subsection{Astrometry}
\begin{figure}[bt]
    \includegraphics[width = 0.5\textwidth]{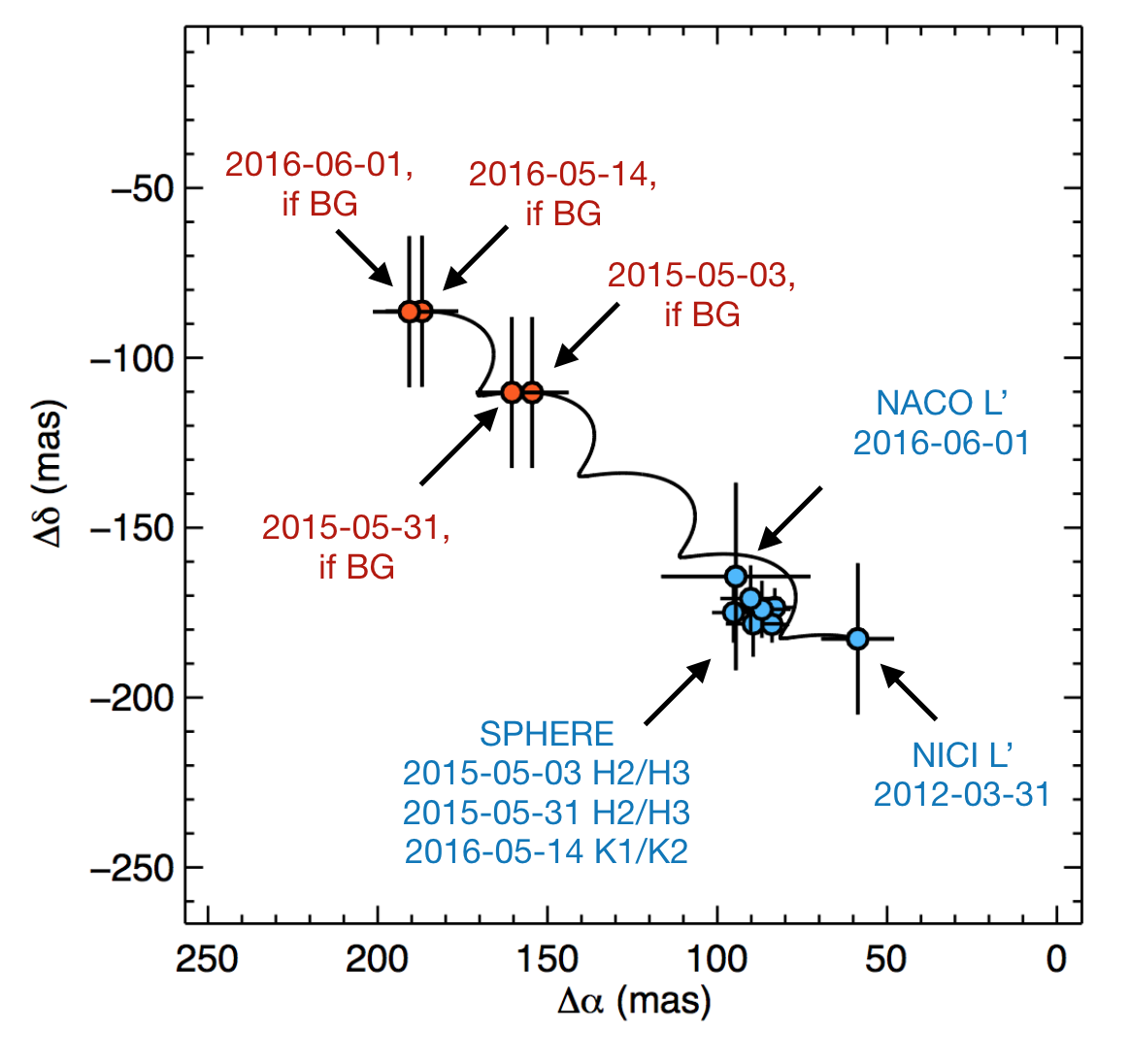}
    \caption{Relative astrometry of the companion. The blue points show the measurements, and the red ones, labelled `BG', the relative position that should have been measured in case the CC detected in the first epoch (NICI) was a stationary background star. }\label{fig:astrometry}
\end{figure}

For the characterisation of the point source we extracted the astrometry and the photometry for all epochs. We detected it with all algorithms (sPCA, ANDROMEDA, PCA-SpeCal and TLOCI), and in the following focus on the analysis of the sPCA reduction. In this sPCA reduction, we divided the image in concentric annuli with a width of 1$\times$FWHM. For each annulus we adjusted the number of modes in such a way that the protection angle was maximised. A maximum number of 20 modes was applied and we set the maximum protection angle to 0.75$\times$FWHM. We extracted the astrometry and photometry by injection of a PSF taken from the unsaturated frames out of the coronagraph with negative flux, as proposed by \cite{2010Sci...329...57L}. Our approach to find the location and flux of the point source consisted of varying the parameters of this negative signal using a predefined grid to minimise the residuals in the resulting sPCA-processed data set. We therefore computed for each parameter set of position and flux the $\chi^2$ value within a segment having a radial and azimuthal extension of 2$\times$FWHM and 4$\times$FWHM around the point source, respectively. To derive uncertainties in the astrometric and photometric values, posterior probability distributions for each parameter were computed following the method described by \cite{2016A&A...591A.108O}. The astrometric uncertainties related to the calibration error take into account the centring accuracy of the stellar position (frame registering was done using the satellite spots for the IRDIS data and fitting a 2D Gaussian to the star in the case of the non-coronagraphic NaCo and NICI data), the detector anamorphism \citep[0.6$\pm$0.02\% in the case of IRDIS,][]{2016SPIE.9908E..34M}, the True North orientation of the images and the uncertainties related to the rotator offset and the pixel scale. The corresponding values are reported in Table \ref{table:astrometric_uncertainties}. We derived the final astrometric uncertainties at each epoch by quadratically summing the errors from these individual contributions. Our astrometric measurements obtained at the different epochs are presented in Table \ref{table:cc_parameters}. As a cross-check, the results of the ANDROMEDA, PCA-SpeCal, and TLOCI reductions are listed in Appendix \ref{app:cc_alt_reductions}. \\
To test whether the point source is part of a physical system with PDS~70, we compared its measured position relative to the star at the different epochs. 
Due to the proper motion \citep[$\mu_{\alpha}\mathrm{cos}\delta$ = -29.7 mas/yr, $\mu_{\delta}$ = -23.8 mas/yr, ][]{2016AA...595A...1G,2018arXiv180409365G}, a stationary background star would have moved by $\sim$160 mas within the given timespan. As the relative motion ($\sim$40 mas during the $\sim$ 4 years observational time span) differs significantly from the prediction for a stationary background object, the astrometric results strongly imply that the point source is comoving with PDS~70. The measurements, together with the expected trajectory for a background star relative to PDS~70, are displayed in Fig. \ref{fig:astrometry}. Further, the probability of detecting at least one background contaminant of similar brightness or brighter within the mean separation of the companion is less than 0.033$\%$ according to the Besan\c{c}on galactic population model \citep{2003A&A...409..523R}. \\
The relative position as measured in the NICI data taken in 2012 does not coincide with the positions derived from the SPHERE and NaCo observations performed in 2015 and 2016 within the 1-$\sigma$ uncertainties. This difference in measured position between the epochs is possibly due to orbital motion. The point source is detected at a mean projected separation of $\sim$195 mas, corresponding to $\sim$22 au. The orbital period of such a bound object, assuming a stellar mass of 0.76 $\mathrm{M_{\odot}}$, would be $\sim$119 years. For a face-on circular orbit, this implies a displacement of $\sim$3$^{\circ}$ per year, resulting in a total change of position angle of 12.5\degr \ within the time covered by our observations, which is in good agreement with the observations. Further, the observed change in position angle is in clockwise direction, which corresponds to the sense of rotation of the disk \citep{2015ApJ...799...43H}. Therefore, this displacement is consistent with an object on a circular face-on orbit rotating in the same sense as the disk. However, regarding the relatively large uncertainties on the astrometry and the short time span covered by our data, detailed orbital fitting exploring the possibility of an inclined and/or eccentric orbit will be performed in a follow-up study on this source \citep{mueller2018}. Although the possibility of the point source being a background star with almost the same proper motion as PDS~70 is very small, only the detection of orbital motion over a significant part of the orbit will allow to fully exclude the background star scenario.

\subsection{Photometry}

\begin{figure*}[hbt]
    \noindent
    \centering
    \begin{minipage}{1.0\textwidth}
    \includegraphics[width =0.33\textwidth]{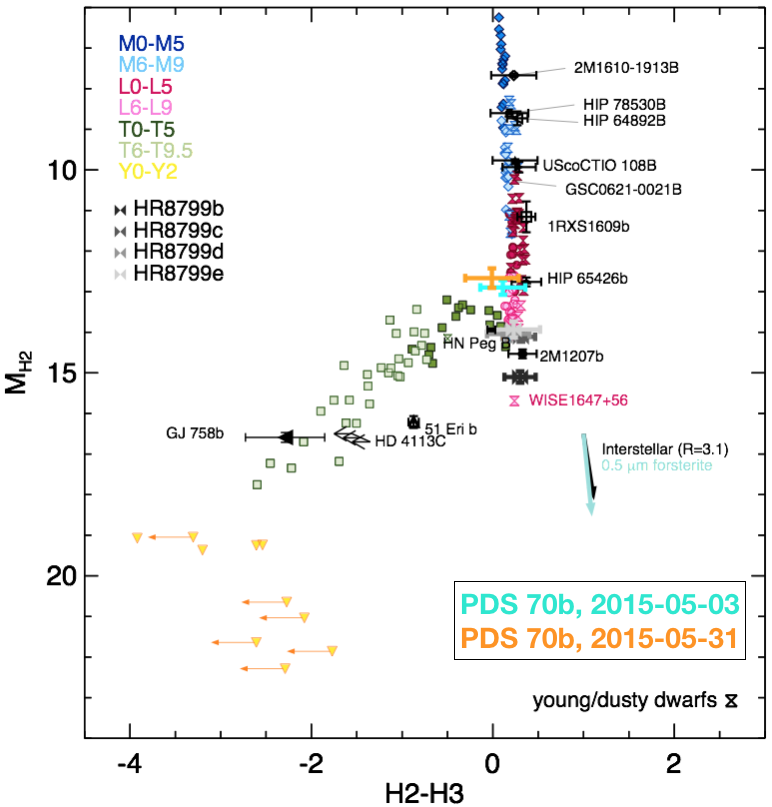}
    \includegraphics[width =0.33\textwidth]{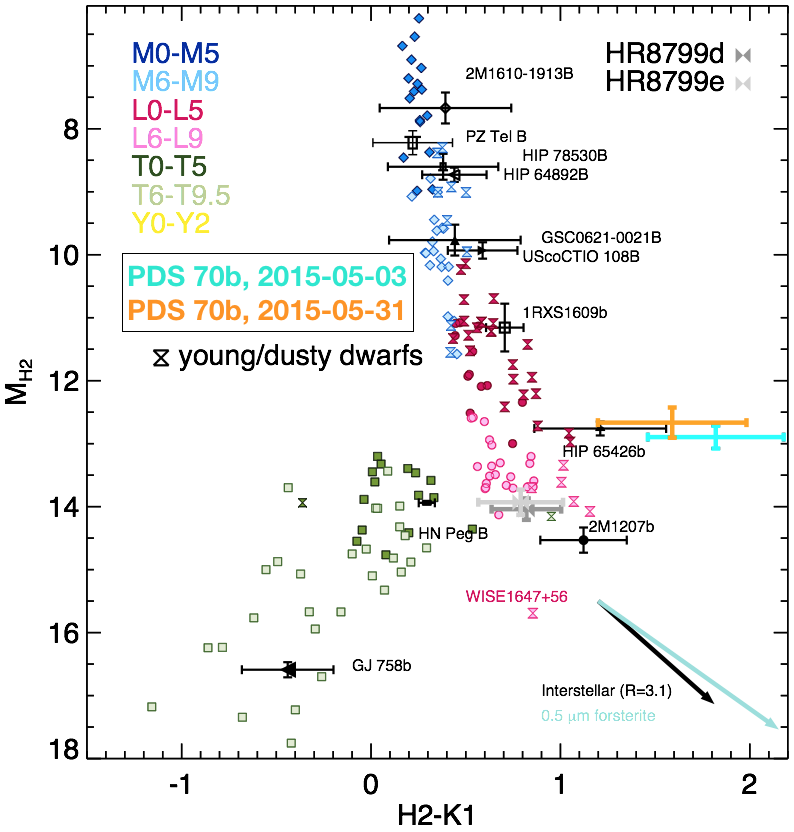}
    \includegraphics[width =0.33\textwidth]{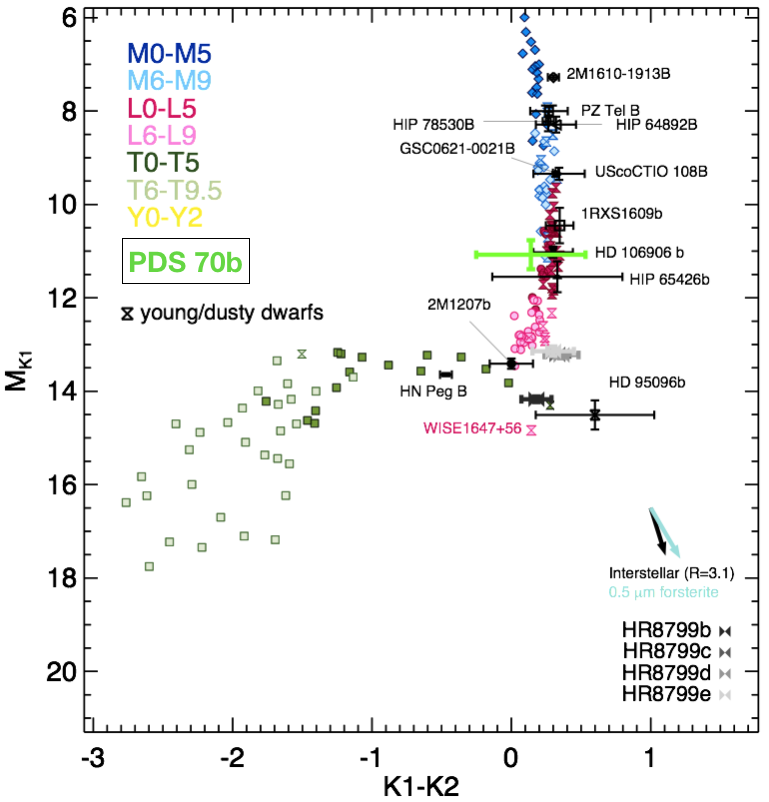}
    \end{minipage}
    \caption{
    Colour-magnitude diagrams considering the SPHERE H2-H3 (\textit{left}), H2-K1 (\textit{centre}), and K1-K2 (\textit{right}) colours, comparing PDS~70b with the photometry of M, L, and T field dwarfs, young companions and red dwarfs. The diagrams are overlaid with reddening vectors from interstellar extinction and 0.5 $\mu$m fosterite grains. See \cite{2018arXiv180105850C} and \cite{2018arXiv180700657B} for details about the CMDs. 
  }\label{fig:CMD}
\end{figure*}

Our current information on the physical properties of the companion candidate relies on the H, K, and L' photometry as derived from our SPHERE/IRDIS, NaCo, and NICI images. It is marginally detected in the IFS data, when the channels corresponding to J-band and H-band are collapsed. Due to the large uncertainties, this data is not considered here. The low S/N detection of the companion candidate in the IFS data can be explained by its faintness and red colours, the larger IFS thermal background (IFS is not cooled contrary to IRDIS), and the smaller IFS pixel scale (7.46 mas/pixel vs $\sim$12.25 mas/pixel). HD~95086~b offers a similar case of a faint companion with red colours for which a detection with IRDIS is achieved in the K-band in individual observation sequences whereas the detection with IFS in the J- and H-bands required the combination of several epochs \citep{2018arXiv180105850C}. \\
The companion has very red colours, with a magnitude difference of H2-K1=1.82$\pm$0.36 mag and 1.59$\pm$0.39 mag, considering the first and second SHINE H-band epochs, respectively. Accordingly, we measured a magnitude difference of H2-L' = 3.67 $\pm$0.46 mag and 3.44$\pm$0.48 mag (considering the NICI L'-band photometry). The properties of the companion are further discussed in Sect. \ref{comp:properties}. 

\subsection{The nature of the point-like source}

\begin{figure*}[hbt]
    \noindent
    \centering
    \begin{minipage}{1.0\textwidth}
    \includegraphics[width = .33\textwidth]{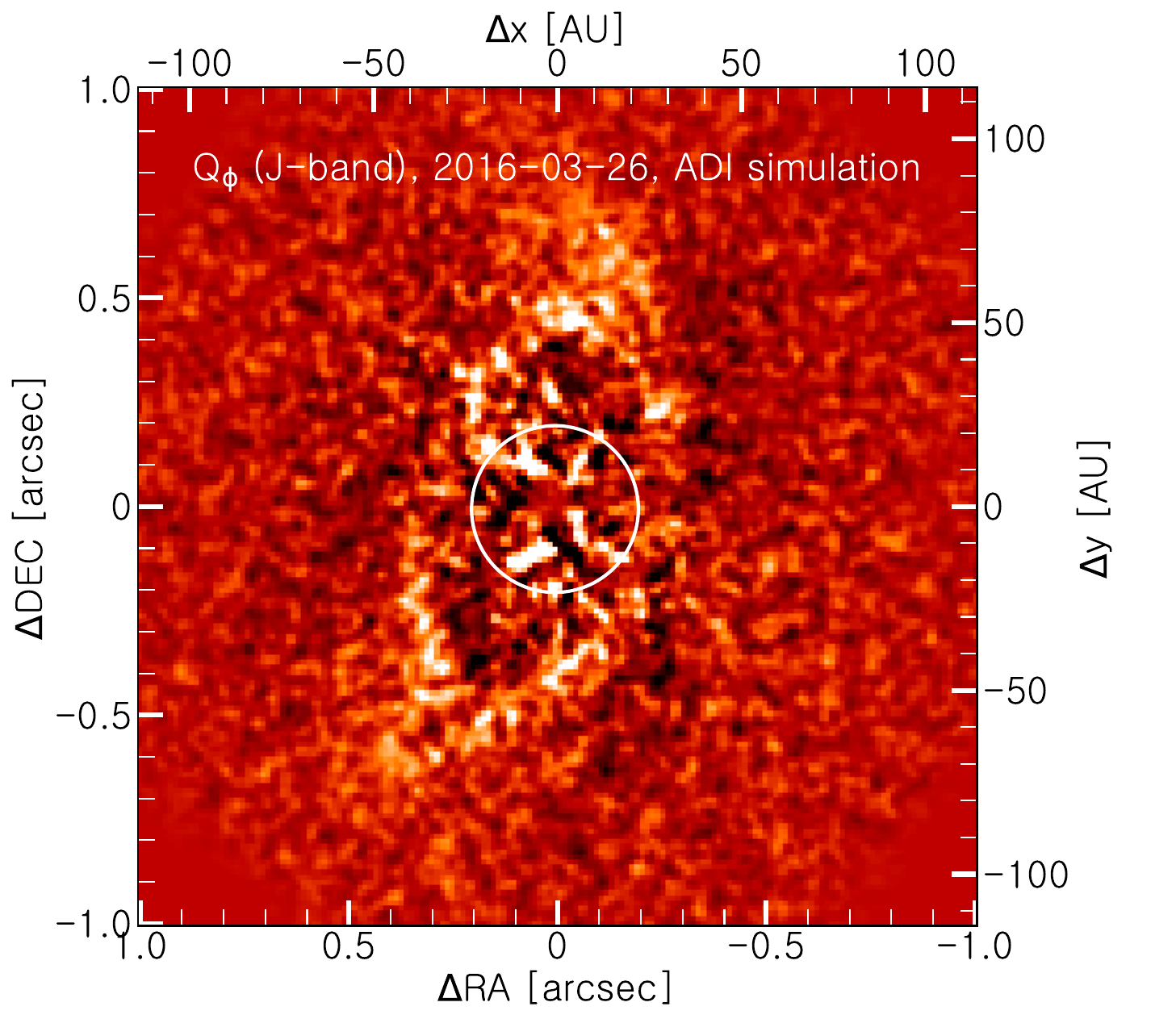}
    \includegraphics[width = .33\textwidth]{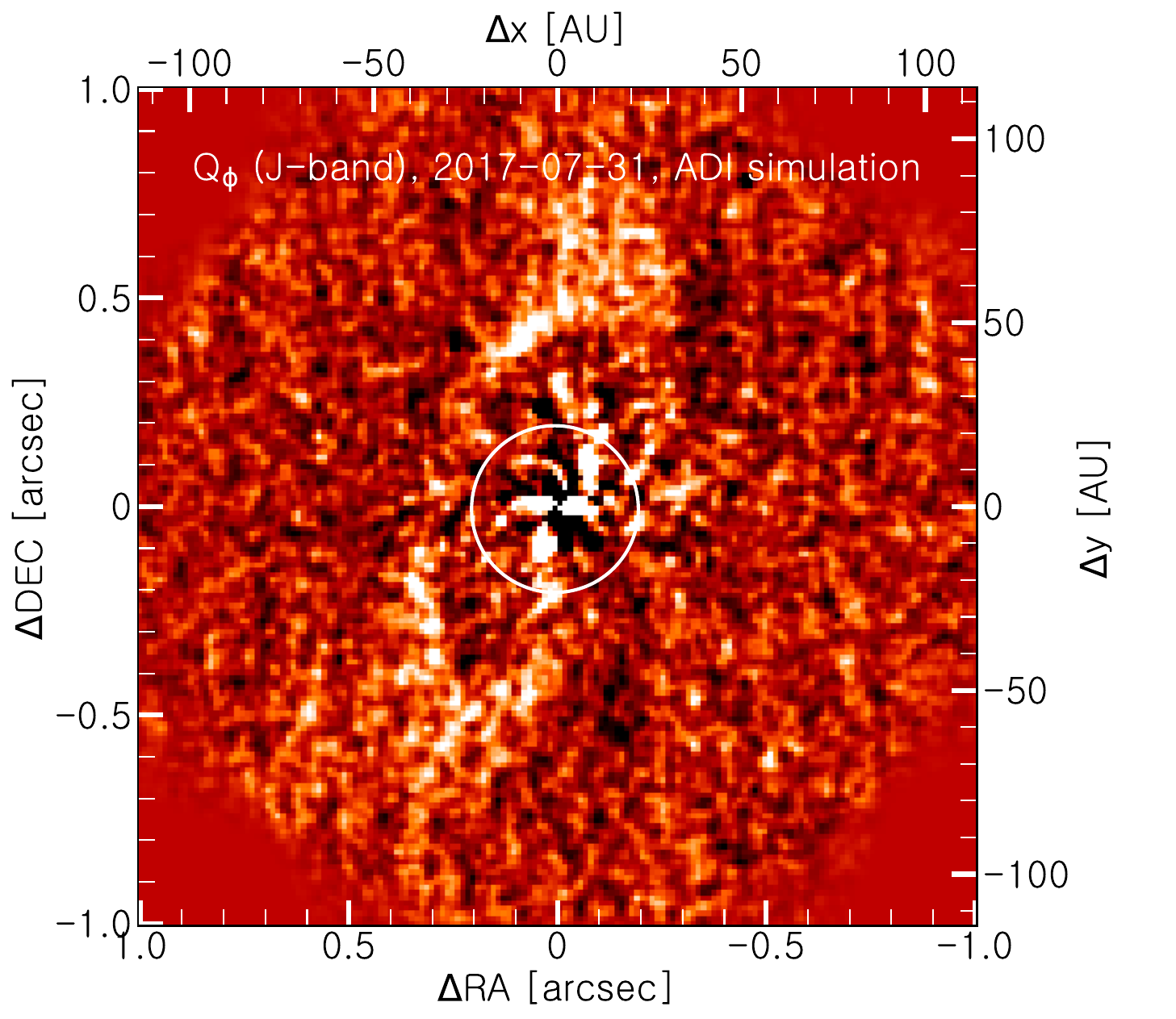}
    \includegraphics[width = .37\textwidth]{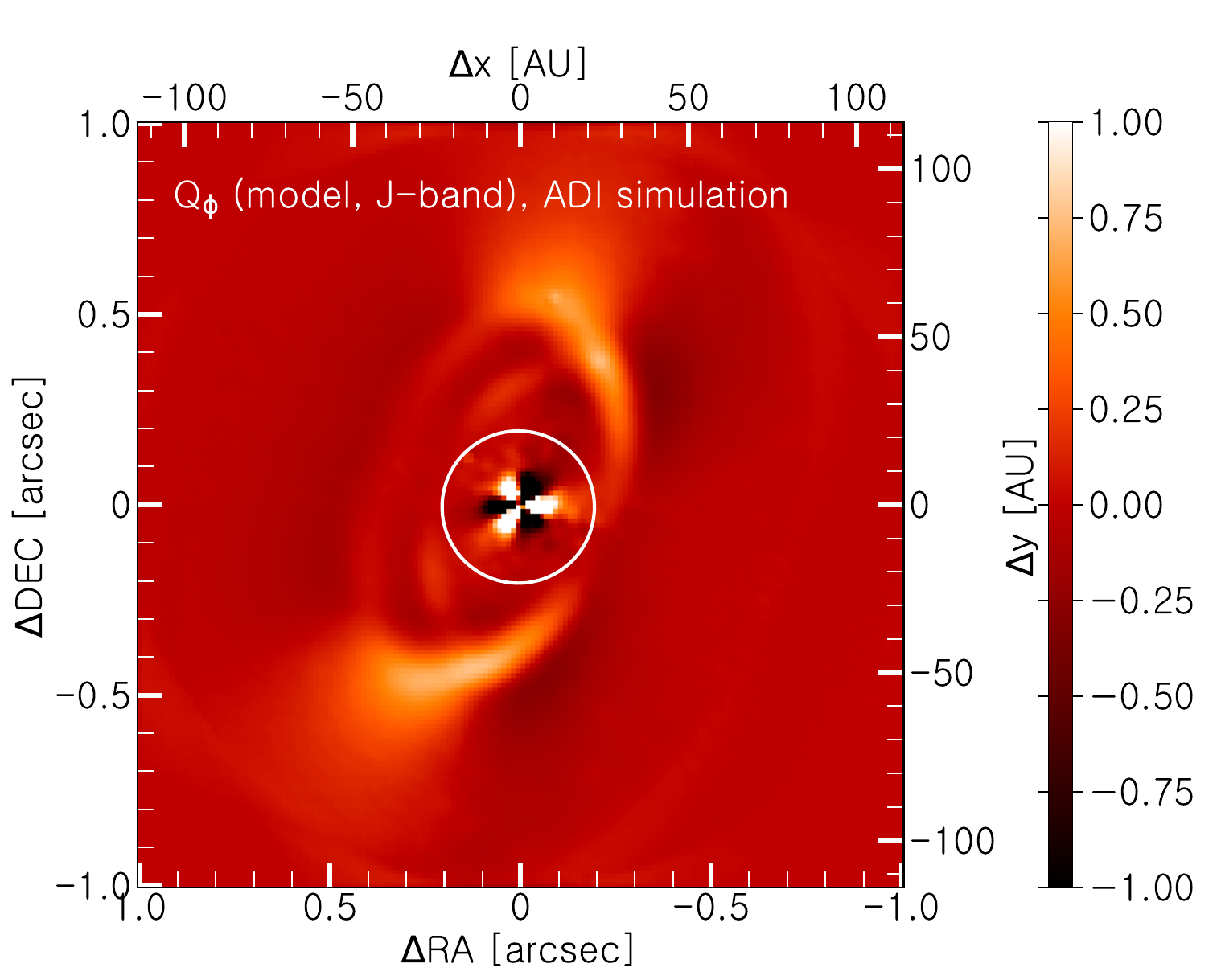}
    \end{minipage}      
    \caption{
cADI simulation for the $\mathrm{Q_{\phi}}$ images of the coronagraphic (left) and non-coronagraphic (middle) J-band observations, after subtracting the central source polarisation. 
The right panel shows the same simulation for our model image. This image was generated by convolving the Stokes Q and U images with a real IRDIS J-band PSF, computing the \Qphi \ and \Uphi \ images, subtracting the central source polarisation by applying a \Uphi \ minimisation, and finally applying the cADI algorithm. The model included the presence of an inner disk with an outer radius of 2 au. The white circle marks a radial distance of 200 mas, approximately the separation of the companion. The colour stretch was adapted individually for visibility purposes. North is up and east is to the left. 
}\label{test_ADI}
\end{figure*}

\begin{figure}[hbt]
    \noindent
    \centering
    \begin{minipage}{0.5\textwidth}
    \includegraphics[width = 1.0\textwidth]{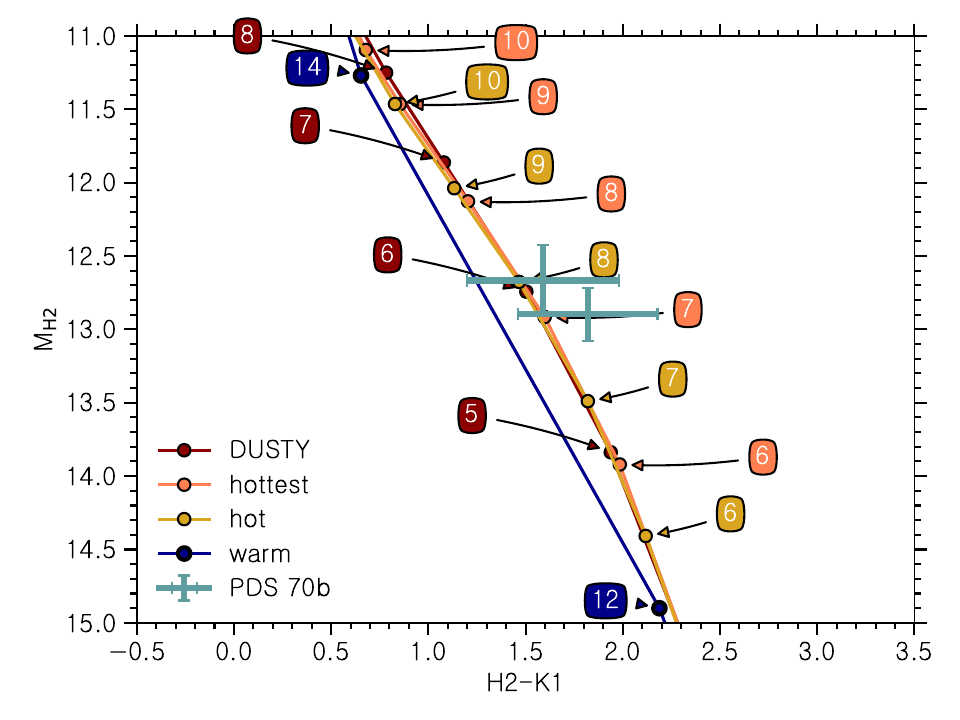}
    \includegraphics[width = 0.96\textwidth]{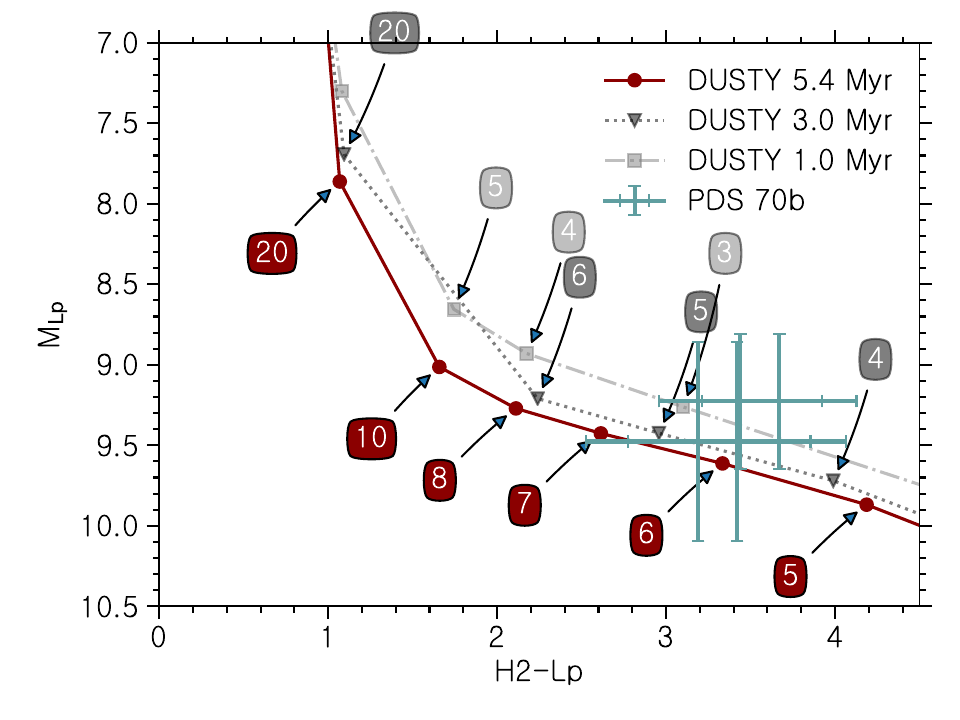}
    \end{minipage}      
    \caption{Photometry of PDS~70b in comparison with evolutionary models, evaluated at an age of 5.4 Myr. The green crosses mark the measurements corresponding to the different H2 and $\mathrm{L^{'}}$ epochs. The upper panel shows the `hottest', `hot', and `warm' models from \cite{2017A&A...608A..72M}, as well as the DUSTY model in a H2 vs. H2-K1 diagram. The lower panel compares the L' vs. H2-L' colour of the DUSTY model with the measured photometry. The corresponding masses in units of \Mjup \ are indicated on the coloured labels. } \label{fig:color_model}
\end{figure}

Due to the detection of the point source at multiple epochs and using several different instruments, filter bands, and image post-processing algorithms at about the same location, we can exclude that the source is due to an instrumental or atmospheric artifact (speckle). Furthermore, the discrepancy of the relative position of the point source with respect to the predicted trajectory of a stationary background star strongly implies a gravitationally bound object. 
Several of the proposed companion candidates within protoplanetary disks are currently under debate, because they are not detected consistently at all available wavelengths. A reason for possible confusion is that the ADI process acts as a spatial frequency filter and enhances sharp and asymmetric disk features; as shown by \cite{2012A&A...545A.111M}, applied on extended disk structures, it can cause distortions or even create artifacts that are not related to physical structures. Disk structures such as rings, spiral arms, or clumps, when processed with ADI, could therefore lead to a misinterpretation of point-like sources. One of the point sources detected around HD~169142 \citep{2014ApJ...792L..22B,2014ApJ...792L..23R} was shown to be related to an inhomogeneous ring structure in the inner region of the disk by \cite{2018MNRAS.473.1774L}. In addition, these authors found that an additional compact structure in that system detected at about 100 mas projected separation could possibly be related to a further ring structure at the given separation. Two companion candidates were also found around HD~100546 \citep{2013ApJ...766L...1Q,2015ApJ...807...64Q,2014ApJ...791..136B,2015ApJ...814L..27C}, but remain debated under consideration of recent GPI and SPHERE observations, as they do not appear point-like at all wavelengths \citep[Sissa et al. subm.]{2017AJ....153..244R,2017AJ....153..264F,2017RNAAS...1...40C}. Regarding LkCa15, three companion candidates have been reported in the literature \citep{2012ApJ...745....5K,2015Natur.527..342S}. \cite{2016ApJ...828L..17T} found that the location of the planet candidates coincides with the bright, near side of the inner component of the LkCa15 disk and conclude that this inner disk might account for at least some of the signal attributed to the protoplanets detected by \cite{2015Natur.527..342S}. Due to the detection of $H_{\alpha}$ emission at the location of LkCa15b, only this candidate makes a convincing case for a protoplanet in this system.  \\
All these debates illustrate that a careful analysis of companion candidates located in protoplanetary disks with respect to a possible link to disk features is required. We therefore address the hypothesis that some asymmetric dust structure at or close to the given location of the point source is responsible for our detection. As stated above, we are not able to resolve the detailed structure of the inner disk with our PDI observations, but suspect the inner disk to be smaller than 17 au in radius. To test the impact of the inner disk signal on the structures seen in ADI, we follow the approach by \cite{2018MNRAS.473.1774L}, and simulate a cADI observation using the IRDIS PDI J-band $Q_{\phi}$-images, after subtracting the central source polarisation. For this purpose, we created a datacube whose 50 frames correspond to identical copies of the PDI \Qphi \ image, rotated by the respective parallactic angles encountered during the ADI epoch of May 31, 2015. We then subtracted the median of this datacube from each single frame, before de-rotating them and computing their median. \\
In addition, we applied the same procedure to the \Qphi \ model image, computed at 1.25 \mum. We convolved our image with the total intensity frame acquired during the non-coronagraphic J-band observations and subtracted the central source polarisation using a \Uphi-minimisation before applying the cADI algorithm. The inner disk in the model configuration used for this test extends out to 2 au. The result is shown in Fig. \ref{test_ADI}. There is no prominent structure appearing at the distance of interest ($\sim$200 mas). While this test does not allow us to completely rule out a disk structure as the cause of this feature, there is at least no obvious polarised inner disk structure that would create this kind of artifact. One further argument against the hypothesis of the companion being a disk feature is the fact that we do not detect strong polarised signal at the location of the companion in the PDI data, which would be consistent with the signal detected in ADI being of thermal origin. We therefore conclude that, given the present data, a physically bound companion is the most plausible explanation for the detected point source, and refer to it as PDS~70b hereafter. 
\footnote{PDS 70 has also recently been observed by MagAO, revealing a potential detection of H$\alpha$ emission at the expected location of PDS70~b (Wagner et al., subm.). If this detection is due to the accretion of gas on the planetary object, this would provide further evidence that the object is neither a disk feature nor a background star.}

\subsection{Companion properties}\label{comp:properties}

Figure \ref{fig:CMD} shows the location of PDS~70b in SPHERE H-band and K-band-based colour-magnitude diagrams (CMD). The diagrams are complemented with the synthetic photometry of M, L, and T dwarfs, as well as with the measurements of young companions and red dwarfs of the Sco-Cen association and other regions. We refer the reader to \cite{2018arXiv180105850C} and \cite{2018arXiv180700657B} for details regarding these diagrams. The diagrams show that the absolute H2 magnitude of the companion is consistent with those of L-type dwarfs.
PDS~70b is located between the $\sim$5-11 Myr-old, $\sim$8-14 \Mjup \ planet 1RXS1609b \citep[][]{2008ApJ...689L.153L,2010ApJ...719..497L,2015ApJ...802...61L} and the 30 Myr-old $\sim$7 \Mjup \ planets HR~8799 c,d,e \citep[see][]{2016A&A...587A..58B}. The location of PDS~70b in the H2-H3 CMD is remarkably close to the recently discovered dusty giant planet HIP~65426b \citep{2017A&A...605L...9C}. Indeed, HIP~65426b's mass (6-12 \Mjup) derived from evolutionary models is similar to the one of PDS~70~b (see below), although significantly older \citep[14$\pm$4 Myr, ][]{2017A&A...605L...9C}. In addition, the K1-K2 diagram reveals a similar photometry to the $11\pm2$ \Mjup \ massive and $13\pm2$ Myr old companion HD~106906b \citep{2014ApJ...780L...4B}.\\  
The colours of PDS~70b are very red. Its H2-K1 colour is redder than most L dwarfs in the field, and is consistent with the very red companions to CD-35 2722 \citep{2011ApJ...729..139W} and 2M1207 \citep{2004A&A...425L..29C}, as well as with HIP~65426 \citep{2017A&A...605L...9C} within the uncertainties. If due to a photosphere, the red colour is only compatible with an L-type object or with a reddened background object, but this latter possibility is very unlikely due to the proper motion test. The absolute L'-band magnitude is brighter than most of the detected companions and consistent with those of late M- to early L-type objects, but again significantly redder than these sources. The H2-L' colour is as red as the $>$ 50 Myr-old, very dusty companion to HD~206893 (H2-L'=3.36$\pm$0.18 mag), which is one of the reddest brown dwarf companions known \citep{2017A&A...597L...2M,2017A&A...608A..79D}. Therefore, the location of PDS~70b on the colour-magnitude diagrams is quite unusual. However, it should be kept in mind that only very few of these objects are of similarly young age as PDS~70, and none of the above objects are detected within the transition disk of its host. PDS~70b might therefore be the only of these objects that is directly observed during its formation process. \\
In order to estimate the mass of the companion, we compared the photometry of PDS~70b to the Bern Exoplanet (BEX) evolution models. These tracks are obtained from the synthetic planetary populations of \cite{2017A&A...608A..72M}, which predict the post-formation planetary luminosity as a function of time, considering different efficiencies of the accretional heating during the formation process and including the effect of deuterium burning.  
The planets formed in the planetary population synthesis are classified in four different populations (`hottest', `hot', `warm' `coldest'), according to their luminosity as a function of mass at the moment when the disk disappears (see \citealt{2017A&A...608A..72M} for details). The planets according to the `hottest' and `coldest' populations have the highest and lowest luminosities, and correspond to the traditional hot- and cold start models \citep[e.g.][]{2007ApJ...655..541M,2000ApJ...542..464C,2003A&A...402..701B}. They describe the two extreme cases where the entire gas accretion shock luminosity is either deposited in the planet's interior, or radiated away during the formation process. These two populations are superseded by the more realistic cases of the `hot' and `warm' populations, which are representative for cases with intermediate initial entropies between the extreme `hot' and `cold' start models. For our comparison we made use of the `hottest', `hot', and `warm' populations, but discarded the `coldest' population. For this scenario, a planet mass larger than 10 \Mjup \ would be required to reproduce the observed magnitudes of PDS~70b. However, to be classified into the `coldest' population requires small core masses which do not develop into planets with such high masses in the planetary population synthesis models \citep[see][Fig. 13]{2017A&A...608A..72M}. In addition, observations suggest that the pure cold start formation is in reality not the preferred formation mechanism and the truth most probably lies somewhere between the two extrema of purely hot and cold start \citep{2014A&A...567L...9B,2017A&A...603A..57S}. From the theoretical side, recent detailed simulations of the accretion shock suggest that hot starts are preferred (e.g. \citealp{2017ApJ...836..221M}, Marleau et al. in prep.). To follow the post-formation cooling, the outcome of the population synthesis was combined with the boundary conditions for the atmospheric structure from the COND models \citep{2003A&A...402..701B}. The synthetic SPHERE magnitudes were then computed using the DUSTY atmospheric model \citep{2000ApJ...542..464C}. These results were linearly interpolated in time to the stellar age (5.4 Myr). Figure \ref{fig:color_model} compares the photometry of PDS~70b with the synthetic colours of the `hottest', `hot', and `warm' tracks from \cite{2017A&A...608A..72M}, as well as with the original DUSTY model of \cite{2000ApJ...542..464C}. We find a mass between 5 and 9 \Mjup \ for the hot start models (`hottest', `hot', `DUSTY'), and a mass between 12 and 14 \Mjup \ for the coldest (`warm') population considered, implying that in the case of lowest entropy, deuterium burning might play a role. It is important to note that the evolutionary tracks do not take into account the time needed for the planet to form. Since this may take up to several million years, the stellar age is only an upper limit on the age of the planet, and consequently, the estimated masses should be considered as conservative upper limits. \\
For completeness, Fig. \ref{fig:color_model} (lower panel) shows the  H2-L' colour of PDS~70b in comparison with the DUSTY tracks, which implies a similar mass estimate (5-8 \Mjup)\footnote{We note that the BEX tracks are currently not available for the NaCo magnitudes, and are therefore not considered for the comparison in the L'-H2L' diagram.}. We also plotted the evolutionary tracks corresponding to a planetary age of 1 and 3 Myr to illustrate the mass range in which the planet would be found in the case that it has formed considerably more recently than the star (down to 2-4 \Mjup \ for a planetary age of 1 Myr). We also compared the H-band photometry to the warm-start models from \cite{2012ApJ...745..174S}, as well as the hot-start COND models from \cite{2003A&A...402..701B}, resulting in a similar finding (5-10 \Mjup \ for the former, and 4-5 \Mjup \ for the latter). However, the colours from the COND model do not match those observed for PDS~70b since they are significantly redder than predicted by the COND models, which suggests the presence of a dusty or cloudy atmosphere. We emphasise that none of these models considers the presence of circumplanetary material, which could affect the observed SED and the corresponding mass estimate. The presence of a circumplanetary disk could also cause an IR excess in the object's SED pushing the photometry towards redder colours. Future ALMA observations will allow us to search for the presence of such material around PDS~70b. \\
Finally, we used the Exoplanet Radiative-convective Equilibrium Model (Exo-REM) to analyse the SED of PDS~70b \citep{2015A&A...582A..83B,2017ApJ...850..150B}. We performed a grid search to determine the parameters that best minimise the $\mathrm{\chi^2}$ taking into account all photometric points from the H2 to the L-band \citep{2015A&A...582A..83B}. All the determined radii were larger than or equal to 2 \Rjup, which is a large value compared to the evolutionary models. Therefore, instead of simply minimising the $\mathrm{\chi^2}$ to find the radius, we determined the minimal radius that gives a spectrum similar to the observation at 1, 3, and 5-$\sigma$ (when applicable). We obtained at 5-$\sigma$ a radius > 1.3 \Rjup, a surface gravity of $\mathrm{log_{10}(g)}$ = 3.9$\pm$0.9 dex, and a temperature of $\mathrm{T_{eff} = 1200\pm200 K}$. Our grid takes into account solar metallicity and high cloud absorption ($\tau_{\mathrm{ref}}$=3), suggested by the fact that the object is redder than usual on the CMD. We note that the photometry is also in good agreement with a simple blackbody with a temperature range between 1150 and 1350 K, further indicating a very dusty or cloudy atmosphere with few spectral features. Also in the blackbody case, large effective radii of several \Rjup \ are needed to fit the absolute flux density, which, again, may be explained by the possible existence of spatially unresolved circumplanetary material contributing to the measured flux. Our atmospheric models are in this respect oversimplified and the possible presence of circumplanetary material would require substantial modifications of the underlying models, which is beyond the scope of our paper, but will be discussed in a forthcoming paper \citep{mueller2018}. 

\section{Summary and conclusions}\label{sect:conclusions}
PDS~70 is a young T-Tauri star hosting a known transition disk with a large gap. Transition disks are thought to host gap-carving planets, and are therefore prime targets to observe ongoing planet formation and planet-disk interactions. \\
We have presented VLT/SPHERE optical and NIR observations in polarimetric differential imaging mode, carried out with SPHERE/ZIMPOL in the VBB band and SPHERE/IRDIS in the J-band. In addition, we have obtained total intensity images with SPHERE/IRDIS in H2H3 and K1K2 dual-band imaging, with simultaneous spectro-imaging using IFS working in YJ-band and YJH-band, respectively. Our observations of the PDS~70 system obtained within this work have been complemented with data taken with VLT/NaCo and archival Gemini/NICI observations in the L'-band using angular differential imaging.
The presented data comprise eight different epochs spanning a time range of five years, leading to the following results: \\

The disk is clearly detected in all data sets presented in this work and resolved in scattered light with high angular resolution. We confirm the previously reported gap with a size of $\sim$ 54 au.
We detect for the first time scattered light from the inner disk. By comparison with our radiative transfer model, we derive that the position angle of the inner disk is approximately the same as the outer disk. We also infer that the inner disk is not seen pole-on and has a maximum outer radius of $<$ 17 au.
The disk's far side is brighter than the near side in PDI (VBB-band, J-band), whereas the disk's near side is brighter in ADI (H-band, K-band, L'-band). We suggest that this can be explained by the flared geometry of the disk in connection with Rayleigh scattering from small, sub-micron-sized grains. \\
We detect a point source at approximately 195 mas separation and 155\degr \ position angle. The detection is achieved at five different epochs, including the SPHERE/IRDIS, Gemini/NICI and VLT/NaCo instruments in the H, K and L'-band filter.
The astrometry of the point source implies that the confusion with a reddened background object is unlikely, and that the object is bound. Due to the astrometric coverage of 4 years, we might see first hints of orbital motion. Astrometric follow-up observations will be performed to confirm the orbital motion and to constrain the orbital parameters. \\
The photometry of the companion shows evidence of very red colours. Comparison with evolutionary models suggests that the photometry is most compatible with a young planetary-mass body with a dusty or cloudy atmosphere.\\


\begin{acknowledgements}
SPHERE is an instrument designed and built by a consortium consisting of IPAG (Grenoble, France), MPIA (Heidelberg, Germany), LAM (Marseille, France), LESIA (Paris, France), Laboratoire Lagrange (Nice, France), INAF -- Osservatorio di Padova (Italy), Observatoire astronomique de l'Universit\'e de Gen\`eve (Switzerland), ETH Zurich (Switzerland), NOVA (Netherlands), ONERA (France), and ASTRON (Netherlands), in collaboration with ESO. SPHERE was funded by ESO, with additional contributions from the CNRS (France), MPIA (Germany), INAF (Italy), FINES (Switzerland), and NOVA (Netherlands). SPHERE also received funding from the European Commission Sixth and Seventh Framework Programs as part of the Optical Infrared Coordination Network for Astronomy (OPTICON) under grant number RII3-Ct-2004-001566 for FP6 (2004--2008), grant number 226604 for FP7 (2009--2012), and grant number 312430 for FP7 (2013--2016). 
This work has made use of the SPHERE Data Centre, jointly operated by OSUG/IPAG (Grenoble), PYTHEAS/LAM/CeSAM (Marseille), OCA/Lagrange (Nice) and Observatoire de Paris/LESIA (Paris) and supported by a grant from Labex OSUG@2020 (Investissements d’avenir – ANR10 LABX56). 
A.M. acknowledges the support of the DFG priority program SPP 1992 "Exploring the Diversity of Extrasolar Planets" (MU 4172/1-1).
This research has made use of NASA's Astrophysics Data System Bibliographic Services of the SIMBAD database, operated at CDS, Strasbourg, France.
This publication makes use of VOSA, developed under the Spanish Virtual Observatory project supported from the Spanish MICINN through grant AyA2011-24052.
This work has made use of data from the European Space Agency (ESA) mission {\it Gaia} (\url{https://www.cosmos.esa.int/gaia}), processed by the {\it Gaia} Data Processing and Analysis Consortium (DPAC,\url{https://www.cosmos.esa.int/web/gaia/dpac/consortium}). Funding for the DPAC has been provided by national institutions, in particular the institutions participating in the {\it Gaia} Multilateral Agreement.
J.-L. B. acknowledges the support of the UK Science and Technology Facilities Council. 
T.B. acknowledges funding from the European Research Council (ERC) under the European Union’s Horizon 2020 research and innovation programme under grant agreement No 714769. 
C.M. acknowledges the support from the Swiss National Science Foundation under grant BSSGI0$\_$155816 ``PlanetsInTime''. Parts of this work have been carried out within the frame of the National Center for Competence in Research PlanetS supported by the SNSF.
J.\,O. acknowledges financial support from the ICM (Iniciativa Cient\'ifica Milenio) via the N\'ucleo Milenio de Formaci\'on Planetaria grant, from the Universidad de Valpara\'iso, and from Fondecyt (grant 1180395). 
We acknowledge support from the “Progetti Premiali”  funding  scheme  of  the  Italian  Ministry  of  Education, University, and Research. 
This work has been supported by the project PRIN-INAF 2016 The Cradle of Life - GENESIS- SKA (General Conditions in Early Planetary Systems for the rise of life with SKA). 
D.M. acknowledges support from the ESO-Government of Chile Joint Comittee program ``Direct imaging and characterization of exoplanets''.
A.Z. acknowledges support from the CONICYT + PAI/ Convocatoria nacional subvenci\'on a la instalaci\'on en la academia, convocatoria 2017 + Folio PAI77170087.

\end{acknowledgements}

\bibliography{bibliography}
\bibliographystyle{aa} 

\appendix

\section{PDS~70b astrometric and photometric results}\label{app:cc_alt_reductions}

\begin{table*}[h]
\centering
\caption{Comparison of photometry and astrometry of the companion candidate, as derived from the sPCA, ANDROMEDA, PCA-SpeCal and TLOCI reductions. }
\begin{tabular}{llccccccc}
\hline \hline
\textbf{Date}      & \textbf{Instrument}& \textbf{Filter}&\textbf{Sep[mas]}&\textbf{PA[deg]}&\textbf{$\Delta$ mag}& \textbf{S/N}  \\
\hline
Results from sPCA \\
\hline
2012-03-31 & NICI  & L' & 191.9$\pm$21.4&162.2$\pm$3.7 &6.59$\pm$0.42& 5.6 \\
2015-05-03 & IRDIS & H2 & 192.3$\pm$4.2 &154.5$\pm$1.2 &9.35$\pm$0.18& 6.3 \\
2015-05-03 & IRDIS & H3 & 197.2$\pm$4.0 &154.9$\pm$1.1 &9.24$\pm$0.17& 8.1 \\
2015-05-31 & IRDIS & H2 & 199.5$\pm$6.9 &153.4$\pm$1.8 &9.12$\pm$0.24& 11.4\\
2015-05-31 & IRDIS & H3 & 194.5$\pm$6.3 &153.5$\pm$1.8 &9.13$\pm$0.16& 6.8 \\
2016-05-14 & IRDIS & K1 & 193.2$\pm$8.3 &152.2$\pm$2.3 &7.81$\pm$0.31& 5.5 \\
2016-05-14 & IRDIS & K2 & 199.2$\pm$7.1 &151.5$\pm$1.6 &7.67$\pm$0.24& 3.6 \\
2016-06-01 & NaCo  & L' & 189.6$\pm$26.3&150.6$\pm$7.1 &6.84$\pm$0.62& 2.7 \\
\hline
Results from ANDROMEDA \\
\hline
2012-03-31 & NICI  & L’ & 211.1$\pm$3.5 &162.7$\pm$0.3 &6.85$\pm$0.32& 4.3\\
2015-05-03 & IRDIS & H2 & 191.7$\pm$3.3 &154.3$\pm$0.2 &9.69$\pm$0.25& 5.5\\
2015-05-03 & IRDIS & H3 & 189.7$\pm$2.6 &154.4$\pm$0.1 &9.47$\pm$0.25& 5.0\\
2015-05-31 & IRDIS & H2 & 200.6$\pm$2.9 &153.1$\pm$0.2 &9.49$\pm$0.20& 6.1\\
2015-05-31 & IRDIS & H3 & 194.3$\pm$2.9 &153.2$\pm$0.2 &9.35$\pm$0.17& 6.9\\
2016-05-14 & IRDIS & K1 & 190.8$\pm$1.6 &152.1$\pm$0.2 &7.81$\pm$0.21& 6.2\\
2016-05-14 & IRDIS & K2 & 195.4$\pm$2.3 &152.0$\pm$0.2 &7.51$\pm$0.25& 4.7\\
2016-06-01 & NaCo  & L' & 148.2$\pm$8.4 &152.0$\pm$0.5 &5.60$\pm$0.32& 3.6\\ 
\hline
Results from PCA (SpeCal) \\
\hline
2012-03-31 & NICI  & L' & 190.3$\pm$12.3&160.6$\pm$3.7 &7.0 $\pm$0.1 & 2.2\\
2015-05-03 & IRDIS & H2 & 206.5$\pm$4.8 &156.8$\pm$1.3 &9.7 $\pm$0.2 & 5.0\\ 
2015-05-03 & IRDIS & H3 & 208.0$\pm$4.9 &156.7$\pm$1.2 &9.6 $\pm$0.2 & 5.1\\
2015-05-31 & IRDIS & H2 & 196.4$\pm$4.4 &155.7$\pm$1.1 &9.1 $\pm$0.1 &14.2\\
2015-05-31 & IRDIS & H3 & 197.0$\pm$6.0 &155.5$\pm$1.6 &8.9 $\pm$0.1 &14.7\\
2016-05-14 & IRDIS & K1 & 198.5$\pm$3.7 &152.5$\pm$1.1 &8.1 $\pm$0.1 &17.1\\
2016-05-14 & IRDIS & K2 & 200.0$\pm$3.0 &152.6$\pm$0.9 &7.5 $\pm$0.1 &17.5\\
2016-06-01 & NaCo  & L' & 181.8$\pm$18.8&148.4$\pm$5.9 &6.9 $\pm$0.5 & 2.4\\
\hline
Results from TLOCI \\
\hline
2012-03-31 & NICI  & L' & 187.7$\pm$35.9&160.5$\pm$10.9&7.1$\pm$0.6& 1.9\\
2015-05-03 & IRDIS & H2 & 198.1$\pm$26.2&154.1$\pm$ 7.5&9.4$\pm$0.8& 1.4\\ 
2015-05-03 & IRDIS & H3 & 195.3$\pm$20.6&154.9$\pm$ 6.0&9.2$\pm$0.6& 1.9\\
2015-05-31 & IRDIS & H2 & 196.5$\pm$14.1&154.9$\pm$ 4.1&9.6$\pm$0.3& 3.4\\
2015-05-31 & IRDIS & H3 & 199.9$\pm$15.4&154.8$\pm$ 4.4&9.5$\pm$0.4& 2.7\\
2016-05-14 & IRDIS & K1 & 192.0$\pm$24.2&151.0$\pm$ 7.2&8.1$\pm$0.5& 2.1\\
2016-05-14 & IRDIS & K2 & 201.0$\pm$27.2&152.2$\pm$ 7.7&7.9$\pm$0.6& 1.8\\
2016-06-01 & NaCo  & L' & 181.7$\pm$54.4&147.8$\pm$17.2&7.1$\pm$1.3& 0.9\\
\hline
\end{tabular}
\tablefoot{The current implementation of the astrometric error estimation for our TLOCI reduction is unreliable for the L' datasets. We therefore used a conservative uncertainty of 2 pixels for the angular separation and the position angle for these datasets. }
\label{app:cc_alt_reductions}
\end{table*}

\begin{figure*}[hbt]
    \noindent
    \centering
    \begin{minipage}[t]{1.0\textwidth}
    \includegraphics[width=1.0\textwidth]{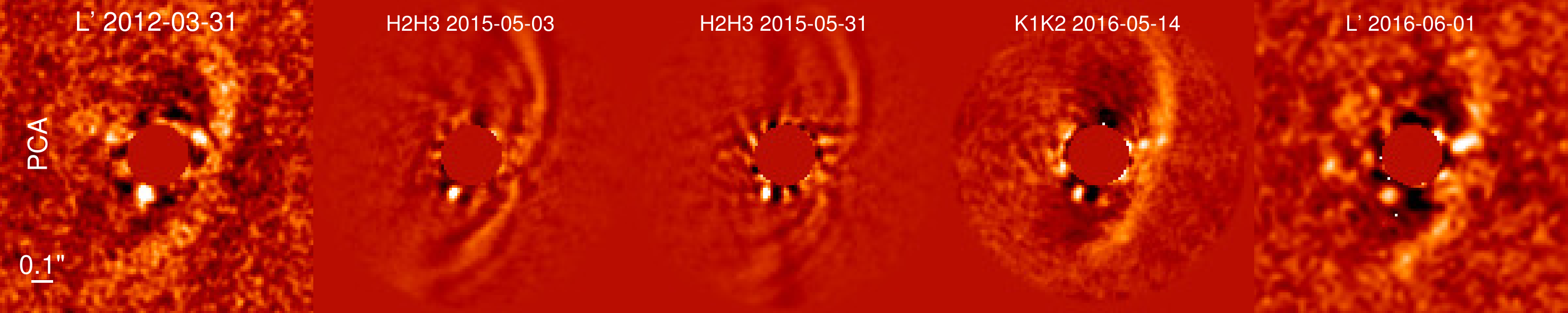}
    \includegraphics[width=1.0\textwidth]{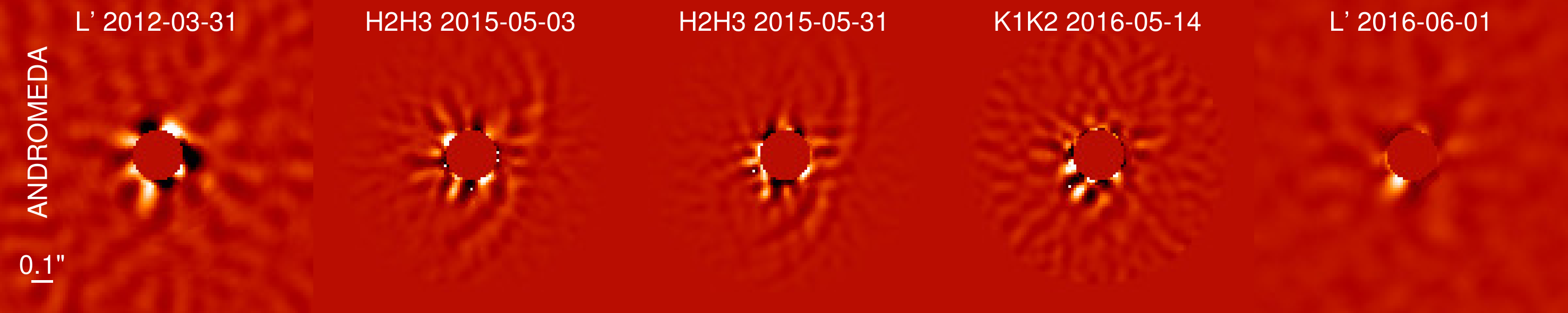}
    \includegraphics[width=1.0\textwidth]{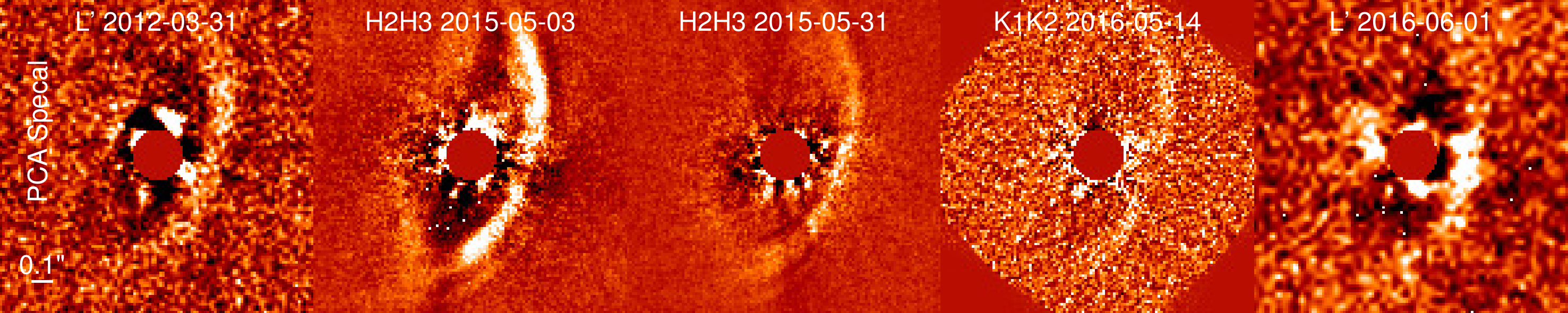}
    \includegraphics[width=1.0\textwidth]{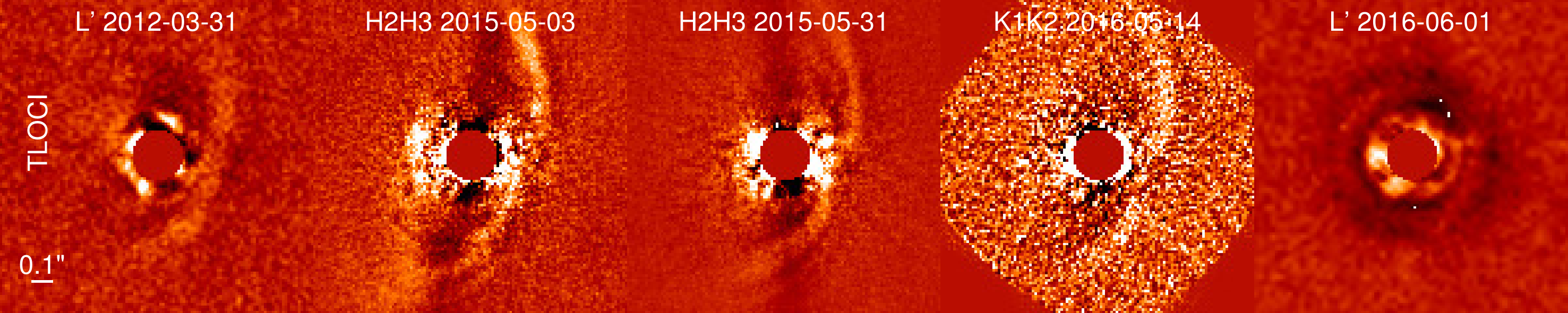}
    \end{minipage}
    \begin{minipage}[t]{1.0\textwidth}
    \end{minipage}
    \caption{Contrast maps of the point source detection as retrieved with the sPCA reduction (first row), ANDROMEDA (second row), PCA SpeCal (third row), and TLOCI (fourth row). From left to right: NICI L'-band (2012-03-31), IRDIS H2H3-band (2015-05-03), IRDIS H2H3-band (2015-05-31), IRDIS K1K2-band (2016-05-14), NaCo L'-band (2016-06-01). The sPCA images were smoothed with a Gaussian kernel of size 0.5$\times$FWHM. North is up and east is to the left. 
The brightness levels were adapted individually for visibility purposes. 
}\label{obs:CC_algos}
\end{figure*}

\begin{figure*}[t]
    \begin{minipage}[tbh]{1.0\textwidth}
    \centering
    \includegraphics[width=1.00\textwidth]{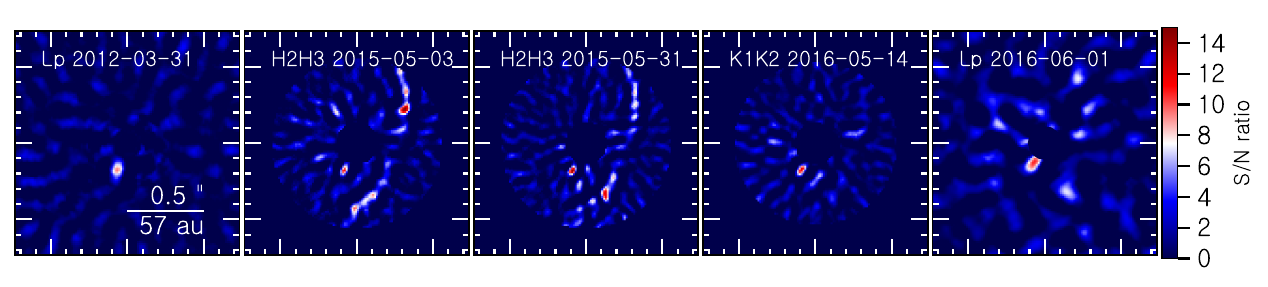}
    \end{minipage}
   \caption{ANDROMEDA S/N maps of the ADI epochs. }\label{fig:ANROMEDA_SNR}
\end{figure*}

\clearpage

\newpage
\newpage

\subsection{Stellar parameters of PDS~70 }\label{stellar_parameters}
\begin{figure}[h]
    \noindent
    \centering
    \begin{minipage}{0.5\textwidth}
    \includegraphics[width = 1.0\textwidth]{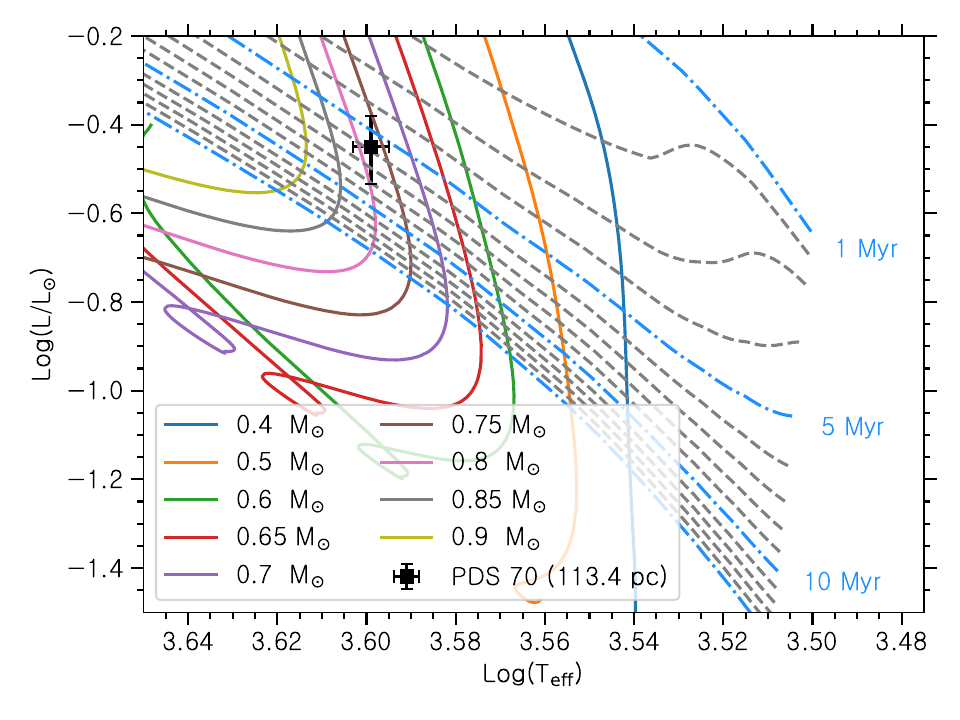}
    \end{minipage}      
    \caption{The location of PDS~70 on a HR diagram in comparison with the PMS evolutionary tracks and isochrones from \cite{2011A&A...533A.109T}. }\label{Tognelli11}
\end{figure}

PDS~70 is located at a distance of $113.43\pm0.52$ pc \citep{2018arXiv180409365G}. 
Using low-resolution optical spectra, \cite{2016MNRAS.461..794P} derived a stellar temperature of 3972 K, corresponding to a K7 spectral type. Assuming a distance of 98.9 pc, these authors derived a luminosity of 0.27 \Lsol \, which, scaled to a distance of 113.4 pc, corresponds to a luminosity of 0.35 \Lsol. \newline
We compare the position of PDS~70 with the PMS evolutionary tracks and isochrones from \cite{2011A&A...533A.109T} in Fig. \ref{Tognelli11} (assuming a metallicity Z~=~0.02, an initial helium abundance of 0.27, a mixing length of 1.68 and a deuterium abundance of $2\times10^{-5}$), which implies a stellar mass of 0.7-0.85 \Msol, and an age of $6\pm2$ Myr. As an additional approach, we fitted the stellar evolutionary models from the MIST project \citep{2016ApJS..222....8D,2016ApJ...823..102C,2011ApJS..192....3P,2013ApJS..208....4P,2015ApJS..220...15P} using a Markov Chain Monte Carlo approach to determine the stellar parameters. This method is described in detail in \citep{mueller2018}. This approach resulted in a stellar age of 5.4 $\pm$1.0 Myr and a mass of 0.76$\pm$0.02 \Msol, which is consistent with the above given values. It is worth noting that our mass estimates are in good agreement with the dynamical mass of PDS~70 \cite[0.6-0.8 \Msol; ][]{2015ApJ...799...43H,2018ApJ...858..112L}.\\

\newpage

\subsection{IRDIS ADI view of the disk}

\begin{figure*}[t]
    \begin{minipage}[tbh]{1.0\textwidth}
    \centering
    \includegraphics[width=0.31\textwidth]{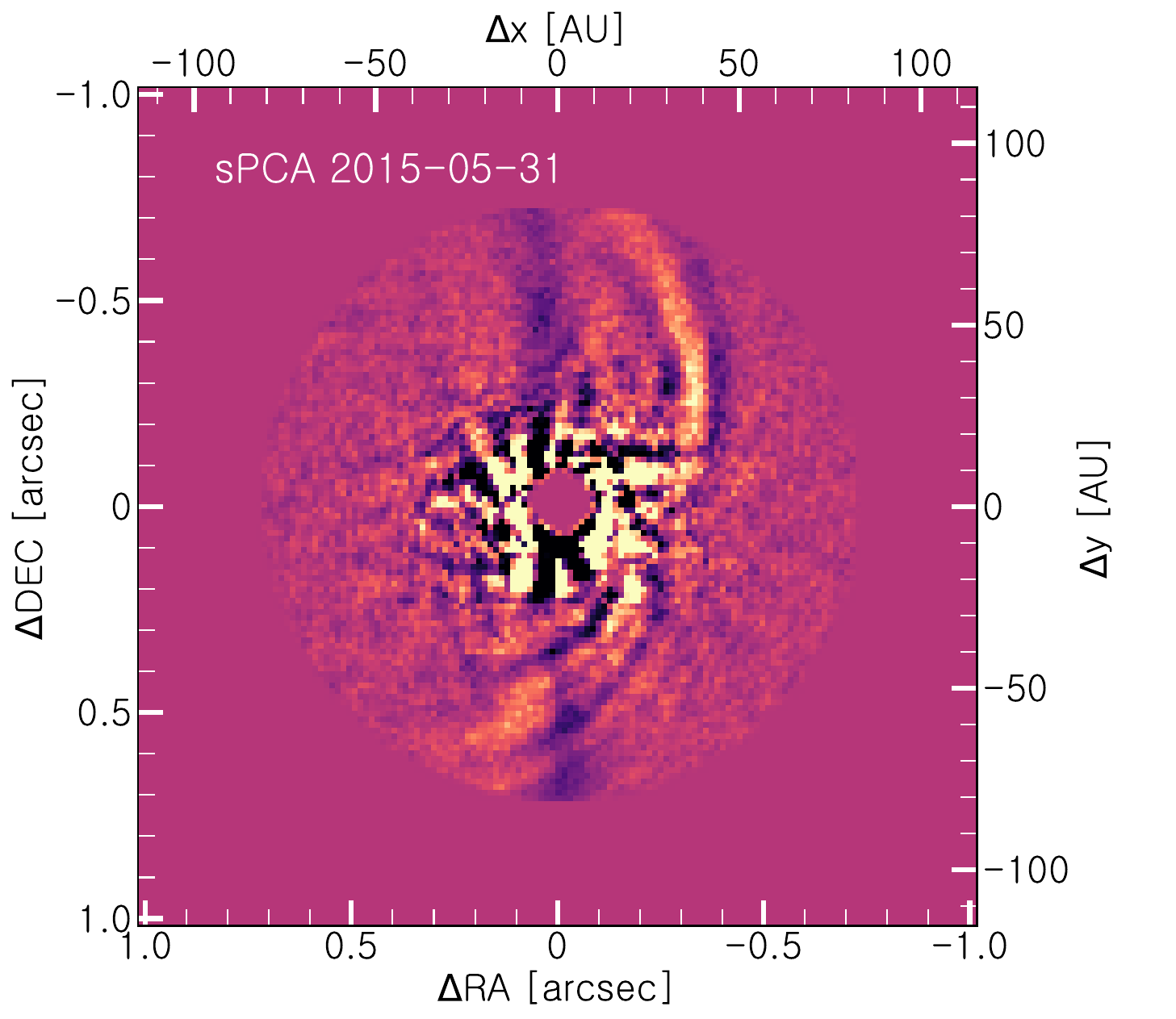}
    \includegraphics[width=0.31\textwidth]{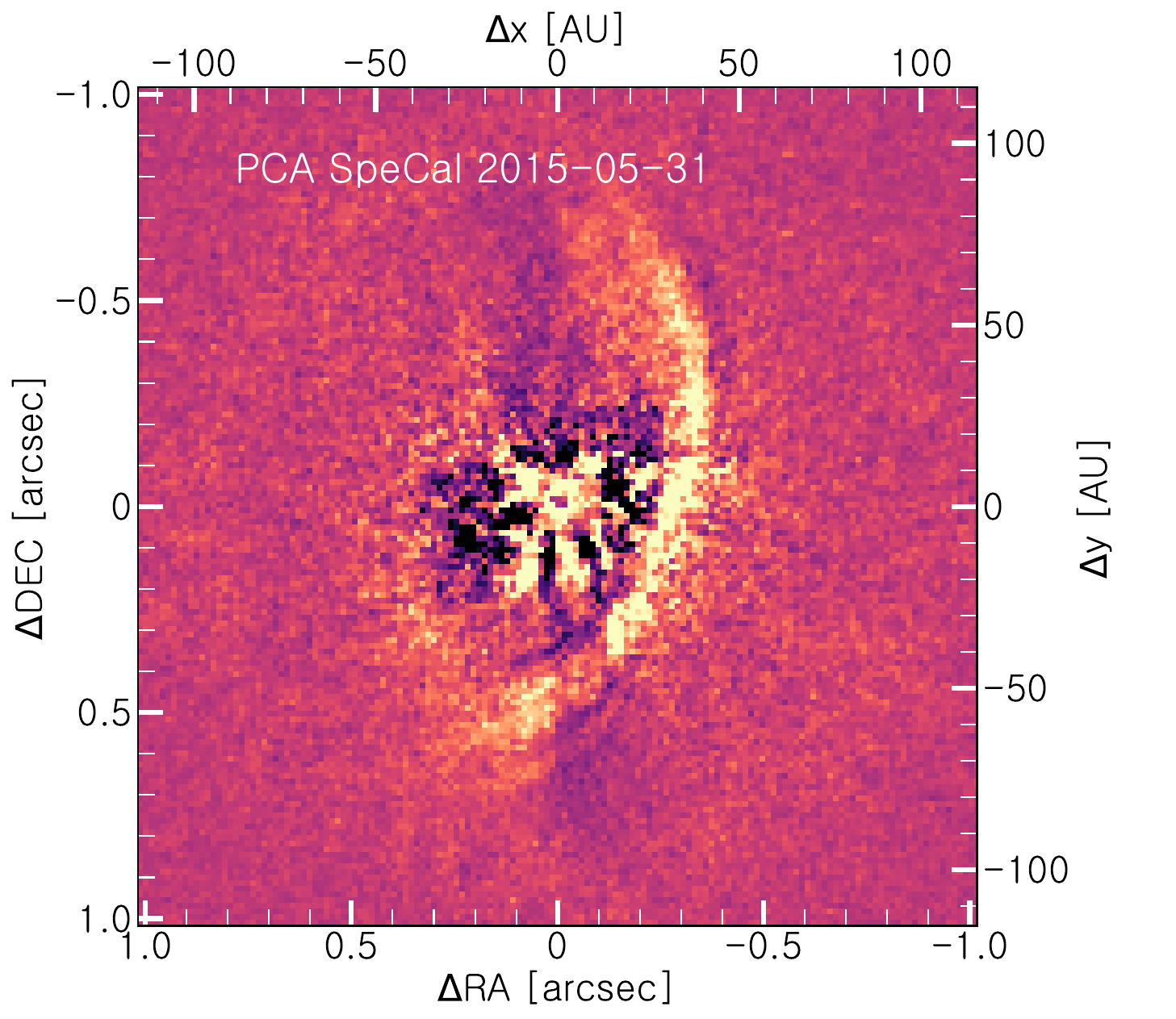}
    \includegraphics[width=0.35\textwidth]{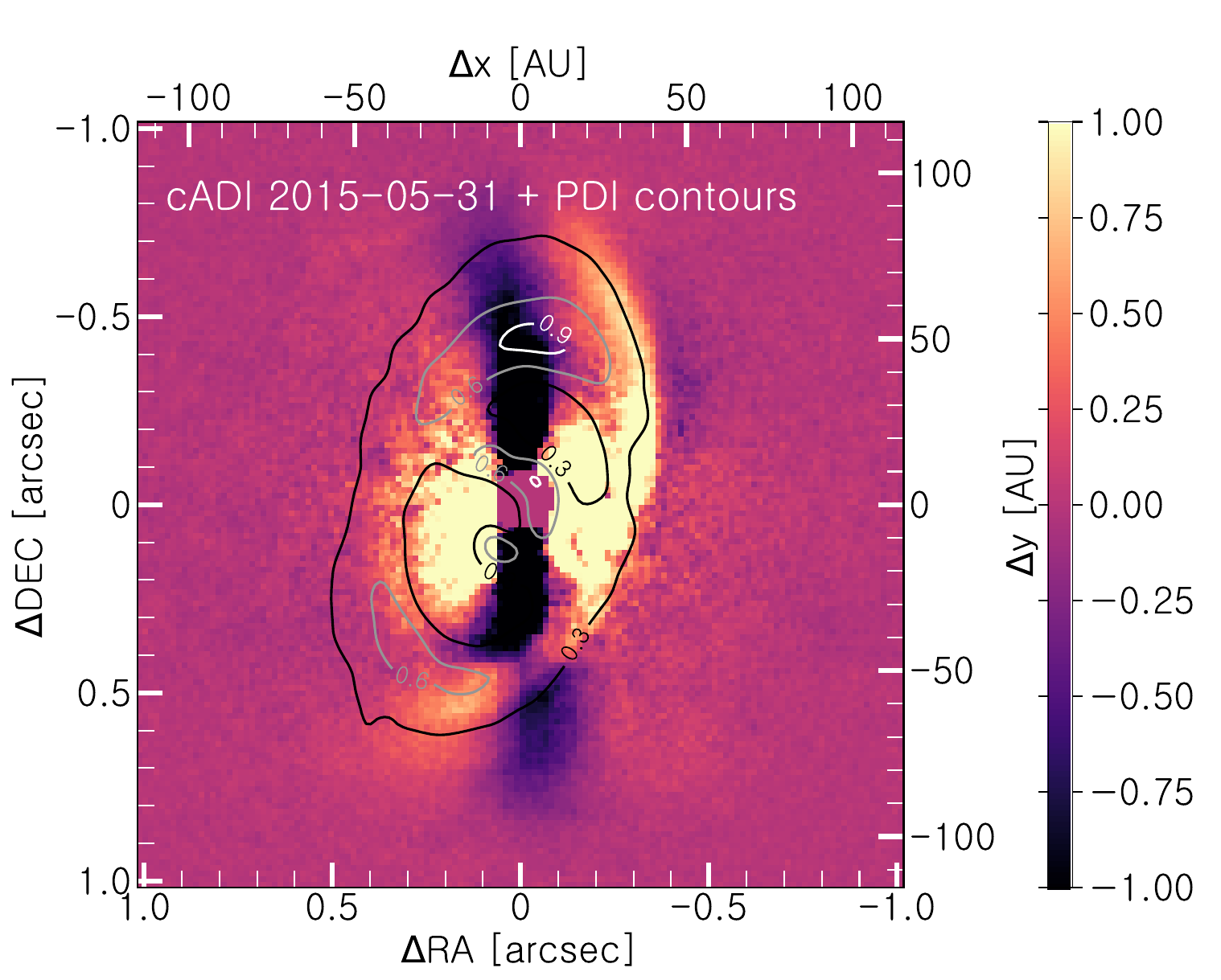}
    \end{minipage}
   \caption{SHINE IRDIS observations of May 31, 2015: sPCA reduction (left), PCA-SpeCal reduction (middle), and cADI with an contour overlay of the PDI coronagraphic J-band image (right). The contours are drawn with respect to the peak value of the PDI image. For visibility purposes, the images are shown on individual colourscales. North is up and east is to the left. }\label{fig:obs_TLOCI}
\end{figure*}


\end{document}